\documentclass[12pt]{article}
\pdfoutput=1
\usepackage{yfonts}
\usepackage{color}
\usepackage{mhchem}
\usepackage{xcolor}
\usepackage{cite}
\usepackage{hyperref}
\hypersetup{colorlinks=true,linkcolor=red,anchorcolor=black,citecolor=green}
\usepackage[toc,page]{appendix}
\usepackage{amsfonts}
\usepackage{bbold}
\usepackage{textcomp}
\usepackage[DIV13]{typearea}
\usepackage{amsmath, amsthm, amssymb, mathtools,empheq,latexsym,dsfont}
\usepackage{bbm}
\usepackage{slashed, simplewick}
\usepackage[T1]{fontenc}
\usepackage[utf8]{inputenc}
\usepackage{graphicx,placeins}
\usepackage{makeidx}
\usepackage[font=small,labelfont=bf]{caption}
\usepackage{nicefrac}
\usepackage{subfigure}
\usepackage{array, bigdelim,multirow,multicol}
\usepackage[integrals]{wasysym}
\usepackage{fancybox}
\usepackage{bm}
\usepackage{float}
\usepackage{rotating}
\usepackage{colortbl}
\usepackage{makecell}
\usepackage{pifont}
\usepackage{longtable}
\usepackage{booktabs}
\usepackage[top=2cm,textwidth=16.6cm,textheight=22.75cm]{geometry}
\usepackage{doi}
\usepackage{fancybox}
\usepackage[compat=1.0.0]{tikz-feynman}
\usepackage{tikz}
\usepackage{empheq}
\allowdisplaybreaks
\newcommand{\ignore}[1]{}
\usepackage{enumerate}
\usepackage{gensymb}
\usepackage{framed}

\usepackage{braket}

\graphicspath{{immagini/}}

\definecolor{Gray}{gray}{0.92}

\DeclareMathAlphabet{\mathscr}{OT1}{pzc}{m}{it}

\renewcommand{\arg}{{\rm Arg}}

\newcommand{\be}{\begin{equation}}
\newcommand{\ee}{\end{equation}}
\newcommand{\bea}{\begin{eqnarray}}
\newcommand{\eea}{\end{eqnarray}}

\makeatletter
\@addtoreset{equation}{section}
\makeatother

\begin{document}

\unitlength = 1mm
\setlength{\extrarowheight}{0.2 cm}
\thispagestyle{empty}
\bigskip
\vskip 1cm

\title{\Large \bf
$\mathbb{Z}_3$ and $S_3$ gaugings of $\mathbb{Z}_N\times\mathbb{Z}'_N$ non-invertible symmetries: quark texture zeros and phenomenology
\\[2mm] }

\date{}

\author{Zheng Jiang$^{a}$\footnote{E-mail: {\tt
jzdyx@mail.ustc.edu.cn}},\
Bu-Yao Qu$^{a}$\footnote{E-mail: {\tt
qubuyao@mail.ustc.edu.cn}},  \
Gui-Jun Ding$^{a,b}$\footnote{E-mail: {\tt
dinggj@ustc.edu.cn}}
\\*[20pt]
\centerline{
\begin{minipage}{\linewidth}
\begin{center}
$^a${\it \small Department of Modern Physics,  and Anhui Center for fundamental sciences in theoretical physics,\\
University of Science and Technology of China, Hefei, Anhui 230026, China}\\[2mm]
$^b${\it \small College of Physics, Guizhou University, Guiyang 550025, China}
\end{center}
\end{minipage}}
\\[10mm]}
\maketitle
\thispagestyle{empty}

\centerline{\large\bf Abstract}

\begin{quote}
\indent

We investigate texture zeros in the quark Yukawa matrices arising from the $\mathbb{Z}_3$ and $S_3$ gauging of $\mathbb{Z}_N\times\mathbb{Z}'_N$, associated with the non-invertible $\Delta(3N^2)$ and $\Delta(6N^2)$ symmetries, respectively. We construct the equivalence classes and their fusion rules and use them as selection rules for the Yukawa interactions. We systematically classify the full-rank quark Yukawa textures generated by the two gaugings for $N=2,3,4,$ and $5$. The two gaugings yield the same texture spectrum for $N=2$ and $N=4$, whereas the $S_3$ gauging produces additional textures for $N=3$ and $N=5$. We further analyze the most restrictive texture pairs compatible with quark data and find that there are at most nine zeros in these texture pairs. For up- and down-quark mass matrices with a total of nine texture zeros, one sector is allowed to contain six zeros while the other contains three, or one contains five zeros while the other contains four. The former case has been studied in previous works. We perform numerical fits for the latter and find it consistent with experimental data. Some viable textures also admit real quark-mass determinants while retaining a physical CKM phase, suggesting a possible connection to the strong CP problem.

\end{quote}

\clearpage

\section{Introduction}

Despite the remarkable success of the Standard Model (SM) in describing elementary-particle interactions, the origin of the observed fermion flavor structure remains one of its most profound open questions. Quark masses span several orders of magnitude from the lightest up quark mass $m_u\approx2.16$ MeV to the heavist top quark $m_t\approx172.60$ GeV, while the Cabibbo–Kobayashi–Maskawa (CKM) matrix exhibits a similarly pronounced hierarchy and contains a single physical CP-violating phase~\cite{ParticleDataGroup:2026mpi}. These features originate in the Yukawa sector, whose couplings are free parameters of the SM and therefore lack a fundamental explanation. Elucidating the structure and origin of the Yukawa matrices is thus essential for understanding the flavor structure of SM and may offer valuable clues to physics beyond the SM.

Among the various attempts to explain fermion masses and flavor mixing, the texture-zero approach has provided an economical and predictive framework. By imposing vanishing entries in fermion mass matrices, i.e., texture zeros, the number of independent Yukawa parameters is reduced and nontrivial correlations among fermion masses, mixing angles, and CP-violating observables are established. The texture-zero framework originated from the pioneering studies in Refs.~\cite{Weinberg:1977hb,Wilczek:1977uh,Fritzsch:1977za} and was subsequently developed for three fermion generations in Refs.~\cite{Fritzsch:1977vd,Fritzsch:1979zq}. It has since been extensively explored in both the quark and lepton sectors, where it provides a predictive description of flavor phenomena~\cite{Fritzsch:1999ee,Gupta:2012fsl}.

From a theoretical standpoint, texture zeros should arise from an underlying symmetry rather than be imposed solely as phenomenological assumptions. A variety of mechanisms, including Abelian flavor symmetries~\cite{Grimus:2004hf,GonzalezFelipe:2014zjk}, non-Abelian flavor symmetries~\cite{Frampton:1995fta,Mondragon:1998gy,Haba:2005ds} and modular flavor symmetries~\cite{Zhang:2019ngf,Lu:2019vgm,Ding:2022aoe,Kikuchi:2022svo}, have been proposed to produce texture-zero structures by forbidding selected Yukawa interactions, see Refs.~\cite{Feruglio:2019ybq,Xing:2020ijf,Ding:2023htn,Ding:2024ozt} for recent review. The systematic classification and phenomenological analysis of texture-zero patterns have further shown that predictive fermion mass matrices can reproduce the observed quark masses, CKM mixing parameters, and CP violation~\cite{Fritzsch:1999ee,Gupta:2012fsl,Ludl:2015lta}. These advances reinforce the significance of texture zeros as a tool for elucidating the origin of fermion flavor structures and identifying possible organizing principles of the Yukawa sector.

In recent years, there has been a revolution in our understanding of symmetries. The notions of global symmetries have been generalized in different directions~\cite{Gaiotto:2014kfa}. The non-invertible symmetries have recently opened up a new avenue for particle-physics model building,  extending the conventional symmetry paradigm through selection rules that differ qualitatively from those associated with ordinary group symmetries. For recent reviews of non-invertible symmetries, see Refs.~\cite{Schafer-Nameki:2023jdn,Brennan:2023mmt,Bhardwaj:2023kri,Shao:2023gho}. Their implications for flavor physics have consequently attracted growing interest. In particular, non-invertible symmetries can realize several texture-zero structures in the quark and lepton sectors that cannot be generated by ordinary flavor symmetries~\cite{Kobayashi:2024cvp,Kobayashi:2025znw,Kobayashi:2025ldi,Chen:2025awz,Jiang:2025psz,Qu:2026omn}, such as the nearest-neighbor interaction texture~\cite{Kobayashi:2024cvp,Kobayashi:2025znw} commonly discussed in the quark sector, and the two-zero texture~\cite{Jiang:2025psz,Qu:2026omn} discussed in the lepton sector, thereby broadening the scope of flavor-model building. These textures can accommodate realistic quark masses, CKM mixing, and CP violation; moreover, certain patterns may furnish an axionless solution to the strong CP problem~\cite{Liang:2025dkm,Kobayashi:2025thd,Kobayashi:2025rpx}. Non-invertible symmetries therefore provide a systematic and predictive route to texture-zero Yukawa matrices and constitute a promising alternative to conventional flavor symmetries for addressing the flavor puzzle.

The automorphism group of the direct product $\mathbb{Z}_N \times \mathbb{Z}'_N$ is isomorphic to the general linear group $GL(2, \mathbb{Z}_N)$, which is the group of two-by-two invertible matrices with entries that are integers modulo $N$. For each $N>1$, the group $GL(2, \mathbb{Z}_N)$ is non-abelian, and it has finite number of elements. The group order is equal to $\left|GL(2, \mathbb{Z}_N)\right|= N^3 \prod_{p\vert N}\left(1-\frac{1}{p^2}\right)$, where the product ranges over the prime divisors $p$ of $N$. The automorphism group $GL(2, \mathbb{Z}_N)$ always has $\mathbb{Z}_3$ and $S_3$ subgroups for $N>1$. The semidirect product of $\mathbb{Z}_3$ and $S_{3}$ automorphism subgroups with $\mathbb{Z}_N \times \mathbb{Z}'_N$ can give rise to the group series $\Delta(3N^2)\cong(\mathbb{Z}_N\times\mathbb{Z}_N)\rtimes\mathbb{Z}_3$~\cite{Luhn:2007uq} and $\Delta(6N^2)\cong(\mathbb{Z}_N\times\mathbb{Z}_N)\rtimes S_3$~\cite{Escobar:2008vc} respectively, which have been extensively studied as flavor symmetries in the literature~\cite{King:2014rwa,Hagedorn:2014wha,Ding:2014ora,Ding:2015rwa}. In this work, we will investigate the $\mathbb{Z}_3$ and $S_3$ gaugings of the $\mathbb{Z}_N\times\mathbb{Z}'_N$ symmetries. We employ the corresponding fusion rules to construct texture structures for the up- and down-type quark Yukawa matrices within the framework of SM. We then perform a numerical analysis to assess the compatibility of the resulting textures with the observed quark mass spectrum, CKM mixing parameters, and CP-violating observables.

The remaining part of this paper is organized as follows.
In section~\ref{sec:Z3-gauging-ZNxZN}, We identify the resulting equivalence classes and establish their fusion rules, and distinguish between the cases where $N$ is and is not divisible by three. In section~\ref{sec:Z3-gauging-ZNXZN-N=2345} we apply these fusion rules as selection rules for SM quark Yukawa interactions. For $N=2,3,4,$ and $5$, we systematically examine the possible assignments of quark and Higgs fields and classify the resulting texture-zero structures of the up- and down-type quark Yukawa matrices. In section~\ref{sec:S3-gauging-ZNxZN}, we extend the construction to the $S_3$ gauging of $\mathbb{Z}_N\times\mathbb{Z}'_N$. We derive the associated equivalence classes and fusion rules and determine the quark Yukawa textures generated for $N=2,3,4,$ and $5$, highlighting their relation to and differences from the $\mathbb{Z}_3$-gauged cases. We then investigate the phenomenological viability of the resulting textures in section~\ref{sec:pheno-analysis}. We analyze the cases with six texture zeros in one quark sector and three in the other, as well as those with five zeros in one sector and four in the other, and confront their predictions with the observed quark masses, CKM mixing parameters, and CP violation. Finally, we summarize our main results and conclusions. In appendix~\ref{app:10-zeros} we provide a detailed classification of quark Yukawa textures with ten zeros and show that these textures containing ten zeros in total are incompatible with current data, so that nine is the maximum viable number of texture zeros. In appendix~\ref{app:strong-cp}, we present parameterizations of the viable nine-zero textures satisfying $\arg\det(Y_uY_d)=0$. These textures can accommodate the measured quark masses and mixing parameters, and may provide a potential solution to the strong CP problem.

\section{\label{sec:Z3-gauging-ZNxZN}$\mathbb{Z}_3$ gauging of $\mathbb{Z}_N\times \mathbb{Z}_N'$ symmetry}

We firstly consider the $\mathbb{Z}_3$ gauging of $\mathbb{Z}_N\times \mathbb{Z}_N'$ symmetry~\cite{Dong:2025jra}. The $\mathbb{Z}_N\times \mathbb{Z}_N'$ symmetry can be generated by their generators $a$ and $a'$ of order $N$, which are commute with each other, namely:
\begin{eqnarray}
\label{eq:ZNxZN-gener}a^N=a'^N=e \,,\quad aa'=a'a \,,
\end{eqnarray}
where $e$ refers to the identity element. The $\mathbb{Z}_3$ is a subgroup of the automorphism group of $\mathbb{Z}_N\times \mathbb{Z}_N'$, we denote its generator by $b$ acting on $a$ and $a'$ as follows,
\begin{eqnarray}
bab^{-1}=a^{-1}a'^{-1} \,,\quad ba'b^{-1}=a \,,\quad  b^3=e \,,
\end{eqnarray}
which implies
\begin{eqnarray}
b\big(a^ka'^\ell\big)b^{-1}&=&a^{-k+\ell}a'^{-k} \,,~~~~
b^2\big(a^ka'^\ell\big)b^{-2}=a^{-\ell}a'^{k-\ell} \,,
\end{eqnarray}
with $k, \ell=0,1, \ldots, N-1$. Thus the three elements $a^ka'^\ell$, $a^{-k+\ell}a'^{-k}$ and $a^{-\ell}a'^{k-\ell}$ are identified by the gauge symmetry $\mathbb{Z}_3$, and they constitute an equivalence class,
\begin{eqnarray}
[k,\ell]_{\mathbb{Z}_3}&=&\{a^ka'^\ell \,,\, a^{-k+\ell}a'^{-k} \,,\, a^{-\ell}a'^{k-\ell}\} \,.\label{eq:eleinckl}
\end{eqnarray}
Since the generators $a$ and $a'$ commute, we can straightforwardly obtain the fusion rules satisfied by these equivalence classes:
\begin{eqnarray}
[k,\ell]_{\mathbb{Z}_3}\cdot [m,n]_{\mathbb{Z}_3}&=&[k+m,\ell+n]_{\mathbb{Z}_3}+[k-m+n,\ell-m]_{\mathbb{Z}_3}+[k-n,\ell-n+m]_{\mathbb{Z}_3} \,.\label{eq:fr-z3gznxzn}
\end{eqnarray}
It is obvious that the above fusion rules are commutable, i.e.
\begin{equation}
[k,\ell]_{\mathbb{Z}_3}\cdot [m,n]_{\mathbb{Z}_3}=[m,n]_{\mathbb{Z}_3}\cdot [k,\ell]_{\mathbb{Z}_3} \,.
\end{equation}

There are several noteworthy points regarding the equivalence classes and their fusion rules. First, we see that the fusion rule in Eq.~\eqref{eq:fr-z3gznxzn} is invariant under the automorphisms $a\leftrightarrow a'$ and $a\leftrightarrow a'^{-1}$ which map the class $[k,\ell]_{\mathbb{Z}_3}$ into $[\ell,k]_{\mathbb{Z}_3}$ and $[-\ell,-k]_{\mathbb{Z}_3}$ respectively. In fact, as long as an automorphism maps each non-$[0,0]_{\mathbb{Z}_3}$ class entirely onto another non-$[0,0]_{\mathbb{Z}_3}$ class, the fusion rules remain invariant.

Although these equivalence classes do not possess inverses in the group-theoretic sense, we may relax this requirement and obtain inverses in a more generalized sense. If the produce of $[k,\ell]_{\mathbb{Z}_3}\cdot [m,n]_{\mathbb{Z}_3}$ includes $[0,0]_{\mathbb{Z}_3}$, we say that $[k,\ell]_{\mathbb{Z}_3}$ and $[m,n]_{\mathbb{Z}_3}$ are inverses of each other. For any class $[k,\ell]_{\mathbb{Z}_3}$, if $[k,\ell]_{\mathbb{Z}_3}\cdot[m,n]_{\mathbb{Z}_3}\supset[0,0]_{\mathbb{Z}_3}$, from Eq.~\eqref{eq:fr-z3gznxzn} we have $(m,n)=(-k,-\ell)$, $(\ell,\ell-k)$ or $(k-\ell,k)$. However, the three classes $[-k,-\ell]_{\mathbb{Z}_3}$, $[\ell,\ell-k]_{\mathbb{Z}_3}$ and $[k-\ell,k]_{\mathbb{Z}_3}$ are esentially the same class, as can be seen from Eq.~\eqref{eq:eleinckl}. Therefore, there exists exactly one inverse class for any class.

In general, the class $[k,\ell]_{\mathbb{Z}_3}$ contains three different elements, as shown in Eq.~\eqref{eq:eleinckl}, but obviously for $k=\ell=0$, the class $[0,0]_{\mathbb{Z}_3}$ contains only one element $a^0a'^0=e$. In addition, the class $[k,\ell]_{\mathbb{Z}_3}$ can collapse to only one element if the indices $k$ and $\ell$ satisfy the following conditions,
\begin{eqnarray}
 k+\ell&=&0 \pmod N\,,~~~~~ 3k=0 \pmod N \,,
\end{eqnarray}
which requires $N$ should be divisible by $3$ with
\begin{eqnarray}
(k,\ell)=\left(\frac{N}{3},\frac{2N}{3}\right) \;\text{or}\; \left(\frac{2N}{3},\frac{N}{3}\right) \,.
\end{eqnarray}
Therefore the equivalence classes of the $\mathbb{Z}_3$ gauging of $\mathbb{Z}_N\times \mathbb{Z}_N'$ symmetry depend on the values of $N$.

When $N$ is not divisible by 3, only the $[0,0]_{\mathbb{Z}_3}$ class contains only one element. All the equivalence classes are divided into two categories:
\begin{eqnarray}
\nonumber    [0,0]_{\mathbb{Z}_3}&=&\{e\} \,,\\
{}[k,\ell]_{\mathbb{Z}_3}&=&\{a^ka'^\ell \,,\, a^{-k+\ell}a'^{-k} \,,\, a^{-\ell}a'^{k-\ell}\} \,, \label{eq:class-Nneq3}
\end{eqnarray}
with $(k,\ell)\neq(0,0)$, and they obey the fusion rules,
\begin{eqnarray}
\nonumber [0,0]_{\mathbb{Z}_3}\cdot [0,0]_{\mathbb{Z}_3}&=&[0,0]_{\mathbb{Z}_3} \,,~~~~ {}[0,0]_{\mathbb{Z}_3}\cdot [k,\ell]_{\mathbb{Z}_3}=[k,\ell]_{\mathbb{Z}_3} \,,\\
\label{eq:fusion-Nneq3}{}[k,\ell]_{\mathbb{Z}_3}\cdot [m,n]_{\mathbb{Z}_3}&=&[k+m,\ell+n]_{\mathbb{Z}_3}+[k-m+n,\ell-m]_{\mathbb{Z}_3}+[k-n,\ell+m-n]_{\mathbb{Z}_3} \,,~~~~\quad ~~~
\end{eqnarray}
with $(k,\ell)\neq(0,0)$ and $(m,n)\neq(0,0)$. Note that if any ones of $(k+m,\ell+n)$, $(k-m+n,\ell-m)$, $(k-n,\ell+m-n)$ are equal to $(0,0)$, $[k+m,\ell+n]_{\mathbb{Z}_3}$, $[k-m+n,\ell-m]_{\mathbb{Z}_3}$ and $[k-n,\ell+m-n]_{\mathbb{Z}_3}$ in Eq.~\eqref{eq:fusion-Nneq3} should be replaced by $3[0,0]_{\mathbb{Z}_3}$.

When $N$ is divisible by 3, there are three classes each containing one element. All the equivalence classes are:
\begin{eqnarray}
&&[r,-r]_{\mathbb{Z}_3}=\{a^{r}a'^{-r}\} \,,\nonumber\\
&&[k,\ell]_{\mathbb{Z}_3}=\{a^ka'^\ell \,,\, a^{-k+\ell}a'^{-k} \,,\, a^{-\ell}a'^{k-\ell}\} \,, \label{eq:class3N}
\end{eqnarray}
with $r=0,N/3,2N/3$ and $(k,\ell)\neq(0,0),(N/3,2N/3),(2N/3,N/3)$. The corresponding fusion rules are given by
\begin{eqnarray}
    &&[r,-r]_{\mathbb{Z}_3}\cdot [s,-s]_{\mathbb{Z}_3}=[r+s,-r-s]_{\mathbb{Z}_3} \,,\nonumber\\
    &&[r,-r]_{\mathbb{Z}_3}\cdot [k,\ell]_{\mathbb{Z}_3}=[r+k,\,-r+\ell]_{\mathbb{Z}_3} \,,\nonumber\\
    {}[k,\ell]_{\mathbb{Z}_3}\cdot [m,n]_{\mathbb{Z}_3}=&&[k+m,\ell+n]_{\mathbb{Z}_3}+[k-m+n,\ell-m]_{\mathbb{Z}_3}+[k-n,\ell+m-n]_{\mathbb{Z}_3} \,,~~\quad ~~\label{eq:fr3N}
\end{eqnarray}
with $s=0,N/3,2N/3$ and $(m,n)\neq(0,0),(N/3,2N/3),(2N/3,N/3)$. If $(k+m,\ell+n)$, $(k-m+n,\ell-m)$, $(k-n,\ell+m-n)$ are equal to $(0,0)$, $(N/3,2N/3)$ or $(2N/3,N/3)$, the classes $[k+m,\ell+n]_{\mathbb{Z}_3}$, $[k-m+n,\ell-m]_{\mathbb{Z}_3}$, $[k-n,\ell+m-n]_{\mathbb{Z}_3}$ in Eq.~\eqref{eq:fr3N} should be replaced by $3[0,0]_{\mathbb{Z}_3}$, $3[N/3,2N/3]_{\mathbb{Z}_3}$ or $3[2N/3,N/3]_{\mathbb{Z}_3}$ respectively.

Similar to $\mathbb{Z}_2$ gauging of $\mathbb{Z}_N$ symmetry, the $\mathbb{Z}_3$ gauging of $\mathbb{Z}_N \times \mathbb{Z}'_{N}$ symmetry could be realized by higher dimensional field theory and string theory. Certain compactification scheme can lead to $\mathbb{Z}_N \times \mathbb{Z}'_{N}$. Under such symmetry, the massless modes $\varphi_{k,\ell}$ transform as
\begin{eqnarray}
    \varphi_{k,\ell} \rightarrow a^{k}\varphi_{k,\ell}
\end{eqnarray}
under $\mathbb{Z}_N$, and
\begin{eqnarray}
    \varphi_{k,\ell} \rightarrow a'^{\ell}\varphi_{k,\ell}
\end{eqnarray}
under $\mathbb{Z}'_{N}$. Here, $k$ and $\ell$ may correspond to two independent discrete momenta, winding numbers, and their mixtures in the compact space. Here, we consider a proper $\mathbb{Z}_3$ orbifolding such that $\mathbb{Z}_3$-invariant mode is written by
\begin{eqnarray}
    \phi_{k,\ell}=\varphi_{k,\ell}+\varphi_{-k+\ell,-k}+\varphi_{-\ell,k-\ell} \,.
\end{eqnarray}
That corresponds to the class $[k,\ell]_{\mathbb{Z}_3}$.

Under the $\mathbb{Z}_3$ gauging of $\mathbb{Z}_N \times \mathbb{Z}'_{N}$ symmetry, the field $\phi_{k,\ell}$ corresponds to the class $[k,\ell]_{\mathbb{Z}_3}$. We say an $n$-point coupling $\prod\phi_{k,\ell}$  is allowed to appear in the Lagrangian if the condition $\prod[k,\ell]_{\mathbb{Z}_3}\supset[0,0]_{\mathbb{Z}_3}$ is satisfied. Due to the presence of mass terms in the Lagrangian, the Hermitian conjugate field $\phi^\dagger_{k,\ell}$ of $\phi_{k,\ell}$ necessarily corresponds to the class $[-k,-\ell]_{\mathbb{Z}_3}$. In addition, the two-point self coupling of a field $\phi_{k,\ell}$ for $(k,\ell)\neq(0, 0)$ is always forbidden in the Lagrangian but the three-point self coupling is always allowed, because
\begin{eqnarray}
    [k,\ell]_{\mathbb{Z}_3}\cdot [k,\ell]_{\mathbb{Z}_3}\stackrel{\eqref{eq:fr-z3gznxzn}}{=}[2k,2\ell]_{\mathbb{Z}_3}+[\ell,\ell-k]_{\mathbb{Z}_3}+[k-\ell,k]_{\mathbb{Z}_3}
\end{eqnarray}
and
\begin{eqnarray}
    \big([k,\ell]_{\mathbb{Z}_3}\big)^3&=&2[0,0]_{\mathbb{Z}_3}+[3k,3\ell]_{\mathbb{Z}_3}+2[k+\ell,2k-\ell]_{\mathbb{Z}_3} \nonumber \\
&&+2[2k-\ell,k+\ell]_{\mathbb{Z}_3}+[2\ell-k,\ell-2k]_{\mathbb{Z}_3}+[k-2\ell,2k-\ell]_{\mathbb{Z}_3} \,.
\end{eqnarray}

\section{Textures of quark Yukawa couplings from $\mathbb{Z}_3$ gauging of $\mathbb{Z}_N\times \mathbb{Z}_N'$ symmetry }\label{sec:Z3-gauging-ZNXZN-N=2345}

The quark Yukawa interactions in standard model can be genrally written as follows
\begin{equation}
\mathcal{L}_{Y}=\bar{Q}_LY_dd_RH+\bar{Q}_LY_uu_R\tilde{H}+\text{h.c.} \,,
\end{equation}
where $Y_u$ and $Y_d$ are the Yukawa matrices of the up-type quarks and down-type quarks respectively, $Q_L$ denote the three left-handed quark doublets, $u_R$ and $d_R$ stand for the right-handed quark fields of up-type and down-type respectively, $H$ is the SM Higgs doublet and $\tilde{H}=i\sigma_2H^*$ with $\sigma_2$ the second Pauli matrix. In the following, we shall analyze the textures of $Y_u$ and $Y_d$ for all possible independent assignments of quark and Higgs fields under the $\mathbb{Z}_3$ gauging of $\mathbb{Z}_N\times \mathbb{Z}_N'$ symmetry.

Note that the up-type Yukawa matrix $Y_u$ couples to the conjugate Higgs field $\widetilde{H}$, whereas the down-type Yukawa matrix $Y_d$ couples to $H$. In sections~\ref{sec:Z3-gauging-ZNXZN-N=2345} and \ref{sec:S3-gauging-ZNxZN}, where a single Yukawa matrix $Y_q$ is considered, $H$ is used generically for the relevant Higgs field for notational simplicity, irrespective of whether $Y_q$ corresponds to the up- or down-type sector. In section~\ref{sec:pheno-analysis}, where $Y_u$ and $Y_d$ are considered simultaneously, the class assignment specified for $H$ refers to the Higgs field appearing in $Y_d$, while the class of $\widetilde{H}$ appearing in $Y_u$ is given by the inverse of this class.

\subsection{$N=2$}

For $N=2$, from Eq.~\eqref{eq:class-Nneq3} we see there are two equivalence classes:
\begin{eqnarray}
[0,0]_{\mathbb{Z}_3}=\{e\} \,,\quad [1,0]_{\mathbb{Z}_3}=\{a,aa',a'\} \,,
\end{eqnarray}
which satisfy the fusion rules in Eq.~\eqref{eq:fusion-Nneq3}, so that we have
\begin{equation}
[0,0]_{\mathbb{Z}_3}\cdot[0,0]_{\mathbb{Z}_3}=[0,0]_{\mathbb{Z}_3} \,,~~ [0,0]_{\mathbb{Z}_3}\cdot[1,0]_{\mathbb{Z}_3}=[1,0]_{\mathbb{Z}_3} \,,~~ [1,0]_{\mathbb{Z}_3}\cdot[1,0]_{\mathbb{Z}_3}=3[0,0]_{\mathbb{Z}_3}+2[1,0]_{\mathbb{Z}_3} \,.
\end{equation}
Consequently we can obtain the following three-point coupling selection rules
\begin{eqnarray}
[0,0]_{\mathbb{Z}_3}[0,0]_{\mathbb{Z}_3}[0,0]_{\mathbb{Z}_3}\supset [0,0]_{\mathbb{Z}_3} &\,,\quad& [0,0]_{\mathbb{Z}_3}[0,0]_{\mathbb{Z}_3}[1,0]_{\mathbb{Z}_3}\not\supset [0,0]_{\mathbb{Z}_3} \,,\nonumber\\
{}[0,0]_{\mathbb{Z}_3}[1,0]_{\mathbb{Z}_3}[1,0]_{\mathbb{Z}_3}\supset [0,0]_{\mathbb{Z}_3} &\,,\quad& [1,0]_{\mathbb{Z}_3}[1,0]_{\mathbb{Z}_3}[1,0]_{\mathbb{Z}_3}\supset [0,0]_{\mathbb{Z}_3} \,.
\end{eqnarray}
In order to obtain full-rank quark Yukawa matrices $Y_u$ and $Y_d$ with texture zeros, the three generations of quark fields $Q_L$, $u_R$ or $d_R$ cannot correspond to the same equivalence class, otherwise at least one row (or one column) of $Y_u$ and $Y_d$ would vanish. Thus we only have the following possible assignments:
\begin{eqnarray}
\nonumber \bar{Q}_L\sim([0,0]_{\mathbb{Z}_3},[0,0]_{\mathbb{Z}_3},[1,0]_{\mathbb{Z}_3}) \quad&\text{or}&\quad ([1,0]_{\mathbb{Z}_3},[1,0]_{\mathbb{Z}_3},[0,0]_{\mathbb{Z}_3}) \,,\\
\nonumber u_R\sim([0,0]_{\mathbb{Z}_3},[0,0]_{\mathbb{Z}_3},[1,0]_{\mathbb{Z}_3}) \quad&\text{or}&\quad ([1,0]_{\mathbb{Z}_3},[1,0]_{\mathbb{Z}_3},[0,0]_{\mathbb{Z}_3}) \,, \\
d_R\sim([0,0]_{\mathbb{Z}_3},[0,0]_{\mathbb{Z}_3},[1,0]_{\mathbb{Z}_3}) \quad&\text{or}&\quad ([1,0]_{\mathbb{Z}_3},[1,0]_{\mathbb{Z}_3},[0,0]_{\mathbb{Z}_3}) \,.
\end{eqnarray}
The Higgs field can correspond to either $[0,0]_{\mathbb{Z}_3}$ or $[1,0]_{\mathbb{Z}_3}$. All possible textures of $Y_q$ ($q=u,d$) which can be reached from the $\mathbb{Z}_3$ gauging of $\mathbb{Z}_2\times \mathbb{Z}_2'$ symmetry are listed in table~\ref{tab:322tex}. A permutation of the class assignments among the three generations of quark fields simply amounts to a corresponding permutation of the rows and columns of the quark mass matrix. For simplicity, these row and column permutations are understood and are not explicitly displayed in table~\ref{tab:322tex} or in what follows.

\begin{table}[h]
\centering
\begin{tabular}{|c|c|c|}
\hline\hline
 & $Y_q \,,\,\, H\sim [0,0]_{\mathbb{Z}_3}$ & $Y_q \,,\,\, H\sim [1,0]_{\mathbb{Z}_3}$ \\ \hline
\makecell{$\bar{Q}_L\sim([0,0]_{\mathbb{Z}_3},[0,0]_{\mathbb{Z}_3},[1,0]_{\mathbb{Z}_3})$ \\ $q_R\sim([0,0]_{\mathbb{Z}_3},[0,0]_{\mathbb{Z}_3},[1,0]_{\mathbb{Z}_3})$} &
$\begin{pmatrix}
\times & \times & 0 \\
\times & \times & 0 \\
0 & 0 & \times
\end{pmatrix}$ &
$\begin{pmatrix}
0 & 0 & \times \\
0 & 0 & \times \\
\times & \times & \times
\end{pmatrix}$ \\ \hline
\makecell{$\bar{Q}_L\sim([0,0]_{\mathbb{Z}_3},[0,0]_{\mathbb{Z}_3},[1,0]_{\mathbb{Z}_3})$ \\ $q_R\sim([0,0]_{\mathbb{Z}_3},[1,0]_{\mathbb{Z}_3},[1,0]_{\mathbb{Z}_3})$} &
$\begin{pmatrix}
\times & 0 & 0 \\
\times & 0 & 0 \\
0 & \times & \times
\end{pmatrix}$ &
$\begin{pmatrix}
0 & \times & \times \\
0 & \times & \times \\
\times & \times & \times
\end{pmatrix}$ \\ \hline
\makecell{$\bar{Q}_L\sim([0,0]_{\mathbb{Z}_3},[1,0]_{\mathbb{Z}_3},[1,0]_{\mathbb{Z}_3})$ \\ $q_R\sim([0,0]_{\mathbb{Z}_3},[0,0]_{\mathbb{Z}_3},[1,0]_{\mathbb{Z}_3})$} &
$\begin{pmatrix}
\times & \times & 0 \\
0 & 0 & \times \\
0 & 0 & \times
\end{pmatrix}$ &
$\begin{pmatrix}
0 & 0 & \times \\
\times & \times & \times \\
\times & \times & \times
\end{pmatrix}$ \\ \hline
\makecell{$\bar{Q}_L\sim([0,0]_{\mathbb{Z}_3},[1,0]_{\mathbb{Z}_3},[1,0]_{\mathbb{Z}_3})$ \\ $q_R\sim([0,0]_{\mathbb{Z}_3},[1,0]_{\mathbb{Z}_3},[1,0]_{\mathbb{Z}_3})$} &
$\begin{pmatrix}
\times & 0 & 0 \\
0 & \times & \times \\
0 & \times & \times
\end{pmatrix}$ &
$\begin{pmatrix}
0 & \times & \times \\
\times & \times & \times \\
\times & \times & \times
\end{pmatrix}$ \\\hline\hline
\end{tabular}
\caption{The possible patterns of $Y_q$ ($q=u,d$) for different assignments of the fields $q_R$ ($q=u,d$), $\bar{Q}_L$ and $H$ in the $\mathbb{Z}_3$ gauging of $\mathbb{Z}_2\times \mathbb{Z}_2'$ symmetry, where the symbols ``$\times$'' denotes a generic non-vanishing elements.}
\label{tab:322tex}
\end{table}

\subsection{\label{subsec:N3-Delta3N2} $N=3$}

When $N=3$, from Eq.~\eqref{eq:class3N} we see there are the following five equivalence classes:
\begin{eqnarray}
&&[0,0]_{\mathbb{Z}_3}=\{e\} \,,\quad [1,2]_{\mathbb{Z}_3}=\{aa'^2\} \,,\quad [2,1]_{\mathbb{Z}_3}=\{a^2a'\} \,,\nonumber\\
&&[1,0]_{\mathbb{Z}_3}=\{a,a^2a'^2,a'\} \,,\quad [2,0]_{\mathbb{Z}_3}=\{a^2,aa',a'^2\} \,,\label{eq:z3N=3class}
\end{eqnarray}
which satisfy the fusion rules in Eq.~\eqref{eq:fr3N}, therefore we have
\begin{eqnarray}
&&[0,0]_{\mathbb{Z}_3}\cdot[k,\ell]_{\mathbb{Z}_3}=[k,\ell]_{\mathbb{Z}_3} \,,\quad [1,2]_{\mathbb{Z}_3}\cdot[1,2]_{\mathbb{Z}_3}=[2,1]_{\mathbb{Z}_3} \,,\quad
[1,2]_{\mathbb{Z}_3}\cdot[2,1]_{\mathbb{Z}_3}=[0,0]_{\mathbb{Z}_3} \,,\nonumber\\
&&[1,2]_{\mathbb{Z}_3}\cdot[1,0]_{\mathbb{Z}_3}=[1,0]_{\mathbb{Z}_3} \,,\quad [1,2]_{\mathbb{Z}_3}\cdot[2,0]_{\mathbb{Z}_3}=[2,0]_{\mathbb{Z}_3} \,,\quad [2,1]_{\mathbb{Z}_3}\cdot[2,1]_{\mathbb{Z}_3}=[1,2]_{\mathbb{Z}_3} \,,\nonumber\\
&&[2,1]_{\mathbb{Z}_3}\cdot[1,0]_{\mathbb{Z}_3}=[1,0]_{\mathbb{Z}_3} \,,\quad [2,1]_{\mathbb{Z}_3}\cdot[2,0]_{\mathbb{Z}_3}=[2,0]_{\mathbb{Z}_3} \,,\quad [1,0]_{\mathbb{Z}_3}\cdot[1,0]_{\mathbb{Z}_3}=3[2,0]_{\mathbb{Z}_3} \,,\nonumber\\
&&[1,0]_{\mathbb{Z}_3}\cdot[2,0]_{\mathbb{Z}_3}=3[0,0]_{\mathbb{Z}_3}+3[1,2]_{\mathbb{Z}_3}+3[2,1]_{\mathbb{Z}_3} \,,\quad [2,0]_{\mathbb{Z}_3}\cdot[2,0]_{\mathbb{Z}_3}=3[1,0]_{\mathbb{Z}_3} \,.\label{eq:frN=3}
\end{eqnarray}
There are two symmetries $[1,0]_{\mathbb{Z}_3}\leftrightarrow [2,0]_{\mathbb{Z}_3}$ and $[1,2]_{\mathbb{Z}_3}\leftrightarrow [2,1]_{\mathbb{Z}_3}$, which leave the fusion rules Eq.~\eqref{eq:frN=3} invariant. The former arises from the automorphism $a\leftrightarrow a'^{-1}$, and the latter corresponds to the automorphism $a\leftrightarrow a'$.

From the fusion rules in Eq.~\eqref{eq:frN=3}, we see that the three-point couplings of the following pattern are invariant under the non-invertible symmetry,
\begin{eqnarray}
&&[0,0]_{\mathbb{Z}_3}[0,0]_{\mathbb{Z}_3}[0,0]_{\mathbb{Z}_3} \,,\, [0,0]_{\mathbb{Z}_3}[1,2]_{\mathbb{Z}_3}[2,1]_{\mathbb{Z}_3} \,,\, [0,0]_{\mathbb{Z}_3}[1,0]_{\mathbb{Z}_3}[2,0]_{\mathbb{Z}_3} \,,\, [1,2]_{\mathbb{Z}_3}[1,2]_{\mathbb{Z}_3}[1,2]_{\mathbb{Z}_3} \,,\nonumber\\
&&[1,2]_{\mathbb{Z}_3}[1,0]_{\mathbb{Z}_3}[2,0]_{\mathbb{Z}_3} \,,\, [2,1]_{\mathbb{Z}_3}[2,1]_{\mathbb{Z}_3}[2,1]_{\mathbb{Z}_3} \,,\, [2,1]_{\mathbb{Z}_3}[1,0]_{\mathbb{Z}_3}[2,0]_{\mathbb{Z}_3} \,,\, [1,0]_{\mathbb{Z}_3}[1,0]_{\mathbb{Z}_3}[1,0]_{\mathbb{Z}_3} \,,\nonumber\\
&&[2,0]_{\mathbb{Z}_3}[2,0]_{\mathbb{Z}_3}[2,0]_{\mathbb{Z}_3} \,.
\end{eqnarray}
For three generations of quark fields $\bar{Q}_L$ and $q_R$ with $q=u ,d$, there are $C_{5+3-1}^3-5=30$ possible assignments, where the cases in which the three generations belong to the same class have already been excluded. Because of the symmetries of the fusion rules, without loss of generality the Higgs field can be assigned to $[0,0]_{\mathbb{Z}_3}$, $[1,2]_{\mathbb{Z}_3}$ or $[1,0]_{\mathbb{Z}_3}$. After considering all possible assignments for the left- and right-handed quarks and Higgs fields, we find that only the following two patterns of the full-rank Yukawa matrix can be achieved from the $\mathbb{Z}_3$ gauging of $\mathbb{Z}_3\times \mathbb{Z}_3'$ symmetry:
\begin{eqnarray}
Y_q=\begin{pmatrix}
 \times & 0 & 0 \\
 0 & \times & 0 \\
 0 & 0 & \times \\
\end{pmatrix} ~ \text{when} ~ (\bar{Q}_L;q_R;H)&\sim& ([{2, 0}]_{\mathbb{Z}_3},[{1, 0}]_{\mathbb{Z}_3},[{0, 0}]_{\mathbb{Z}_3};[{1, 0}]_{\mathbb{Z}_3},[{2, 0}]_{\mathbb{Z}_3},[{0, 0}]_{\mathbb{Z}_3};[{0, 0}]_{\mathbb{Z}_3}) \nonumber\\[-22pt]
&\text{or}&([{2, 0}]_{\mathbb{Z}_3},[{1, 0}]_{\mathbb{Z}_3},[{0, 0}]_{\mathbb{Z}_3};[{1, 0}]_{\mathbb{Z}_3},[{2, 0}]_{\mathbb{Z}_3},[{2, 1}]_{\mathbb{Z}_3};[{1, 2}]_{\mathbb{Z}_3}) \nonumber\\
&\text{or}&([{2, 0}]_{\mathbb{Z}_3},[{1, 0}]_{\mathbb{Z}_3},[{0, 0}]_{\mathbb{Z}_3};[{0, 0}]_{\mathbb{Z}_3},[{1, 0}]_{\mathbb{Z}_3},[{2, 0}]_{\mathbb{Z}_3};[{1, 0}]_{\mathbb{Z}_3}) \dots\nonumber\\
\end{eqnarray}

\begin{eqnarray}
Y_q=\begin{pmatrix}
 \times & \times & 0 \\
 \times & \times & 0 \\
 0 & 0 & \times \\
\end{pmatrix} ~ \text{when} ~ (\bar{Q}_L;q_R;H)&\sim& ([{0, 0}]_{\mathbb{Z}_3},[{0, 0}]_{\mathbb{Z}_3},[{1, 0}]_{\mathbb{Z}_3};[{0, 0}]_{\mathbb{Z}_3},[{0, 0}]_{\mathbb{Z}_3},[{2, 0}]_{\mathbb{Z}_3};[{0, 0}]_{\mathbb{Z}_3}) \nonumber\\[-22pt]
&\text{or}&([{0, 0}]_{\mathbb{Z}_3},[{0, 0}]_{\mathbb{Z}_3},[{1, 2}]_{\mathbb{Z}_3};[{2, 1}]_{\mathbb{Z}_3},[{2, 1}]_{\mathbb{Z}_3},[{1, 2}]_{\mathbb{Z}_3};[{1, 2}]_{\mathbb{Z}_3}) \nonumber\\
&\text{or}&([{0, 0}]_{\mathbb{Z}_3},[{0, 0}]_{\mathbb{Z}_3},[{1, 0}]_{\mathbb{Z}_3};[{2, 0}]_{\mathbb{Z}_3},[{2, 0}]_{\mathbb{Z}_3},[{1, 0}]_{\mathbb{Z}_3};[{1, 0}]_{\mathbb{Z}_3}) \dots\nonumber\\
\end{eqnarray}

\subsection{$N=4$}

As shown in Eq.~\eqref{eq:class-Nneq3}, there are six independent equivalence classes for this  case,
\begin{eqnarray}
&&[0,0]_{\mathbb{Z}_3}=\{e\} \,,~~ [1,0]_{\mathbb{Z}_3}=\{a,a^3a'^3,a'\} \,,~~  [2,0]_{\mathbb{Z}_3}=\{a^2,a^2a'^2,a'^2\} \,,\nonumber\\
{}[3,0]_{\mathbb{Z}_3}&=&\{a^3,aa',a'^3\} \,,~~  [2,1]_{\mathbb{Z}_3}=\{a^2a',a^3a'^2,a^3a'\} \,,~~  [1,2]_{\mathbb{Z}_3}=\{aa'^2,aa'^3,a^2a'^3\} \,.~~
\end{eqnarray}
The corresponding fusion rules can be obtained from Eq.~\eqref{eq:fusion-Nneq3} with $N=4$, and they read as,
\begin{eqnarray}
&&[0,0]_{\mathbb{Z}_3}\cdot[k,\ell]_{\mathbb{Z}_3}=[k,\ell]_{\mathbb{Z}_3} \,,\quad [1,0]_{\mathbb{Z}_3}\cdot[1,0]_{\mathbb{Z}_3}=[2,0]_{\mathbb{Z}_3}+2[3,0]_{\mathbb{Z}_3} \,,\nonumber\\
&&[1,0]_{\mathbb{Z}_3}\cdot[2,0]_{\mathbb{Z}_3}=[3,0]_{\mathbb{Z}_3}+[2,1]_{\mathbb{Z}_3}+[1,2]_{\mathbb{Z}_3} \,,\quad [1,0]_{\mathbb{Z}_3}\cdot[3,0]_{\mathbb{Z}_3}=3[0,0]_{\mathbb{Z}_3}+[2,1]_{\mathbb{Z}_3}+[1,2]_{\mathbb{Z}_3} \,,\nonumber\\
&&[1,0]_{\mathbb{Z}_3}\cdot[2,1]_{\mathbb{Z}_3}=[2,1]_{\mathbb{Z}_3}+[2,0]_{\mathbb{Z}_3}+[1,0]_{\mathbb{Z}_3} \,,\quad [1,0]_{\mathbb{Z}_3}\cdot[1,2]_{\mathbb{Z}_3}=[2,0]_{\mathbb{Z}_3}+[1,2]_{\mathbb{Z}_3}+[1,0]_{\mathbb{Z}_3} \,,\nonumber\\
&&[2,0]_{\mathbb{Z}_3}\cdot[2,0]_{\mathbb{Z}_3}=3[0,0]_{\mathbb{Z}_3}+2[2,0]_{\mathbb{Z}_3} \,,\quad [2,0]_{\mathbb{Z}_3}\cdot[3,0]_{\mathbb{Z}_3}=[1,0]_{\mathbb{Z}_3}+[2,1]_{\mathbb{Z}_3}+[1,2]_{\mathbb{Z}_3} \,,\nonumber\\
&&[2,0]_{\mathbb{Z}_3}\cdot[2,1]_{\mathbb{Z}_3}=[1,0]_{\mathbb{Z}_3}+[1,2]_{\mathbb{Z}_3}+[3,0]_{\mathbb{Z}_3} \,,\quad [2,0]_{\mathbb{Z}_3}\cdot[1,2]_{\mathbb{Z}_3}=[2,1]_{\mathbb{Z}_3}+[1,0]_{\mathbb{Z}_3}+[3,0]_{\mathbb{Z}_3} \,,\nonumber\\
&&[3,0]_{\mathbb{Z}_3}\cdot[3,0]_{\mathbb{Z}_3}=[2,0]_{\mathbb{Z}_3}+2[1,0]_{\mathbb{Z}_3} \,,\quad [3,0]_{\mathbb{Z}_3}\cdot[2,1]_{\mathbb{Z}_3}=[3,0]_{\mathbb{Z}_3}+[2,0]_{\mathbb{Z}_3}+[2,1]_{\mathbb{Z}_3} \,,\nonumber\\
&&[3,0]_{\mathbb{Z}_3}\cdot[1,2]_{\mathbb{Z}_3}=[2,0]_{\mathbb{Z}_3}+[3,0]_{\mathbb{Z}_3}+[1,2]_{\mathbb{Z}_3} \,,\quad [2,1]_{\mathbb{Z}_3}\cdot[2,1]_{\mathbb{Z}_3}=[2,0]_{\mathbb{Z}_3}+2[1,2]_{\mathbb{Z}_3} \,,\nonumber\\
&&[2,1]_{\mathbb{Z}_3}\cdot[1,2]_{\mathbb{Z}_3}=[1,0]_{\mathbb{Z}_3}+[3,0]_{\mathbb{Z}_3}+3[0,0]_{\mathbb{Z}_3} \,,\quad [1,2]_{\mathbb{Z}_3}\cdot[1,2]_{\mathbb{Z}_3}=[2,0]_{\mathbb{Z}_3}+2[2,1]_{\mathbb{Z}_3} \,. \label{eq:FRs-N4}
\end{eqnarray}
These fusion rules remain invariant under four distinct transformations,
\begin{eqnarray}
\nonumber && [1,0]_{\mathbb{Z}_3}\leftrightarrow [3,0]_{\mathbb{Z}_3},\quad  [1,2]_{\mathbb{Z}_3}\leftrightarrow [2,1]_{\mathbb{Z}_3}, \\
&& [1,0]_{\mathbb{Z}_3}\leftrightarrow [2,1]_{\mathbb{Z}_3}~~\text{together with}~~[3,0]_{\mathbb{Z}_3}\leftrightarrow [1,2]_{\mathbb{Z}_3}\,, \label{eq:FR-sym1-N4}
\end{eqnarray}
and the cyclic transformation
\begin{eqnarray}
[1,0]_{\mathbb{Z}_3}\rightarrow[2,1]_{\mathbb{Z}_3}\rightarrow[3,0]_{\mathbb{Z}_3}\rightarrow[1,2]_{\mathbb{Z}_3}\rightarrow[1,0]_{\mathbb{Z}_3}\,.\label{eq:FR-sym2-N4}
\end{eqnarray}
These transformations are induced, respectively, by the automorphisms
\begin{eqnarray}
a\leftrightarrow a'^{-1},\quad  a\leftrightarrow a', \quad (a, a')\rightarrow (a^2a', a^3a'), \quad (a, a')\rightarrow (a^2a', a^3a'^2)\,.
\end{eqnarray}

The fusion rules in Eq.~\eqref{eq:FRs-N4} imply that the following combinations of the three equivalence classes can fuse to the identity class $[0,0]_{\mathbb{Z}_3}$,
\begin{eqnarray}
&&[0,0]_{\mathbb{Z}_3}[0,0]_{\mathbb{Z}_3}[0,0]_{\mathbb{Z}_3} \,,\, [0,0]_{\mathbb{Z}_3}[2,0]_{\mathbb{Z}_3}[2,0]_{\mathbb{Z}_3} \,,\, [0,0]_{\mathbb{Z}_3}[1,0]_{\mathbb{Z}_3}[3,0]_{\mathbb{Z}_3} \,,\, [0,0]_{\mathbb{Z}_3}[1,2]_{\mathbb{Z}_3}[2,1]_{\mathbb{Z}_3} \,,\nonumber\\
&&[1,0]_{\mathbb{Z}_3}[1,0]_{\mathbb{Z}_3}[1,0]_{\mathbb{Z}_3} \,,\, [1,0]_{\mathbb{Z}_3}[1,0]_{\mathbb{Z}_3}[2,0]_{\mathbb{Z}_3} \,,\, [1,0]_{\mathbb{Z}_3}[2,0]_{\mathbb{Z}_3}[2,1]_{\mathbb{Z}_3} \,,\, [1,0]_{\mathbb{Z}_3}[3,0]_{\mathbb{Z}_3}[2,1]_{\mathbb{Z}_3} \,,\nonumber\\
&&[1,0]_{\mathbb{Z}_3}[2,0]_{\mathbb{Z}_3}[1,2]_{\mathbb{Z}_3} \,,\, [1,0]_{\mathbb{Z}_3}[3,0]_{\mathbb{Z}_3}[1,2]_{\mathbb{Z}_3} \,,\, [1,0]_{\mathbb{Z}_3}[2,1]_{\mathbb{Z}_3}[1,2]_{\mathbb{Z}_3} \,,\, [2,0]_{\mathbb{Z}_3}[2,0]_{\mathbb{Z}_3}[2,0]_{\mathbb{Z}_3} \,,\nonumber\\
&&[2,0]_{\mathbb{Z}_3}[3,0]_{\mathbb{Z}_3}[3,0]_{\mathbb{Z}_3} \,,\, [2,0]_{\mathbb{Z}_3}[3,0]_{\mathbb{Z}_3}[2,1]_{\mathbb{Z}_3} \,,\, [2,0]_{\mathbb{Z}_3}[2,1]_{\mathbb{Z}_3}[2,1]_{\mathbb{Z}_3} \,,\, [2,0]_{\mathbb{Z}_3}[3,0]_{\mathbb{Z}_3}[1,2]_{\mathbb{Z}_3} \,,\nonumber\\
&&[2,0]_{\mathbb{Z}_3}[1,2]_{\mathbb{Z}_3}[1,2]_{\mathbb{Z}_3} \,,\, [3,0]_{\mathbb{Z}_3}[3,0]_{\mathbb{Z}_3}[3,0]_{\mathbb{Z}_3} \,,\, [3,0]_{\mathbb{Z}_3}[1,2]_{\mathbb{Z}_3}[2,1]_{\mathbb{Z}_3} \,,\, [2,1]_{\mathbb{Z}_3}[2,1]_{\mathbb{Z}_3}[2,1]_{\mathbb{Z}_3} \,,\nonumber\\
&&[1,2]_{\mathbb{Z}_3}[1,2]_{\mathbb{Z}_3}[1,2]_{\mathbb{Z}_3} \,.
\end{eqnarray}

The four symmetries of the fusion rules, listed in Eqs.~\eqref{eq:FR-sym1-N4} and~\eqref{eq:FR-sym2-N4}, allow us to restrict the Higgs-field assignment to the three classes
$[0,0]_{\mathbb{Z}_3}$, $[1,0]_{\mathbb{Z}_3}$ and $[2,0]_{\mathbb{Z}_3}$. We then systematically examine all possible assignments of the three generations of quark fields
$\bar{Q}_L$, $q_R$, as well as the Higgs field $H$, and obtain the fifteen full-rank quark Yukawa matrices listed in table~\ref{tab:z3n4-yukawa-textures}. For each viable and inequivalent assignment of the Higgs fields, we provide a corresponding quark-field assignment as an illustrative example.

\begin{table}[hptb!]
\centering
\setlength{\extrarowheight}{0pt}
\begin{tabular}{|>{\scriptsize}c|c|c|c|}
\hline\hline
\rule{0pt}{3.5ex}{\normalsize $Y_q$} & $\bar{Q}_L$ & $q_R$ & $H$ \\
\hline
\multirow{3}{*}{$\begin{pmatrix} \times & 0 & 0 \\ 0 & \times & 0 \\ 0 & 0 & \times \\ \end{pmatrix}$} & $[{2, 0}]_{\mathbb{Z}_3},[{1, 0}]_{\mathbb{Z}_3},[{0, 0}]_{\mathbb{Z}_3}$ & $[{2, 0}]_{\mathbb{Z}_3},[{3, 0}]_{\mathbb{Z}_3},[{0, 0}]_{\mathbb{Z}_3}$ & $[{0, 0}]_{\mathbb{Z}_3}$ \\
\cline{2-4}
 & $[{0, 0}]_{\mathbb{Z}_3},[{1, 0}]_{\mathbb{Z}_3},[{2, 0}]_{\mathbb{Z}_3}$ & $[{3, 0}]_{\mathbb{Z}_3},[{2, 0}]_{\mathbb{Z}_3},[{2, 1}]_{\mathbb{Z}_3}$ & $[{1, 0}]_{\mathbb{Z}_3}$ \\
\cline{2-4}
 & $[{3, 0}]_{\mathbb{Z}_3},[{1, 0}]_{\mathbb{Z}_3},[{0, 0}]_{\mathbb{Z}_3}$ & $[{3, 0}]_{\mathbb{Z}_3},[{1, 0}]_{\mathbb{Z}_3},[{2, 0}]_{\mathbb{Z}_3}$ & $[{2, 0}]_{\mathbb{Z}_3}$ \\
\hline
\multirow{2}{*}{$\begin{pmatrix} \times & \times & 0 \\ 0 & \times & 0 \\ 0 & 0 & \times \\ \end{pmatrix}$} & $[{1, 0}]_{\mathbb{Z}_3},[{2, 0}]_{\mathbb{Z}_3},[{0, 0}]_{\mathbb{Z}_3}$ & $[{2, 0}]_{\mathbb{Z}_3},[{1, 0}]_{\mathbb{Z}_3},[{3, 0}]_{\mathbb{Z}_3}$ & $[{1, 0}]_{\mathbb{Z}_3}$ \\
\cline{2-4}
 & $[{2, 0}]_{\mathbb{Z}_3},[{0, 0}]_{\mathbb{Z}_3},[{1, 0}]_{\mathbb{Z}_3}$ & $[{0, 0}]_{\mathbb{Z}_3},[{2, 0}]_{\mathbb{Z}_3},[{1, 0}]_{\mathbb{Z}_3}$ & $[{2, 0}]_{\mathbb{Z}_3}$ \\
\hline
\multirow{3}{*}{$\begin{pmatrix} \times & \times & 0 \\ \times & \times & 0 \\ 0 & 0 & \times \\ \end{pmatrix}$} & $[{0, 0}]_{\mathbb{Z}_3},[{0, 0}]_{\mathbb{Z}_3},[{1, 0}]_{\mathbb{Z}_3}$ & $[{0, 0}]_{\mathbb{Z}_3},[{0, 0}]_{\mathbb{Z}_3},[{3, 0}]_{\mathbb{Z}_3}$ & $[{0, 0}]_{\mathbb{Z}_3}$ \\
\cline{2-4}
 & $[{0, 0}]_{\mathbb{Z}_3},[{0, 0}]_{\mathbb{Z}_3},[{1, 0}]_{\mathbb{Z}_3}$ & $[{3, 0}]_{\mathbb{Z}_3},[{3, 0}]_{\mathbb{Z}_3},[{1, 0}]_{\mathbb{Z}_3}$ & $[{1, 0}]_{\mathbb{Z}_3}$ \\
\cline{2-4}
 & $[{0, 0}]_{\mathbb{Z}_3},[{0, 0}]_{\mathbb{Z}_3},[{1, 0}]_{\mathbb{Z}_3}$ & $[{2, 0}]_{\mathbb{Z}_3},[{2, 0}]_{\mathbb{Z}_3},[{1, 0}]_{\mathbb{Z}_3}$ & $[{2, 0}]_{\mathbb{Z}_3}$ \\
\hline
$\begin{pmatrix} \times & \times & 0 \\ 0 & \times & \times \\ 0 & 0 & \times \\ \end{pmatrix}$ & $[{1, 0}]_{\mathbb{Z}_3},[{2, 1}]_{\mathbb{Z}_3},[{0, 0}]_{\mathbb{Z}_3}$ & $[{1, 0}]_{\mathbb{Z}_3},[{2, 0}]_{\mathbb{Z}_3},[{3, 0}]_{\mathbb{Z}_3}$ & $[{1, 0}]_{\mathbb{Z}_3}$ \\
\hline
$\begin{pmatrix} \times & \times & \times \\ 0 & \times & 0 \\ 0 & 0 & \times \\ \end{pmatrix}$ & $[{2, 1}]_{\mathbb{Z}_3},[{1, 0}]_{\mathbb{Z}_3},[{0, 0}]_{\mathbb{Z}_3}$ & $[{1, 2}]_{\mathbb{Z}_3},[{2, 0}]_{\mathbb{Z}_3},[{3, 0}]_{\mathbb{Z}_3}$ & $[{1, 0}]_{\mathbb{Z}_3}$ \\
\hline
$\begin{pmatrix} \times & \times & 0 \\ 0 & \times & 0 \\ 0 & \times & \times \\ \end{pmatrix}$ & $[{1, 2}]_{\mathbb{Z}_3},[{0, 0}]_{\mathbb{Z}_3},[{2, 1}]_{\mathbb{Z}_3}$ & $[{2, 1}]_{\mathbb{Z}_3},[{3, 0}]_{\mathbb{Z}_3},[{1, 2}]_{\mathbb{Z}_3}$ & $[{1, 0}]_{\mathbb{Z}_3}$ \\
\hline
\multirow{2}{*}{$\begin{pmatrix} \times & \times & 0 \\ 0 & \times & \times \\ \times & 0 & \times \\ \end{pmatrix}$} & $[{2, 1}]_{\mathbb{Z}_3},[{1, 0}]_{\mathbb{Z}_3},[{2, 0}]_{\mathbb{Z}_3}$ & $[{1, 2}]_{\mathbb{Z}_3},[{2, 0}]_{\mathbb{Z}_3},[{1, 0}]_{\mathbb{Z}_3}$ & $[{1, 0}]_{\mathbb{Z}_3}$ \\
\cline{2-4}
 & $[{3, 0}]_{\mathbb{Z}_3},[{1, 0}]_{\mathbb{Z}_3},[{2, 1}]_{\mathbb{Z}_3}$ & $[{3, 0}]_{\mathbb{Z}_3},[{1, 2}]_{\mathbb{Z}_3},[{1, 0}]_{\mathbb{Z}_3}$ & $[{2, 0}]_{\mathbb{Z}_3}$ \\
\hline
$\begin{pmatrix} \times & \times & \times \\ 0 & \times & \times \\ 0 & 0 & \times \\ \end{pmatrix}$ & $[{1, 2}]_{\mathbb{Z}_3},[{2, 1}]_{\mathbb{Z}_3},[{0, 0}]_{\mathbb{Z}_3}$ & $[{2, 1}]_{\mathbb{Z}_3},[{2, 0}]_{\mathbb{Z}_3},[{3, 0}]_{\mathbb{Z}_3}$ & $[{1, 0}]_{\mathbb{Z}_3}$ \\
\hline
\multirow{2}{*}{$\begin{pmatrix} \times & \times & \times \\ \times & \times & 0 \\ 0 & 0 & \times \\ \end{pmatrix}$} & $[{2, 1}]_{\mathbb{Z}_3},[{1, 0}]_{\mathbb{Z}_3},[{0, 0}]_{\mathbb{Z}_3}$ & $[{2, 0}]_{\mathbb{Z}_3},[{2, 0}]_{\mathbb{Z}_3},[{3, 0}]_{\mathbb{Z}_3}$ & $[{1, 0}]_{\mathbb{Z}_3}$ \\
\cline{2-4}
 & $[{2, 1}]_{\mathbb{Z}_3},[{1, 0}]_{\mathbb{Z}_3},[{3, 0}]_{\mathbb{Z}_3}$ & $[{1, 0}]_{\mathbb{Z}_3},[{1, 0}]_{\mathbb{Z}_3},[{3, 0}]_{\mathbb{Z}_3}$ & $[{2, 0}]_{\mathbb{Z}_3}$ \\
\hline
\multirow{2}{*}{$\begin{pmatrix} \times & \times & 0 \\ \times & \times & 0 \\ \times & 0 & \times \\ \end{pmatrix}$} & $[{1, 0}]_{\mathbb{Z}_3},[{1, 0}]_{\mathbb{Z}_3},[{2, 0}]_{\mathbb{Z}_3}$ & $[{1, 0}]_{\mathbb{Z}_3},[{2, 0}]_{\mathbb{Z}_3},[{2, 1}]_{\mathbb{Z}_3}$ & $[{1, 0}]_{\mathbb{Z}_3}$ \\
\cline{2-4}
 & $[{1, 0}]_{\mathbb{Z}_3},[{1, 0}]_{\mathbb{Z}_3},[{3, 0}]_{\mathbb{Z}_3}$ & $[{2, 1}]_{\mathbb{Z}_3},[{1, 0}]_{\mathbb{Z}_3},[{3, 0}]_{\mathbb{Z}_3}$ & $[{2, 0}]_{\mathbb{Z}_3}$ \\
\hline
$\begin{pmatrix} \times & \times & \times \\ \times & \times & 0 \\ \times & 0 & \times \\ \end{pmatrix}$ & $[{2, 1}]_{\mathbb{Z}_3},[{1, 0}]_{\mathbb{Z}_3},[{3, 0}]_{\mathbb{Z}_3}$ & $[{2, 1}]_{\mathbb{Z}_3},[{1, 0}]_{\mathbb{Z}_3},[{3, 0}]_{\mathbb{Z}_3}$ & $[{2, 0}]_{\mathbb{Z}_3}$ \\
\hline
\multirow{2}{*}{$\begin{pmatrix} \times & \times & \times \\ \times & \times & \times \\ 0 & 0 & \times \\ \end{pmatrix}$} & $[{2, 1}]_{\mathbb{Z}_3},[{2, 1}]_{\mathbb{Z}_3},[{0, 0}]_{\mathbb{Z}_3}$ & $[{2, 0}]_{\mathbb{Z}_3},[{2, 0}]_{\mathbb{Z}_3},[{3, 0}]_{\mathbb{Z}_3}$ & $[{1, 0}]_{\mathbb{Z}_3}$ \\
\cline{2-4}
 & $[{2, 0}]_{\mathbb{Z}_3},[{2, 0}]_{\mathbb{Z}_3},[{0, 0}]_{\mathbb{Z}_3}$ & $[{0, 0}]_{\mathbb{Z}_3},[{0, 0}]_{\mathbb{Z}_3},[{2, 0}]_{\mathbb{Z}_3}$ & $[{2, 0}]_{\mathbb{Z}_3}$ \\
\hline
\multirow{2}{*}{$\begin{pmatrix} \times & \times & \times \\ 0 & \times & \times \\ 0 & \times & \times \\ \end{pmatrix}$} & $[{2, 1}]_{\mathbb{Z}_3},[{0, 0}]_{\mathbb{Z}_3},[{0, 0}]_{\mathbb{Z}_3}$ & $[{2, 0}]_{\mathbb{Z}_3},[{3, 0}]_{\mathbb{Z}_3},[{3, 0}]_{\mathbb{Z}_3}$ & $[{1, 0}]_{\mathbb{Z}_3}$ \\
\cline{2-4}
 & $[{2, 0}]_{\mathbb{Z}_3},[{0, 0}]_{\mathbb{Z}_3},[{0, 0}]_{\mathbb{Z}_3}$ & $[{0, 0}]_{\mathbb{Z}_3},[{2, 0}]_{\mathbb{Z}_3},[{2, 0}]_{\mathbb{Z}_3}$ & $[{2, 0}]_{\mathbb{Z}_3}$ \\
\hline
\multirow{2}{*}{$\begin{pmatrix} \times & \times & \times \\ \times & \times & \times \\ \times & 0 & \times \\ \end{pmatrix}$} & $[{2, 1}]_{\mathbb{Z}_3},[{2, 1}]_{\mathbb{Z}_3},[{0, 0}]_{\mathbb{Z}_3}$ & $[{3, 0}]_{\mathbb{Z}_3},[{2, 0}]_{\mathbb{Z}_3},[{3, 0}]_{\mathbb{Z}_3}$ & $[{1, 0}]_{\mathbb{Z}_3}$ \\
\cline{2-4}
 & $[{2, 0}]_{\mathbb{Z}_3},[{2, 0}]_{\mathbb{Z}_3},[{0, 0}]_{\mathbb{Z}_3}$ & $[{2, 0}]_{\mathbb{Z}_3},[{0, 0}]_{\mathbb{Z}_3},[{2, 0}]_{\mathbb{Z}_3}$ & $[{2, 0}]_{\mathbb{Z}_3}$ \\
\hline
\multirow{2}{*}{$\begin{pmatrix} \times & \times & \times \\ \times & \times & \times \\ \times & \times & \times \\ \end{pmatrix}$} & $[{3, 0}]_{\mathbb{Z}_3},[{2, 0}]_{\mathbb{Z}_3},[{2, 0}]_{\mathbb{Z}_3}$ & $[{1, 2}]_{\mathbb{Z}_3},[{2, 1}]_{\mathbb{Z}_3},[{2, 1}]_{\mathbb{Z}_3}$ & $[{1, 0}]_{\mathbb{Z}_3}$ \\
\cline{2-4}
 & $[{1, 0}]_{\mathbb{Z}_3},[{1, 0}]_{\mathbb{Z}_3},[{3, 0}]_{\mathbb{Z}_3}$ & $[{1, 2}]_{\mathbb{Z}_3},[{2, 1}]_{\mathbb{Z}_3},[{2, 1}]_{\mathbb{Z}_3}$ & $[{2, 0}]_{\mathbb{Z}_3}$ \\
\hline
\hline
\end{tabular}
\caption{The field assignments and the corresponding quark Yukawa textures which can be generated from the $\mathbb{Z}_3$ gauging of the $\mathbb{Z}_4\times\mathbb{Z}'_4$ symmetry.}
\label{tab:z3n4-yukawa-textures}
\end{table}

\subsection{$N=5$}

The $\mathbb{Z}_3$ gauging of $\mathbb{Z}_5\times \mathbb{Z}_5'$ symmetry has the
following nine equivalence classes:
\begin{small}
\begin{eqnarray}
&&[0,0]_{\mathbb{Z}_3}=\{e\} \,,~~ [1,0]_{\mathbb{Z}_3}=\{a,a^4a'^4,a'\} \,,~~ [2,0]_{\mathbb{Z}_3}=\{a^2,a^3a'^3,a'^2\} \,,\nonumber\\
{}[3,0]_{\mathbb{Z}_3}&=&\{a^3,a^2a'^2,a'^3\} \,,~~ [4,0]_{\mathbb{Z}_3}=\{a^4,aa',a'^4\} \,,~~ [2,1]_{\mathbb{Z}_3}=\{a^2a',a^4a'^3,a^4a'\} \,,\nonumber\\
{}[3,1]_{\mathbb{Z}_3}&=&\{a^3a',a^3a'^2,a^4a'^2\} \,,~~ [1,2]_{\mathbb{Z}_3}=\{aa'^2,aa'^4,a^3a'^4\} \,,~~ [1,3]_{\mathbb{Z}_3}=\{aa'^3,a^2a'^4,a^2a'^3\} \,.~~~~~
\end{eqnarray}
\end{small}
Using Eq.~\eqref{eq:fusion-Nneq3}, the explicit fusion rules can be obtained, which are invariant under the following cyclic transformations:
\begin{small}
\begin{eqnarray}
\nonumber s_1:&&[1,0]_{\mathbb{Z}_3}\rightarrow[2,1]_{\mathbb{Z}_3}\rightarrow[2,0]_{\mathbb{Z}_3}\rightarrow[3,1]_{\mathbb{Z}_3}\rightarrow[4,0]_{\mathbb{Z}_3}\rightarrow[1,2]_{\mathbb{Z}_3}\rightarrow[3,0]_{\mathbb{Z}_3}\rightarrow[1,3]_{\mathbb{Z}_3}\rightarrow[1,0]_{\mathbb{Z}_3} \,,\\
s_2: &&[1,0]_{\mathbb{Z}_3}\rightarrow[2,1]_{\mathbb{Z}_3}\rightarrow[3,0]_{\mathbb{Z}_3}\rightarrow[1,3]_{\mathbb{Z}_3}\rightarrow[4,0]_{\mathbb{Z}_3}\rightarrow[1,2]_{\mathbb{Z}_3}\rightarrow[2,0]_{\mathbb{Z}_3}\rightarrow[3,1]_{\mathbb{Z}_3}\rightarrow[1,0]_{\mathbb{Z}_3} \,.~~~~~~~~~\label{eq:n=5symmetry2}
\end{eqnarray}
\end{small}
They arise from the automorphisms $(a, a')\rightarrow (a^2a', a^4a')$ and $(a, a')\rightarrow (a^2a', a^4a'^3)$ respectively. The fusion rules possess two manifest automorphisms, $a\leftrightarrow a'$ and $a\leftrightarrow a'^{-1}$, both of which can be expressed in terms of the generators $s_1$ and $s_2$. The first automorphism, $a\leftrightarrow a'$, acts on the equivalence classes as
\begin{equation}
[1,2]_{\mathbb{Z}_3}\leftrightarrow[2,1]_{\mathbb{Z}_3},,\qquad
[1,3]_{\mathbb{Z}_3}\leftrightarrow[3,1]_{\mathbb{Z}_3},,
\end{equation}
and it can be generated by appropriate compositions of $s_1$ and $s_2$ such as $s_2s^3_1$. The second automorphism, $a\leftrightarrow a'^{-1}$, acts as
\begin{equation}
[1,0]_{\mathbb{Z}_3}\leftrightarrow[4,0]_{\mathbb{Z}_3},,\qquad
[2,0]_{\mathbb{Z}_3}\leftrightarrow[3,0]_{\mathbb{Z}_3},,
\end{equation}
and is generated by the composition $s_1s_2s_1^2$. Therefore, exploiting the symmetries discussed above, the fusion rules can, without loss of generality, be represented by the following subset involving $[0,0]_{\mathbb{Z}_3}$ and $[1,0]_{\mathbb{Z}_3}$,
\begin{eqnarray}\label{eq:n=5partfr}
&&[0,0]_{\mathbb{Z}_3}\cdot[k,\ell]_{\mathbb{Z}_3}=[k,\ell]_{\mathbb{Z}_3} \,,\quad [1,0]_{\mathbb{Z}_3}\cdot[1,0]_{\mathbb{Z}_3}=[2,0]_{\mathbb{Z}_3}+2[4,0]_{\mathbb{Z}_3} \,,\nonumber\\
&&[1,0]_{\mathbb{Z}_3}\cdot[2,0]_{\mathbb{Z}_3}=[3,0]_{\mathbb{Z}_3}+[2,1]_{\mathbb{Z}_3}+[1,2]_{\mathbb{Z}_3} \,,\quad [1,0]_{\mathbb{Z}_3}\cdot[3,0]_{\mathbb{Z}_3}=[4,0]_{\mathbb{Z}_3}+[3,1]_{\mathbb{Z}_3}+[1,3]_{\mathbb{Z}_3} \,,\nonumber\\
&&[1,0]_{\mathbb{Z}_3}\cdot[4,0]_{\mathbb{Z}_3}=3[0,0]_{\mathbb{Z}_3}+[2,1]_{\mathbb{Z}_3}+[1,2]_{\mathbb{Z}_3} \,,\quad [1,0]_{\mathbb{Z}_3}\cdot[2,1]_{\mathbb{Z}_3}=[3,1]_{\mathbb{Z}_3}+[3,0]_{\mathbb{Z}_3}+[1,0]_{\mathbb{Z}_3} \,,\nonumber\\
&&[1,0]_{\mathbb{Z}_3}\cdot[3,1]_{\mathbb{Z}_3}=[2,1]_{\mathbb{Z}_3}+[3,1]_{\mathbb{Z}_3}+[2,0]_{\mathbb{Z}_3} \,,\quad [1,0]_{\mathbb{Z}_3}\cdot[1,2]_{\mathbb{Z}_3}=[3,0]_{\mathbb{Z}_3}+[1,3]_{\mathbb{Z}_3}+[1,0]_{\mathbb{Z}_3} \,,\nonumber\\
&&[1,0]_{\mathbb{Z}_3}\cdot[1,3]_{\mathbb{Z}_3}=[1,3]_{\mathbb{Z}_3}+[1,2]_{\mathbb{Z}_3}+[2,0]_{\mathbb{Z}_3} \,.
\end{eqnarray}
The remaining fusion rules can be obtained from Eq.~\eqref{eq:n=5partfr} by making use of the symmetries in Eq.~\eqref{eq:n=5symmetry2}. To construct a three-point coupling containing $[0,0]_{\mathbb{Z}_3}$, it is sufficient to first consider a two-point fusion and then fuse the resulting class with the inverse of either class appearing in the two-point fusion product.

For three generations of quarks $\bar{Q}_L$ or $q_R$, there are $C_{9+3-1}^3-9=156$ possible class assignments, after excluding the cases in which all three generations belong to the same class. Owing to the symmetries of the fusion rules described above, the Higgs field can, without loss of generality, be assigned to either
$[0,0]_{\mathbb{Z}_3}$ or $[1,0]_{\mathbb{Z}_3}$. By enumerating all possible class assignments for $\bar{Q}_L$, $q_R$ and $H$, we obtain fourteen full-rank Yukawa matrix textures and list them in table~\ref{tab:z3n5-yukawa-textures}. Compared with the $N=4$ case, the two-zero texture:
\begin{eqnarray}
\begin{pmatrix}
 \times & \times & \times \\
 \times & \times & 0 \\
 \times & 0 & \times \\
\end{pmatrix}
\end{eqnarray}
is not realizable through the $\mathbb{Z}_3$ gauging of $\mathbb{Z}_5\times\mathbb{Z}'_5$, while all the other textures can be obtained.

\begin{table}[hptb!]
\centering
\setlength{\extrarowheight}{0pt}
\begin{tabular}{|>{\scriptsize}c|c|c|c|}
\hline\hline
\rule{0pt}{3.5ex}{\normalsize $Y_q$} & $\bar{Q}_L$ & $q_R$ & $H$ \\
\hline
\multirow{2}{*}{$\begin{pmatrix} \times & 0 & 0 \\ 0 & \times & 0 \\ 0 & 0 & \times \\ \end{pmatrix}$} & $[{2, 0}]_{\mathbb{Z}_3},[{1, 0}]_{\mathbb{Z}_3},[{0, 0}]_{\mathbb{Z}_3}$ & $[{3, 0}]_{\mathbb{Z}_3},[{4, 0}]_{\mathbb{Z}_3},[{0, 0}]_{\mathbb{Z}_3}$ & $[{0, 0}]_{\mathbb{Z}_3}$ \\
\cline{2-4}
 & $[{2, 0}]_{\mathbb{Z}_3},[{1, 0}]_{\mathbb{Z}_3},[{0, 0}]_{\mathbb{Z}_3}$ & $[{2, 0}]_{\mathbb{Z}_3},[{1, 0}]_{\mathbb{Z}_3},[{4, 0}]_{\mathbb{Z}_3}$ & $[{1, 0}]_{\mathbb{Z}_3}$ \\
\hline
$\begin{pmatrix} \times & \times & 0 \\ 0 & \times & 0 \\ 0 & 0 & \times \\ \end{pmatrix}$ & $[{1, 0}]_{\mathbb{Z}_3},[{3, 0}]_{\mathbb{Z}_3},[{0, 0}]_{\mathbb{Z}_3}$ & $[{3, 0}]_{\mathbb{Z}_3},[{1, 0}]_{\mathbb{Z}_3},[{4, 0}]_{\mathbb{Z}_3}$ & $[{1, 0}]_{\mathbb{Z}_3}$ \\
\hline
\multirow{2}{*}{$\begin{pmatrix} \times & \times & 0 \\ \times & \times & 0 \\ 0 & 0 & \times \\ \end{pmatrix}$} & $[{0, 0}]_{\mathbb{Z}_3},[{0, 0}]_{\mathbb{Z}_3},[{1, 0}]_{\mathbb{Z}_3}$ & $[{0, 0}]_{\mathbb{Z}_3},[{0, 0}]_{\mathbb{Z}_3},[{4, 0}]_{\mathbb{Z}_3}$ & $[{0, 0}]_{\mathbb{Z}_3}$ \\
\cline{2-4}
 & $[{0, 0}]_{\mathbb{Z}_3},[{0, 0}]_{\mathbb{Z}_3},[{1, 0}]_{\mathbb{Z}_3}$ & $[{4, 0}]_{\mathbb{Z}_3},[{4, 0}]_{\mathbb{Z}_3},[{1, 0}]_{\mathbb{Z}_3}$ & $[{1, 0}]_{\mathbb{Z}_3}$ \\
\hline
$\begin{pmatrix} \times & \times & 0 \\ 0 & \times & \times \\ 0 & 0 & \times \\ \end{pmatrix}$ & $[{2, 0}]_{\mathbb{Z}_3},[{2, 1}]_{\mathbb{Z}_3},[{0, 0}]_{\mathbb{Z}_3}$ & $[{2, 1}]_{\mathbb{Z}_3},[{2, 0}]_{\mathbb{Z}_3},[{4, 0}]_{\mathbb{Z}_3}$ & $[{1, 0}]_{\mathbb{Z}_3}$ \\
\hline
$\begin{pmatrix} \times & \times & \times \\ 0 & \times & 0 \\ 0 & 0 & \times \\ \end{pmatrix}$ & $[{2, 1}]_{\mathbb{Z}_3},[{0, 0}]_{\mathbb{Z}_3},[{2, 0}]_{\mathbb{Z}_3}$ & $[{1, 3}]_{\mathbb{Z}_3},[{4, 0}]_{\mathbb{Z}_3},[{2, 0}]_{\mathbb{Z}_3}$ & $[{1, 0}]_{\mathbb{Z}_3}$ \\
\hline
$\begin{pmatrix} \times & \times & 0 \\ 0 & \times & 0 \\ 0 & \times & \times \\ \end{pmatrix}$ & $[{1, 2}]_{\mathbb{Z}_3},[{0, 0}]_{\mathbb{Z}_3},[{2, 1}]_{\mathbb{Z}_3}$ & $[{3, 1}]_{\mathbb{Z}_3},[{4, 0}]_{\mathbb{Z}_3},[{1, 3}]_{\mathbb{Z}_3}$ & $[{1, 0}]_{\mathbb{Z}_3}$ \\
\hline
$\begin{pmatrix} \times & \times & 0 \\ 0 & \times & \times \\ \times & 0 & \times \\ \end{pmatrix}$ & $[{3, 0}]_{\mathbb{Z}_3},[{1, 0}]_{\mathbb{Z}_3},[{3, 1}]_{\mathbb{Z}_3}$ & $[{1, 3}]_{\mathbb{Z}_3},[{1, 0}]_{\mathbb{Z}_3},[{3, 0}]_{\mathbb{Z}_3}$ & $[{1, 0}]_{\mathbb{Z}_3}$ \\
\hline
$\begin{pmatrix} \times & \times & \times \\ 0 & \times & \times \\ 0 & 0 & \times \\ \end{pmatrix}$ & $[{1, 2}]_{\mathbb{Z}_3},[{2, 1}]_{\mathbb{Z}_3},[{0, 0}]_{\mathbb{Z}_3}$ & $[{3, 1}]_{\mathbb{Z}_3},[{2, 0}]_{\mathbb{Z}_3},[{4, 0}]_{\mathbb{Z}_3}$ & $[{1, 0}]_{\mathbb{Z}_3}$ \\
\hline
$\begin{pmatrix} \times & \times & \times \\ \times & \times & 0 \\ 0 & 0 & \times \\ \end{pmatrix}$ & $[{2, 1}]_{\mathbb{Z}_3},[{2, 0}]_{\mathbb{Z}_3},[{0, 0}]_{\mathbb{Z}_3}$ & $[{2, 0}]_{\mathbb{Z}_3},[{2, 0}]_{\mathbb{Z}_3},[{4, 0}]_{\mathbb{Z}_3}$ & $[{1, 0}]_{\mathbb{Z}_3}$ \\
\hline
$\begin{pmatrix} \times & \times & 0 \\ \times & \times & 0 \\ \times & 0 & \times \\ \end{pmatrix}$ & $[{1, 0}]_{\mathbb{Z}_3},[{1, 0}]_{\mathbb{Z}_3},[{3, 0}]_{\mathbb{Z}_3}$ & $[{1, 0}]_{\mathbb{Z}_3},[{3, 0}]_{\mathbb{Z}_3},[{3, 1}]_{\mathbb{Z}_3}$ & $[{1, 0}]_{\mathbb{Z}_3}$ \\
\hline
$\begin{pmatrix} \times & \times & \times \\ \times & \times & \times \\ 0 & 0 & \times \\ \end{pmatrix}$ & $[{2, 1}]_{\mathbb{Z}_3},[{2, 1}]_{\mathbb{Z}_3},[{0, 0}]_{\mathbb{Z}_3}$ & $[{2, 0}]_{\mathbb{Z}_3},[{2, 0}]_{\mathbb{Z}_3},[{4, 0}]_{\mathbb{Z}_3}$ & $[{1, 0}]_{\mathbb{Z}_3}$ \\
\hline
$\begin{pmatrix} \times & \times & \times \\ 0 & \times & \times \\ 0 & \times & \times \\ \end{pmatrix}$ & $[{2, 1}]_{\mathbb{Z}_3},[{0, 0}]_{\mathbb{Z}_3},[{0, 0}]_{\mathbb{Z}_3}$ & $[{2, 0}]_{\mathbb{Z}_3},[{4, 0}]_{\mathbb{Z}_3},[{4, 0}]_{\mathbb{Z}_3}$ & $[{1, 0}]_{\mathbb{Z}_3}$ \\
\hline
$\begin{pmatrix} \times & \times & \times \\ \times & \times & \times \\ \times & 0 & \times \\ \end{pmatrix}$ & $[{2, 1}]_{\mathbb{Z}_3},[{2, 1}]_{\mathbb{Z}_3},[{0, 0}]_{\mathbb{Z}_3}$ & $[{4, 0}]_{\mathbb{Z}_3},[{2, 0}]_{\mathbb{Z}_3},[{4, 0}]_{\mathbb{Z}_3}$ & $[{1, 0}]_{\mathbb{Z}_3}$ \\
\hline
$\begin{pmatrix} \times & \times & \times \\ \times & \times & \times \\ \times & \times & \times \\ \end{pmatrix}$ & $[{4, 0}]_{\mathbb{Z}_3},[{2, 0}]_{\mathbb{Z}_3},[{2, 0}]_{\mathbb{Z}_3}$ & $[{1, 2}]_{\mathbb{Z}_3},[{2, 1}]_{\mathbb{Z}_3},[{2, 1}]_{\mathbb{Z}_3}$ & $[{1, 0}]_{\mathbb{Z}_3}$ \\
\hline
\hline
\end{tabular}
\caption{The field assignments and the corresponding quark Yukawa textures which can be   generated from the $\mathbb{Z}_3$ gauging of the $\mathbb{Z}_5\times\mathbb{Z}'_5$ symmetry.}
\label{tab:z3n5-yukawa-textures}
\end{table}

\section{\label{sec:S3-gauging-ZNxZN}$S_3$ gauging of $\mathbb{Z}_N\times\mathbb{Z}_N'$ symmetry }

We now turn to the gauging of the $S_3$ of $\mathbb{Z}_N\times\mathbb{Z}_N'$ symmetry. As in section~\ref{sec:Z3-gauging-ZNxZN}, we use $a$ and $a'$ to denote the generators of $\mathbb{Z}_N$ and $\mathbb{Z'_N}$ respectively,
satisfying the multiplication relations in Eq.~\eqref{eq:ZNxZN-gener}. The $S_3$ group is generated by the $Z_3$ generator $b$ and an additional $Z_2$ generator $c$, with the following defining relations:
\begin{eqnarray}
b^3=c^2=(bc)^2=e\,.
\end{eqnarray}
The $S_3$ group has six elements including $e$, $b$, $b^2$, $c$, $bc$ and $b^2c$. The $S_3$ is an automorphism subgroup of $\mathbb{Z}_N\times\mathbb{Z}_N'$ and its action is given by
\begin{eqnarray}
\nonumber bab^{-1}=a^{-1}a'^{-1} \,,\quad ba'b^{-1}=a \,,\\
cac^{-1}=a'^{-1} \,,\quad ca'c^{-1}=a^{-1} \,.\label{eq:S3-gauge-rules}
\end{eqnarray}
Therefore the $S_3$ gauging identifies the following six elements,
\begin{eqnarray}
[k,\ell]_{S_3}\equiv \{a^ka'^\ell \,,\, a^{-k+\ell}a'^{-k} \,,\, a^{-\ell}a'^{k-\ell} \,,\, a^{-\ell}a'^{-k} \,,\, a^{-k+\ell}a'^{\ell} \,,\, a^{k}a'^{k-\ell}\} \,, \label{eq:delta6n2class}
\end{eqnarray}
for each integer $k$ and $\ell$. Since $\mathbb{Z}_3$ is a subgroup of $S_3$, compared to the $\mathbb{Z}_3$ gauging of $\mathbb{Z}_N\times\mathbb{Z}_N'$ symmetry, the $S_3$ gauging of $\mathbb{Z}_N\times\mathbb{Z}_N'$ class $[k,\ell]_{S_3}$ contains more elements than the $\mathbb{Z}_3$ invariant class $[k,\ell]_{\mathbb{Z}_3}$. More specifically, the class $[k,\ell]_{S_3}$ contains the elements of both $[k,\ell]_{\mathbb{Z}_3}$ and $[-\ell,-k]_{\mathbb{Z}_3}$, i.e.
\begin{eqnarray}
\label{eq:S3gauging-doble-Z3Gaug}[k,\ell]_{S_3}=[k,\ell]_{\mathbb{Z}_3}+[-\ell,-k]_{\mathbb{Z}_3}\,.
\end{eqnarray}
From Eq.~\eqref{eq:delta6n2class}, we know that the fusion rules between different equivalence classes are given by
\begin{eqnarray}
[k,\ell]_{S_3}\cdot[m,n]_{S_3}
&=&[k+m,\ell+n]_{S_3}+[k-m+n,\ell-m]_{S_3}+[k-n,\ell+m-n]_{S_3} \nonumber\\
&&+[k-n,\ell-m]_{S_3}+[k+n-m,\ell+n]_{S_3}+[k+m,\ell-n+m]_{S_3} \,.
\end{eqnarray}
As can be seen from Eq.~\eqref{eq:delta6n2class}, an equivalence class $[k,\ell]_{S_3}$ generally consists of six elements. This class contains only three elements if $k+\ell=0 \pmod N$ or $2k-\ell=0 \pmod N$ or $k-2\ell=0 \pmod N$. It can further reduce to a single element for $k=-\ell=N/3$ or $2N/3$ $\pmod{N}$, when $N$ is divisible by 3. Thus if $N/3$ is not an integer, there are three types of equivalence classes:
\begin{eqnarray}
\nonumber [0,0]_{S_3}&=&\{e\} \,,\\
\nonumber {}[p,-p]_{S_3}&=&\{a^{p}a'^{-p} \,,\, a^{-2p}a'^{-p} \,,\, a^{p}a'^{2p}\} \,,\\
{}[k,\ell]_{S_3}&=&\{a^ka'^\ell \,,\, a^{-k+\ell}a'^{-k} \,,\, a^{-\ell}a'^{k-\ell} \,,\, a^{-\ell}a'^{-k} \,,\, a^{-k+\ell}a'^{\ell} \,,\, a^{k}a'^{k-\ell}\} \,,\label{eq:s3class-Nneq3int}
\end{eqnarray}
where $p\neq0 \pmod{N}$ and $\ell\neq-k$, $2k$, $k/2$ or $(k+N)/2 \pmod{N}$. The corresponding fusion rules are
\begin{eqnarray}
&&[0,0]_{S_3}\cdot[0,0]_{S_3}=[0,0]_{S_3} \,,\quad [0,0]_{S_3}\cdot[k,-k]_{S_3}=[k,-k]_{S_3} \,,\quad [0,0]_{S_3}\cdot[k,\ell]_{S_3}=[k,\ell]_{S_3} \,,\nonumber\\
&&[p,-p]_{S_3}\cdot[q,-q]_{S_3}=[p+q,-p-q]_{S_3}+[p-2q,-p-q]_{S_3} \,,\nonumber\\
&&[p,-p]_{S_3}\cdot[k,\ell]_{S_3}=[p+k,-p+\ell]_{S_3}+[p-k+\ell,-p-k]_{S_3}+[p-\ell,-p+k-\ell]_{S_3} \,,\nonumber\\
&&[k,\ell]_{S_3}\cdot[m,n]_{S_3}=[k+m,\ell+n]_{S_3}+[k-m+n,\ell-m]_{S_3}+[k-n,\ell+m-n]_{S_3} \nonumber\\
&&\qquad\qquad\qquad+[k-n,\ell-m]_{S_3}+[k-m+n,\ell+n]_{S_3}+[k+m,\ell+m-n]_{S_3} \,. \label{eq:N/3neqintfr}
\end{eqnarray}
where $p,q\neq0 \pmod{N}$, $\ell\notin\{-k, 2k, k/2, (k+N)/2\}\pmod{N}$ and $n\notin\{-m, 2m, m/2,(m+N)/2\}\pmod{N}$. When $N/3$ is an integer, the equivalence classes can be classified into four types:
\begin{eqnarray}
\nonumber [0,0]_{S_3}&=&\{e\} \,,\\
\nonumber {}[r,-r]_{S_3}&=&\{a^{r}a'^{-r}\} \,,\\
\nonumber {}[p,-p]_{S_3}&=&\{a^{p}a'^{-p} \,,\, a^{-2p}a'^{-p} \,,\, a^{p}a'^{2p}\} \,,\\
{}[k,\ell]_{S_3}&=&\{a^ka'^\ell \,,\, a^{-k+\ell}a'^{-k} \,,\, a^{-\ell}a'^{k-\ell} \,,\, a^{-\ell}a'^{-k} \,,\, a^{-k+\ell}a'^{\ell} \,,\, a^{k}a'^{k-\ell}\} \,,\label{eq:s3class-Neq3int}
\end{eqnarray}
where $r=N/3$ or $2N/3\pmod{N}$, while $p$ and $\ell$ satisfy $p\neq0,N/3,2N/3\pmod{N}$ and $\ell\neq-k$, $2k$, $k/2$ or $(k+N)/2\pmod{N}$. The corresponding fusion rules are given by:
\begin{eqnarray}
&&[0,0]_{S_3}\cdot[0,0]_{S_3}=[0,0]_{S_3} \,,\quad [0,0]_{S_3}\cdot[r,-r]_{S_3}=[r,-r]_{S_3} \,,\nonumber\\
&&[0,0]_{S_3}\cdot[p,-p]_{S_3}=[p,-p]_{S_3} \,,\quad [0,0]_{S_3}\cdot[k,\ell]_{S_3}=[k,\ell]_{S_3} \,,\nonumber\\
&&[r,-r]_{S_3}\cdot[s,-s]_{S_3}=[r+s,-r-s]_{S_3}=\left\{\begin{array}{cc}
[2r,-2r] & (r=s) \\
{}[0,0] & (r=-s)\,,
\end{array}\right. \nonumber\\
&&[r,-r]_{S_3}\cdot[p,-p]_{S_3}=[r+p,-r-p]_{S_3} \,,\nonumber\\
&&[r,-r]_{S_3}\cdot[k,\ell]_{S_3}=[r+k,-r+\ell] \,,\nonumber\\
&&[p,-p]_{S_3}\cdot[q,-q]_{S_3}=[p+q,-p-q]_{S_3}+[p-2q,-p-q]_{S_3} \,,\nonumber\\
&&[p,-p]_{S_3}\cdot[k,\ell]_{S_3}=[p+k,-p+\ell]_{S_3}+[p-k+\ell,-p-k]_{S_3}+[p-\ell,-p+k-\ell]_{S_3} \,,\nonumber\\
&&[k,\ell]_{S_3}\cdot[m,n]_{S_3}=[k+m,\ell+n]_{S_3}+[k-m+n,\ell-m]_{S_3}+[k-n,\ell+m-n]_{S_3} \nonumber\\
&&\qquad\qquad\qquad+[k-n,\ell-m]_{S_3}+[k-m+n,\ell+n]_{S_3}+[k+m,\ell+m-n]_{S_3} \,. \label{eq:N/3=intfr}
\end{eqnarray}
where $r,s=N/3$ or $2N/3\pmod{N}$, $p,q\neq0,N/3,2N/3\pmod{N}$, $\ell\neq-k$, $2k$, $k/2$ or $(k+N)/2\pmod{N}$ and $n\neq-m$, $2m$, $m/2$ or $(m+N)/2\pmod{N}$. For specific values of $k$, $\ell$, $m$ and $n$, the equivalence class on the right-hand side of the fusion rules in Eqs.~\eqref{eq:N/3neqintfr} and~\eqref{eq:N/3=intfr} may contain fewer elements than the generic six. In such cases, an appropriate multiplicity factor must be included: a factor of 2 when a six-element class reduces to a three-element class, a factor of 6 when it reduces to a single element, and a factor of 3 when a three-element class reduces to a single element.

From Eq.~\eqref{eq:S3gauging-doble-Z3Gaug}, we observe that each $S_3$ equivalence class $[k,\ell]_{S_3}$ contains twice as many elements as its $\mathbb{Z}_3$ counterpart $[k,\ell]_{\mathbb{Z}_3}$. This difference has an important consequence for the Yukawa textures. The fusion rules of the $\mathbb{Z}_3$ gauging are more restrictive and therefore generate fewer allowed interactions than those of the $S_3$ gauging. Hence, an interaction forbidden under the $S_3$ gauging cannot become allowed under the corresponding $\mathbb{Z}_3$ gauging. In particular, any texture zero generated by the $S_3$ gauging necessarily survives in the corresponding Yukawa matrix of the $\mathbb{Z}_3$ gauging, provided the quark and Higgs fields are assigned to the corresponding classes~\footnote{ $[k,\ell]_{S_3}$ corresponds to either $[k,\ell]_{\mathbb{Z}_3}$ or $[-\ell, -k]_{\mathbb{Z}_3}$.}. In the following, we systematically explore all quark Yukawa textures that can be realized through the $S_3$ gauging of $\mathbb{Z}_N\times\mathbb{Z}'_{N}$ symmetry for $N=2,3,4,5$, considering all inequivalent assignments of the quark and Higgs fields.

\subsection{$N=2$}

For $N=2$, the $S_3$ gauging of $\mathbb{Z}_2\times\mathbb{Z}'_2$ symmetry gives rise to two equivalence classes,
\begin{eqnarray}
    [0,0]_{S_3}=\{e\} \,,\quad [1,0]_{S_3}=\{a,aa',a'\} \,.
\end{eqnarray}
The fusion rules of the two classes are
\begin{eqnarray}
    [0,0]_{S_3}\cdot[k,\ell]_{S_3}=[k,\ell]_{S_3} \,,\quad [1,0]_{S_3}\cdot[1,0]_{S_3}=3[0,0]_{S_3}+2[1,0]_{S_3} \,.
\end{eqnarray}
After examining all possible class assignments of the left- and right-handed quarks and the Higgs field, we find that four distinct full-rank Yukawa matrix textures can be realized:
\begin{eqnarray}
Y_q=\begin{pmatrix}
 \times & \times & 0 \\
 \times & \times & 0 \\
 0 & 0 & \times \\
\end{pmatrix} ~ \text{when} ~ (\bar{Q}_L;q_R;H)&\sim& ([{0, 0}]_{S_3},[{0, 0}]_{S_3},[{1, 0}]_{S_3};[{0, 0}]_{S_3},[{0, 0}]_{S_3},[{1, 0}]_{S_3};[{0, 0}]_{S_3})\,,\nonumber\\[-22pt]
&&([{1, 0}]_{S_3},[{1, 0}]_{S_3},[{0, 0}]_{S_3};[{1, 0}]_{S_3},[{1, 0}]_{S_3},[{0, 0}]_{S_3};[{0, 0}]_{S_3})\,,\,\nonumber\\
\end{eqnarray}

\begin{eqnarray}
Y_q=\begin{pmatrix}
 \times & \times & \times \\
 \times & \times & \times \\
 0 & 0 & \times \\
\end{pmatrix} ~ \text{when} ~ (\bar{Q}_L;q_R;H)&\sim& ([{1, 0}]_{S_3},[{1, 0}]_{S_3},[{0, 0}]_{S_3};[{0, 0}]_{S_3},[{0, 0}]_{S_3},[{1, 0}]_{S_3};[{1, 0}]_{S_3})\,,\,\nonumber\\[-22pt]
\end{eqnarray}

\begin{eqnarray}
Y_q=\begin{pmatrix}
 \times & \times & \times \\
 0 & \times & \times \\
 0 & \times & \times \\
\end{pmatrix} ~ \text{when} ~ (\bar{Q}_L;q_R;H)&\sim& ([{1, 0}]_{S_3},[{0, 0}]_{S_3},[{0, 0}]_{S_3};[{0, 0}]_{S_3},[{1, 0}]_{S_3},[{1, 0}]_{S_3};[{1, 0}]_{S_3})\,,\,\nonumber\\[-22pt]
\end{eqnarray}

\begin{eqnarray}
Y_q=\begin{pmatrix}
 \times & \times & \times \\
 \times & \times & \times \\
 \times & 0 & \times \\
\end{pmatrix} ~ \text{when} ~ (\bar{Q}_L;q_R;H)&\sim& ([{1, 0}]_{S_3},[{1, 0}]_{S_3},[{0, 0}]_{S_3};[{1, 0}]_{S_3},[{0, 0}]_{S_3},[{1, 0}]_{S_3};[{1, 0}]_{S_3})\,,\nonumber\\[-22pt]
\end{eqnarray}
where $q=u$ or $d$. It should be noted that, after excluding the cases with non-full-rank mass matrices, the textures generated by the $\mathbb{Z}_3$ gauging of $\mathbb{Z}_2\times\mathbb{Z}'_2$
in table~\ref{tab:322tex} coincide with those obtained from the $S_3$ gauging of $\mathbb{Z}_2\times\mathbb{Z}'_2$ considered here up to row and column permutations.

\subsection{\label{subsec:N3-Delta6N2}$N=3$}

For $N=3$, the nine elements of $\mathbb{Z}_3\times\mathbb{Z}_3$ form four equivalence classes under the action of the $S_3$ gauging,
\begin{eqnarray}
&&[0,0]_{S_3}=\{e\} \,,\quad [1,0]_{S_3}=\{a,aa',a',a^2,a^2a'^2,a'^2\} \,,\nonumber\\
&&[2,1]_{S_3}=\{a^2a'\} \,,\quad [1,2]_{S_3}=\{aa'^2\} \,.\label{eq:s3N=3class}
\end{eqnarray}
The corresponding fusion rules of these classes are summarized as follows:
\begin{eqnarray}
&&[0,0]_{S_3}\cdot[k,\ell]_{S_3}=[k,\ell]_{S_3} \,,\quad [1,0]_{S_3}\cdot[1,0]_{S_3}=6[0,0]_{S_3}+3[1,0]_{S_3}+6[2,1]_{S_3}+6[1,2]_{S_3} \,,\nonumber\\
&&[1,0]_{S_3}\cdot[2,1]_{S_3}=[1,0]_{S_3} \,,\quad [1,0]_{S_3}\cdot[1,2]_{S_3}=[1,0]_{S_3} \,,\quad [2,1]_{S_3}\cdot[2,1]_{S_3}=[1,2]_{S_3} \nonumber\\
&&[2,1]_{S_3}\cdot[1,2]_{S_3}=[0,0]_{S_3} \,,\quad [1,2]_{S_3}\cdot[1,2]_{S_3}=[2,1]_{S_3} \,.
\end{eqnarray}
The automorphism $a\leftrightarrow a'$ induces an exchange symmetry between the two classes $[2,1]_{S_3}\leftrightarrow[1,2]_{S_3}$, with respect to which the fusion rules are invariant. As a consequence of this symmetry, the classification of Higgs-field assignments can be restricted to the three classes $[0,0]_{S_3}$, $[1,0]_{S_3}$ and $[2,1]_{S_3}$. After systematically analyzing all possible class combinations associated with the left- and right-handed quarks and Higgs fields, we obtain five full-rank Yukawa matrix textures:
\begin{eqnarray}
Y_q=\begin{pmatrix}
 \times & 0 & 0 \\
 0 & \times & 0 \\
 0 & 0 & \times \\
\end{pmatrix} ~ \text{when} ~ (\bar{Q}_L;q_R;H)&\sim& ([{2, 1}]_{S_3},[{1, 0}]_{S_3},[{0, 0}]_{S_3};[{1, 2}]_{S_3},[{1, 0}]_{S_3},[{0, 0}]_{S_3};[{0, 0}]_{S_3})\,,\nonumber\\[-22pt]
&&([{2, 1}]_{S_3},[{1, 0}]_{S_3},[{0, 0}]_{S_3};[{2, 1}]_{S_3},[{1, 0}]_{S_3},[{1, 2}]_{S_3};[{2, 1}]_{S_3})\dots\nonumber\\
\end{eqnarray}

\begin{eqnarray}
Y_q=\begin{pmatrix}
 \times & \times & 0 \\
 \times & \times & 0 \\
 0 & 0 & \times \\
\end{pmatrix} ~ \text{when} ~ (\bar{Q}_L;q_R;H)&\sim& ([{0, 0}]_{S_3},[{0, 0}]_{S_3},[{1, 0}]_{S_3};[{0, 0}]_{S_3},[{0, 0}]_{S_3},[{1, 0}]_{S_3};[{0, 0}]_{S_3})\,,\nonumber\\[-22pt]
&&([{0, 0}]_{S_3},[{0, 0}]_{S_3},[{1, 0}]_{S_3};[{1, 2}]_{S_3},[{1, 2}]_{S_3},[{1, 0}]_{S_3};[{2, 1}]_{S_3}) \dots\nonumber\\
\end{eqnarray}

\begin{eqnarray}\label{eq:s3z3z3201}
Y_q=\begin{pmatrix}
 \times & \times & \times \\
 \times & \times & \times \\
 0 & 0 & \times \\
\end{pmatrix} ~ \text{when} ~ (\bar{Q}_L;q_R;H)&\sim& ([{1, 0}]_{S_3},[{1, 0}]_{S_3},[{0, 0}]_{S_3};[{0, 0}]_{S_3},[{0, 0}]_{S_3},[{1, 0}]_{S_3};[{1, 0}]_{S_3})\,\dots\nonumber\\[-22pt]
\end{eqnarray}

\begin{eqnarray}\label{eq:s3z3z3202}
Y_q=\begin{pmatrix}
 \times & \times & \times \\
 0 & \times & \times \\
 0 & \times & \times \\
\end{pmatrix} ~ \text{when} ~ (\bar{Q}_L;q_R;H)&\sim& ([{1, 0}]_{S_3},[{0, 0}]_{S_3},[{0, 0}]_{S_3};[{0, 0}]_{S_3},[{1, 0}]_{S_3},[{1, 0}]_{S_3};[{1, 0}]_{S_3})\,\dots\nonumber\\[-22pt]
\end{eqnarray}

\begin{eqnarray}\label{eq:s3z3z310}
Y_q=\begin{pmatrix}
 \times & \times & \times \\
 \times & \times & \times \\
 \times & 0 & \times \\
\end{pmatrix} ~ \text{when} ~ (\bar{Q}_L;q_R;H)&\sim& ([{1, 0}]_{S_3},[{1, 0}]_{S_3},[{0, 0}]_{S_3};[{1, 0}]_{S_3},[{0, 0}]_{S_3},[{1, 0}]_{S_3};[{1, 0}]_{S_3})\,\dots\nonumber\\[-22pt]
\end{eqnarray}
In comparison with the $\mathbb{Z}_3$ gauging of $\mathbb{Z}_3\times\mathbb{Z}'_3$, the $S_3$
gauging of $\mathbb{Z}_3\times\mathbb{Z}'_3$ allows for additional two-zero textures, given in Eqs.~\eqref{eq:s3z3z3201} and~\eqref{eq:s3z3z3202}, as well as the one-zero texture in Eq.~\eqref{eq:s3z3z310}.

\subsection{$N=4$}

For $N=4$, the following five equivalence classes are identified,
\begin{eqnarray}
&&[0,0]_{S_3}=\{e\} \,,\quad [1,0]_{S_3}=\{a,aa',a',a^3,a^3a'^3,a'^3\} \,,\nonumber\\
&&[2,0]_{S_3}=\{a^2,a^2a'^2,a'^2\} \,,\quad [2,1]_{S_3}=\{a^2a',a^3a',a^3a'^2\} \,,\nonumber\\
&&[1,2]_{S_3}=\{aa'^2,aa'^3,a^2a'^3\} \,.
\end{eqnarray}
The fusion rules involving these classes can be written as
\begin{eqnarray}
&&[0,0]_{S_3}\cdot[k,\ell]_{S_3}=[k,\ell]_{S_3} \,,\quad [1,0]_{S_3}\cdot[1,0]_{S_3}=6[0,0]_{S_3}+2[1,0]_{S_3}+2[2,0]_{S_3}+2[2,1]_{S_3}+2[1,2]_{S_3} \,,\nonumber\\
&&[1,0]_{S_3}\cdot[2,0]_{S_3}=[1,0]_{S_3}+2[2,1]_{S_3}+2[1,2]_{S_3} \,,\quad [1,0]_{S_3}\cdot[2,1]_{S_3}=[1,0]_{S_3}+2[2,0]_{S_3}+2[2,1]_{S_3} \,,\nonumber\\
&&[1,0]_{S_3}\cdot[1,2]_{S_3}=[1,0]_{S_3}+2[2,0]_{S_3}+2[1,2]_{S_3} \,,\quad [2,0]_{S_3}\cdot[2,0]_{S_3}=3[0,0]_{S_3}+2[2,0]_{S_3} \,,\nonumber\\
&&[2,0]_{S_3}\cdot[2,1]_{S_3}=[1,0]_{S_3}+[1,2]_{S_3} \,,\quad [2,0]_{S_3}\cdot[1,2]_{S_3}=[1,0]_{S_3}+[2,1]_{S_3} \,,\nonumber\\
&&[2,1]_{S_3}\cdot[2,1]_{S_3}=[2,0]_{S_3}+2[1,2]_{S_3} \,,\quad [2,1]_{S_3}\cdot[1,2]_{S_3}=3[0,0]_{S_3}+[1,0]_{S_3} \nonumber\\
&&[1,2]_{S_3}\cdot[1,2]_{S_3}=[2,0]_{S_3}+2[2,1]_{S_3}\,.
\end{eqnarray}
The automorphism $a\leftrightarrow a'$ gives rise to the class transformation $[2,1]_{S_3}\leftrightarrow[1,2]_{S_3}$ which leaves the fusion rules invariant. Therefore, without loss of generality, it suffices to consider the four classes $[0,0]_{S_3}$, $[1,0]_{S_3}$, $[2,0]_{S_3}$ and $[2,1]_{S_3}$ for the Higgs-field assignment. A systematic analysis of all possible class assignments for the left- and right-handed quarks and the Higgs field yields the fifteen full-rank Yukawa matrix textures which are listed in table~\ref{tab:s3n4-yukawa-textures}. We see that, up to row and column permutations, the $S_3$ gauging of $\mathbb{Z}_4\times\mathbb{Z}'_4$ symmetry generates the same set of all 15 textures as the corresponding $\mathbb{Z}_3$ gauging.

\begin{table}[hptb!]
\centering
\setlength{\extrarowheight}{0pt}
\begin{tabular}{|>{\scriptsize}c|c|c|c|}
\hline\hline
\rule{0pt}{3.5ex}{\normalsize $Y_q$} & $\bar{Q}_L$ & $q_R$ & $H$ \\
\hline
\multirow{3}{*}{$\begin{pmatrix} \times & 0 & 0 \\ 0 & \times & 0 \\ 0 & 0 & \times \\ \end{pmatrix}$} & $[{2, 0}]_{S_3},[{1, 0}]_{S_3},[{0, 0}]_{S_3}$ & $[{2, 0}]_{S_3},[{1, 0}]_{S_3},[{0, 0}]_{S_3}$ & $[{0, 0}]_{S_3}$ \\
\cline{2-4}
 & $[{0, 0}]_{S_3},[{2, 1}]_{S_3},[{1, 2}]_{S_3}$ & $[{2, 0}]_{S_3},[{2, 1}]_{S_3},[{1, 2}]_{S_3}$ & $[{2, 0}]_{S_3}$ \\
\cline{2-4}
 & $[{2, 1}]_{S_3},[{2, 0}]_{S_3},[{0, 0}]_{S_3}$ & $[{2, 0}]_{S_3},[{1, 0}]_{S_3},[{1, 2}]_{S_3}$ & $[{2, 1}]_{S_3}$ \\
\hline
\multirow{2}{*}{$\begin{pmatrix} \times & \times & 0 \\ 0 & \times & 0 \\ 0 & 0 & \times \\ \end{pmatrix}$} & $[{2, 0}]_{S_3},[{0, 0}]_{S_3},[{1, 0}]_{S_3}$ & $[{0, 0}]_{S_3},[{2, 0}]_{S_3},[{1, 0}]_{S_3}$ & $[{2, 0}]_{S_3}$ \\
\cline{2-4}
 & $[{1, 0}]_{S_3},[{0, 0}]_{S_3},[{2, 0}]_{S_3}$ & $[{2, 0}]_{S_3},[{1, 2}]_{S_3},[{2, 1}]_{S_3}$ & $[{2, 1}]_{S_3}$ \\
\hline
\multirow{4}{*}{$\begin{pmatrix} \times & \times & 0 \\ \times & \times & 0 \\ 0 & 0 & \times \\ \end{pmatrix}$} & $[{0, 0}]_{S_3},[{0, 0}]_{S_3},[{1, 0}]_{S_3}$ & $[{0, 0}]_{S_3},[{0, 0}]_{S_3},[{1, 0}]_{S_3}$ & $[{0, 0}]_{S_3}$ \\
\cline{2-4}
 & $[{2, 0}]_{S_3},[{2, 0}]_{S_3},[{2, 1}]_{S_3}$ & $[{2, 1}]_{S_3},[{2, 1}]_{S_3},[{2, 0}]_{S_3}$ & $[{1, 0}]_{S_3}$ \\
\cline{2-4}
 & $[{0, 0}]_{S_3},[{0, 0}]_{S_3},[{1, 0}]_{S_3}$ & $[{2, 0}]_{S_3},[{2, 0}]_{S_3},[{1, 0}]_{S_3}$ & $[{2, 0}]_{S_3}$ \\
\cline{2-4}
 & $[{0, 0}]_{S_3},[{0, 0}]_{S_3},[{2, 0}]_{S_3}$ & $[{1, 2}]_{S_3},[{1, 2}]_{S_3},[{1, 0}]_{S_3}$ & $[{2, 1}]_{S_3}$ \\
\hline
$\begin{pmatrix} \times & \times & 0 \\ 0 & \times & \times \\ 0 & 0 & \times \\ \end{pmatrix}$ & $[{2, 0}]_{S_3},[{1, 0}]_{S_3},[{0, 0}]_{S_3}$ & $[{2, 1}]_{S_3},[{1, 0}]_{S_3},[{1, 2}]_{S_3}$ & $[{2, 1}]_{S_3}$ \\
\hline
\multirow{2}{*}{$\begin{pmatrix} \times & \times & \times \\ 0 & \times & 0 \\ 0 & 0 & \times \\ \end{pmatrix}$} & $[{1, 0}]_{S_3},[{2, 0}]_{S_3},[{2, 1}]_{S_3}$ & $[{0, 0}]_{S_3},[{2, 1}]_{S_3},[{2, 0}]_{S_3}$ & $[{1, 0}]_{S_3}$ \\
\cline{2-4}
 & $[{1, 0}]_{S_3},[{0, 0}]_{S_3},[{2, 0}]_{S_3}$ & $[{2, 0}]_{S_3},[{1, 2}]_{S_3},[{1, 0}]_{S_3}$ & $[{2, 1}]_{S_3}$ \\
\hline
\multirow{2}{*}{$\begin{pmatrix} \times & \times & 0 \\ 0 & \times & 0 \\ 0 & \times & \times \\ \end{pmatrix}$} & $[{2, 0}]_{S_3},[{0, 0}]_{S_3},[{2, 1}]_{S_3}$ & $[{2, 1}]_{S_3},[{1, 0}]_{S_3},[{2, 0}]_{S_3}$ & $[{1, 0}]_{S_3}$ \\
\cline{2-4}
 & $[{1, 0}]_{S_3},[{2, 0}]_{S_3},[{1, 2}]_{S_3}$ & $[{2, 0}]_{S_3},[{1, 0}]_{S_3},[{0, 0}]_{S_3}$ & $[{2, 1}]_{S_3}$ \\
\hline
\multirow{2}{*}{$\begin{pmatrix} \times & \times & 0 \\ 0 & \times & \times \\ \times & 0 & \times \\ \end{pmatrix}$} & $[{2, 1}]_{S_3},[{2, 0}]_{S_3},[{1, 2}]_{S_3}$ & $[{2, 0}]_{S_3},[{1, 2}]_{S_3},[{2, 1}]_{S_3}$ & $[{1, 0}]_{S_3}$ \\
\cline{2-4}
 & $[{2, 0}]_{S_3},[{2, 1}]_{S_3},[{1, 0}]_{S_3}$ & $[{1, 0}]_{S_3},[{2, 1}]_{S_3},[{2, 0}]_{S_3}$ & $[{2, 1}]_{S_3}$ \\
\hline
$\begin{pmatrix} \times & \times & \times \\ 0 & \times & \times \\ 0 & 0 & \times \\ \end{pmatrix}$ & $[{1, 0}]_{S_3},[{2, 0}]_{S_3},[{0, 0}]_{S_3}$ & $[{0, 0}]_{S_3},[{2, 1}]_{S_3},[{1, 0}]_{S_3}$ & $[{1, 0}]_{S_3}$ \\
\hline
\multirow{3}{*}{$\begin{pmatrix} \times & \times & \times \\ \times & \times & 0 \\ 0 & 0 & \times \\ \end{pmatrix}$} & $[{1, 0}]_{S_3},[{2, 1}]_{S_3},[{2, 0}]_{S_3}$ & $[{2, 0}]_{S_3},[{2, 0}]_{S_3},[{2, 1}]_{S_3}$ & $[{1, 0}]_{S_3}$ \\
\cline{2-4}
 & $[{1, 0}]_{S_3},[{2, 1}]_{S_3},[{1, 2}]_{S_3}$ & $[{2, 1}]_{S_3},[{2, 1}]_{S_3},[{1, 2}]_{S_3}$ & $[{2, 0}]_{S_3}$ \\
\cline{2-4}
 & $[{1, 0}]_{S_3},[{2, 0}]_{S_3},[{0, 0}]_{S_3}$ & $[{1, 0}]_{S_3},[{1, 0}]_{S_3},[{1, 2}]_{S_3}$ & $[{2, 1}]_{S_3}$ \\
\hline
\multirow{3}{*}{$\begin{pmatrix} \times & \times & 0 \\ \times & \times & 0 \\ \times & 0 & \times \\ \end{pmatrix}$} & $[{2, 0}]_{S_3},[{2, 0}]_{S_3},[{2, 1}]_{S_3}$ & $[{1, 0}]_{S_3},[{2, 1}]_{S_3},[{2, 0}]_{S_3}$ & $[{1, 0}]_{S_3}$ \\
\cline{2-4}
 & $[{2, 1}]_{S_3},[{2, 1}]_{S_3},[{1, 2}]_{S_3}$ & $[{1, 0}]_{S_3},[{2, 1}]_{S_3},[{1, 2}]_{S_3}$ & $[{2, 0}]_{S_3}$ \\
\cline{2-4}
 & $[{1, 0}]_{S_3},[{1, 0}]_{S_3},[{2, 0}]_{S_3}$ & $[{1, 0}]_{S_3},[{2, 0}]_{S_3},[{2, 1}]_{S_3}$ & $[{2, 1}]_{S_3}$ \\
\hline
\multirow{2}{*}{$\begin{pmatrix} \times & \times & \times \\ \times & \times & 0 \\ \times & 0 & \times \\ \end{pmatrix}$} & $[{1, 0}]_{S_3},[{2, 0}]_{S_3},[{2, 1}]_{S_3}$ & $[{1, 0}]_{S_3},[{2, 1}]_{S_3},[{2, 0}]_{S_3}$ & $[{1, 0}]_{S_3}$ \\
\cline{2-4}
 & $[{1, 0}]_{S_3},[{1, 2}]_{S_3},[{2, 1}]_{S_3}$ & $[{1, 0}]_{S_3},[{1, 2}]_{S_3},[{2, 1}]_{S_3}$ & $[{2, 0}]_{S_3}$ \\
\hline
\multirow{3}{*}{$\begin{pmatrix} \times & \times & \times \\ \times & \times & \times \\ 0 & 0 & \times \\ \end{pmatrix}$} & $[{1, 0}]_{S_3},[{1, 0}]_{S_3},[{0, 0}]_{S_3}$ & $[{0, 0}]_{S_3},[{0, 0}]_{S_3},[{1, 0}]_{S_3}$ & $[{1, 0}]_{S_3}$ \\
\cline{2-4}
 & $[{2, 0}]_{S_3},[{2, 0}]_{S_3},[{0, 0}]_{S_3}$ & $[{0, 0}]_{S_3},[{0, 0}]_{S_3},[{2, 0}]_{S_3}$ & $[{2, 0}]_{S_3}$ \\
\cline{2-4}
 & $[{1, 0}]_{S_3},[{1, 0}]_{S_3},[{0, 0}]_{S_3}$ & $[{1, 0}]_{S_3},[{1, 0}]_{S_3},[{1, 2}]_{S_3}$ & $[{2, 1}]_{S_3}$ \\
\hline
\multirow{3}{*}{$\begin{pmatrix} \times & \times & \times \\ 0 & \times & \times \\ 0 & \times & \times \\ \end{pmatrix}$} & $[{1, 0}]_{S_3},[{0, 0}]_{S_3},[{0, 0}]_{S_3}$ & $[{0, 0}]_{S_3},[{1, 0}]_{S_3},[{1, 0}]_{S_3}$ & $[{1, 0}]_{S_3}$ \\
\cline{2-4}
 & $[{2, 0}]_{S_3},[{0, 0}]_{S_3},[{0, 0}]_{S_3}$ & $[{0, 0}]_{S_3},[{2, 0}]_{S_3},[{2, 0}]_{S_3}$ & $[{2, 0}]_{S_3}$ \\
\cline{2-4}
 & $[{1, 0}]_{S_3},[{0, 0}]_{S_3},[{0, 0}]_{S_3}$ & $[{1, 0}]_{S_3},[{1, 2}]_{S_3},[{1, 2}]_{S_3}$ & $[{2, 1}]_{S_3}$ \\
\hline
\multirow{3}{*}{$\begin{pmatrix} \times & \times & \times \\ \times & \times & \times \\ \times & 0 & \times \\ \end{pmatrix}$} & $[{1, 0}]_{S_3},[{1, 0}]_{S_3},[{0, 0}]_{S_3}$ & $[{1, 0}]_{S_3},[{0, 0}]_{S_3},[{1, 0}]_{S_3}$ & $[{1, 0}]_{S_3}$ \\
\cline{2-4}
 & $[{2, 0}]_{S_3},[{2, 0}]_{S_3},[{0, 0}]_{S_3}$ & $[{2, 0}]_{S_3},[{0, 0}]_{S_3},[{2, 0}]_{S_3}$ & $[{2, 0}]_{S_3}$ \\
\cline{2-4}
 & $[{1, 0}]_{S_3},[{1, 0}]_{S_3},[{0, 0}]_{S_3}$ & $[{1, 2}]_{S_3},[{1, 0}]_{S_3},[{1, 2}]_{S_3}$ & $[{2, 1}]_{S_3}$ \\
\hline
\multirow{2}{*}{$\begin{pmatrix} \times & \times & \times \\ \times & \times & \times \\ \times & \times & \times \\ \end{pmatrix}$} & $[{2, 0}]_{S_3},[{1, 0}]_{S_3},[{1, 0}]_{S_3}$ & $[{2, 1}]_{S_3},[{1, 0}]_{S_3},[{1, 0}]_{S_3}$ & $[{1, 0}]_{S_3}$ \\
\cline{2-4}
 & $[{2, 1}]_{S_3},[{1, 0}]_{S_3},[{1, 0}]_{S_3}$ & $[{2, 1}]_{S_3},[{1, 0}]_{S_3},[{1, 0}]_{S_3}$ & $[{2, 0}]_{S_3}$ \\
\hline
\hline
\end{tabular}
\caption{The quark Yukawa textures generated by the $S_3$ gauging of the $\mathbb{Z}_4\times\mathbb{Z}'_4$ symmetry and the corresponding field assignments.}
\label{tab:s3n4-yukawa-textures}
\end{table}

\subsection{$N=5$}

For $N=5$, the $S_3$ gauging action partitions the twenty-five elements of $\mathbb{Z}_5\times\mathbb{Z}_5$ into seven equivalence classes,
\begin{eqnarray}
    &&[0,0]_{S_3}=\{e\} \,,\quad [1,0]_{S_3}=\{a,aa',a',a^4,a^4a'^4,a'^4\} \,,\nonumber\\
    &&[2,0]_{S_3}=\{a^2,a^2a'^2,a'^2,a^3,a^3a'^3,a'^3\} \,,\quad [2,1]_{S_3}=\{a^2a',a^4a',a^4a'^3\} \,,\nonumber\\
    &&[1,2]_{S_3}=\{aa'^2,aa'^4,a^3a'^4\} \,,\quad [3,1]_{S_3}=\{a^3a',a^3a'^2,a^4a'^2\} \,,\nonumber\\
    &&[1,3]_{S_3}=\{aa'^3,a^2a'^3,a^2a'^4\} \,.
\end{eqnarray}
The automorphism $(a, a')\rightarrow(a^2,a'^2)$ defines a symmetry of the fusion rules, under which the equivalence classes transform as follows
\begin{eqnarray}
[1,0]_{S_3}\leftrightarrow[2,0]_{S_3} \,,\quad [2,1]_{S_3}\rightarrow[3,1]_{S_3}\rightarrow[1,2]_{S_3}\rightarrow[1,3]_{S_3}\rightarrow[2,1]_{S_3} \,.\label{eq:sym-FR-S3N5}
\end{eqnarray}
After taking into account the symmetries of the fusion rules, it is sufficient to study the fusions involving at least one of the equivalence classes $[1,0]_{S_3}$ or $[2,1]_{S_3}$. The corresponding fusion rules are
\begin{eqnarray}
&&[0,0]_{S_3}\cdot[k,\ell]_{S_3}=[k,\ell]_{S_3} \,,\quad [1,0]_{S_3}\cdot[1,0]_{S_3}=[2,0]_{S_3}+2[1,0]_{S_3}+2[1,2]_{S_3}+6[0,0]_{S_3}+2[2,1]_{S_3} \,,\nonumber\\
&&[1,0]_{S_3}\cdot[2,0]_{S_3}=[2,0]_{S_3}+2[2,1]_{S_3}+2[1,2]_{S_3}+2[1,3]_{S_3}+[1,0]_{S_3}+2[3,1]_{S_3} \,,\nonumber\\
&&[1,0]_{S_3}\cdot[2,1]_{S_3}=[1,0]_{S_3}+[2,0]_{S_3}+2[3,1]_{S_3} \,,\quad [1,0]_{S_3}\cdot[3,1]_{S_3}=2[3,1]_{S_3}+2[2,1]_{S_3}+[2,0]_{S_3} \,,\nonumber\\
&&[1,0]_{S_3}\cdot[1,2]_{S_3}=2[1,3]_{S_3}+[2,0]_{S_3}+[1,0]_{S_3} \,,\quad [1,0]_{S_3}\cdot[1,3]_{S_3}=[2,0]_{S_3}+2[1,2]_{S_3}+2[1,3]_{S_3} \,,\nonumber\\
&&[2,1]_{S_3}\cdot[2,1]_{S_3}=[3,1]_{S_3}+2[1,2]_{S_3} \,,\quad
    [2,1]_{S_3}\cdot[3,1]_{S_3}=[1,3]_{S_3}+[2,0]_{S_3} \,,\nonumber\\
&&[2,1]_{S_3}\cdot[1,2]_{S_3}=3[0,0]_{S_3}+[2,0]_{S_3} \,,\quad [2,1]_{S_3}\cdot[1,3]_{S_3}=[1,2]_{S_3}+[1,0]_{S_3} \,. \label{eq:fusion-rules-S3N5}
\end{eqnarray}
The fusion rules for the remaining pairs of equivalence classes can be derived from Eq.~\eqref{eq:fusion-rules-S3N5} by making use of the symmetry given in Eq.~\eqref{eq:sym-FR-S3N5}. Therefore, without loss of generality, the Higgs field needs to be assigned only to $[0,0]_{S_3}$, $[1,0]_{S_3}$ or $[2,1]_{S_3}$. Considering all possible class assignments of the quark and Higgs fields, we obtain the fifteen full-rank Yukawa matrix textures in table~\ref{tab:s3n5-yukawa-textures}. We note that these textures are identical to those derived from the $S_3$ gauging of $\mathbb{Z}_4\times\mathbb{Z}'_4$ symmetry. Furthermore, the $S_3$ gauging of $\mathbb{Z}_5\times\mathbb{Z}'_5$ symmetry yields an additional two-zero texture $\begin{pmatrix} \times & \times & \times \\ \times & \times & 0 \\ \times & 0 & \times \\ \end{pmatrix}$, which cannot be obtained in the corresponding $\mathbb{Z}_3$ gauging case.

\begin{table}[hptb!]
\centering
\setlength{\extrarowheight}{0pt}
\begin{tabular}{|>{\scriptsize}c|c|c|c|}
\hline\hline
\rule{0pt}{3.5ex}{\normalsize $Y_q$} & $\bar{Q}_L$ & $q_R$ & $H$ \\
\hline
\multirow{3}{*}{$\begin{pmatrix} \times & 0 & 0 \\ 0 & \times & 0 \\ 0 & 0 & \times \\ \end{pmatrix}$} & $[{2, 0}]_{S_3},[{1, 0}]_{S_3},[{0, 0}]_{S_3}$ & $[{2, 0}]_{S_3},[{1, 0}]_{S_3},[{0, 0}]_{S_3}$ & $[{0, 0}]_{S_3}$ \\
\cline{2-4}
 & $[{0, 0}]_{S_3},[{1, 3}]_{S_3},[{3, 1}]_{S_3}$ & $[{1, 0}]_{S_3},[{2, 1}]_{S_3},[{1, 2}]_{S_3}$ & $[{1, 0}]_{S_3}$ \\
\cline{2-4}
 & $[{2, 1}]_{S_3},[{1, 0}]_{S_3},[{0, 0}]_{S_3}$ & $[{2, 1}]_{S_3},[{1, 0}]_{S_3},[{1, 2}]_{S_3}$ & $[{2, 1}]_{S_3}$ \\
\hline
\multirow{2}{*}{$\begin{pmatrix} \times & \times & 0 \\ 0 & \times & 0 \\ 0 & 0 & \times \\ \end{pmatrix}$} & $[{1, 0}]_{S_3},[{0, 0}]_{S_3},[{3, 1}]_{S_3}$ & $[{0, 0}]_{S_3},[{1, 0}]_{S_3},[{1, 3}]_{S_3}$ & $[{1, 0}]_{S_3}$ \\
\cline{2-4}
 & $[{2, 0}]_{S_3},[{0, 0}]_{S_3},[{1, 0}]_{S_3}$ & $[{3, 1}]_{S_3},[{1, 2}]_{S_3},[{2, 0}]_{S_3}$ & $[{2, 1}]_{S_3}$ \\
\hline
\multirow{3}{*}{$\begin{pmatrix} \times & \times & 0 \\ \times & \times & 0 \\ 0 & 0 & \times \\ \end{pmatrix}$} & $[{0, 0}]_{S_3},[{0, 0}]_{S_3},[{1, 0}]_{S_3}$ & $[{0, 0}]_{S_3},[{0, 0}]_{S_3},[{1, 0}]_{S_3}$ & $[{0, 0}]_{S_3}$ \\
\cline{2-4}
 & $[{0, 0}]_{S_3},[{0, 0}]_{S_3},[{3, 1}]_{S_3}$ & $[{1, 0}]_{S_3},[{1, 0}]_{S_3},[{2, 0}]_{S_3}$ & $[{1, 0}]_{S_3}$ \\
\cline{2-4}
 & $[{0, 0}]_{S_3},[{0, 0}]_{S_3},[{1, 0}]_{S_3}$ & $[{1, 2}]_{S_3},[{1, 2}]_{S_3},[{1, 0}]_{S_3}$ & $[{2, 1}]_{S_3}$ \\
\hline
\multirow{2}{*}{$\begin{pmatrix} \times & \times & 0 \\ 0 & \times & \times \\ 0 & 0 & \times \\ \end{pmatrix}$} & $[{3, 1}]_{S_3},[{1, 0}]_{S_3},[{0, 0}]_{S_3}$ & $[{1, 3}]_{S_3},[{2, 0}]_{S_3},[{1, 0}]_{S_3}$ & $[{1, 0}]_{S_3}$ \\
\cline{2-4}
 & $[{1, 0}]_{S_3},[{2, 0}]_{S_3},[{0, 0}]_{S_3}$ & $[{2, 0}]_{S_3},[{1, 0}]_{S_3},[{1, 2}]_{S_3}$ & $[{2, 1}]_{S_3}$ \\
\hline
\multirow{2}{*}{$\begin{pmatrix} \times & \times & \times \\ 0 & \times & 0 \\ 0 & 0 & \times \\ \end{pmatrix}$} & $[{1, 0}]_{S_3},[{3, 1}]_{S_3},[{0, 0}]_{S_3}$ & $[{0, 0}]_{S_3},[{2, 0}]_{S_3},[{1, 0}]_{S_3}$ & $[{1, 0}]_{S_3}$ \\
\cline{2-4}
 & $[{2, 0}]_{S_3},[{1, 0}]_{S_3},[{0, 0}]_{S_3}$ & $[{3, 1}]_{S_3},[{1, 0}]_{S_3},[{1, 2}]_{S_3}$ & $[{2, 1}]_{S_3}$ \\
\hline
\multirow{2}{*}{$\begin{pmatrix} \times & \times & 0 \\ 0 & \times & 0 \\ 0 & \times & \times \\ \end{pmatrix}$} & $[{2, 0}]_{S_3},[{0, 0}]_{S_3},[{1, 0}]_{S_3}$ & $[{3, 1}]_{S_3},[{1, 0}]_{S_3},[{0, 0}]_{S_3}$ & $[{1, 0}]_{S_3}$ \\
\cline{2-4}
 & $[{1, 3}]_{S_3},[{2, 0}]_{S_3},[{1, 0}]_{S_3}$ & $[{2, 1}]_{S_3},[{1, 0}]_{S_3},[{2, 0}]_{S_3}$ & $[{2, 1}]_{S_3}$ \\
\hline
\multirow{2}{*}{$\begin{pmatrix} \times & \times & 0 \\ 0 & \times & \times \\ \times & 0 & \times \\ \end{pmatrix}$} & $[{2, 1}]_{S_3},[{3, 1}]_{S_3},[{1, 0}]_{S_3}$ & $[{1, 0}]_{S_3},[{1, 3}]_{S_3},[{1, 2}]_{S_3}$ & $[{1, 0}]_{S_3}$ \\
\cline{2-4}
 & $[{3, 1}]_{S_3},[{1, 0}]_{S_3},[{2, 0}]_{S_3}$ & $[{3, 1}]_{S_3},[{2, 0}]_{S_3},[{1, 0}]_{S_3}$ & $[{2, 1}]_{S_3}$ \\
\hline
$\begin{pmatrix} \times & \times & \times \\ 0 & \times & \times \\ 0 & 0 & \times \\ \end{pmatrix}$ & $[{1, 0}]_{S_3},[{2, 0}]_{S_3},[{0, 0}]_{S_3}$ & $[{0, 0}]_{S_3},[{2, 0}]_{S_3},[{1, 0}]_{S_3}$ & $[{1, 0}]_{S_3}$ \\
\hline
\multirow{2}{*}{$\begin{pmatrix} \times & \times & \times \\ \times & \times & 0 \\ 0 & 0 & \times \\ \end{pmatrix}$} & $[{1, 0}]_{S_3},[{0, 0}]_{S_3},[{3, 1}]_{S_3}$ & $[{1, 0}]_{S_3},[{1, 0}]_{S_3},[{2, 0}]_{S_3}$ & $[{1, 0}]_{S_3}$ \\
\cline{2-4}
 & $[{2, 0}]_{S_3},[{1, 0}]_{S_3},[{0, 0}]_{S_3}$ & $[{1, 0}]_{S_3},[{1, 0}]_{S_3},[{1, 2}]_{S_3}$ & $[{2, 1}]_{S_3}$ \\
\hline
\multirow{2}{*}{$\begin{pmatrix} \times & \times & 0 \\ \times & \times & 0 \\ \times & 0 & \times \\ \end{pmatrix}$} & $[{1, 0}]_{S_3},[{1, 0}]_{S_3},[{2, 0}]_{S_3}$ & $[{1, 0}]_{S_3},[{0, 0}]_{S_3},[{3, 1}]_{S_3}$ & $[{1, 0}]_{S_3}$ \\
\cline{2-4}
 & $[{1, 0}]_{S_3},[{1, 0}]_{S_3},[{2, 0}]_{S_3}$ & $[{1, 0}]_{S_3},[{2, 0}]_{S_3},[{1, 2}]_{S_3}$ & $[{2, 1}]_{S_3}$ \\
\hline
$\begin{pmatrix} \times & \times & \times \\ \times & \times & 0 \\ \times & 0 & \times \\ \end{pmatrix}$ & $[{2, 0}]_{S_3},[{2, 1}]_{S_3},[{1, 0}]_{S_3}$ & $[{1, 0}]_{S_3},[{1, 3}]_{S_3},[{2, 1}]_{S_3}$ & $[{1, 0}]_{S_3}$ \\
\hline
\multirow{2}{*}{$\begin{pmatrix} \times & \times & \times \\ \times & \times & \times \\ 0 & 0 & \times \\ \end{pmatrix}$} & $[{1, 0}]_{S_3},[{1, 0}]_{S_3},[{0, 0}]_{S_3}$ & $[{0, 0}]_{S_3},[{0, 0}]_{S_3},[{1, 0}]_{S_3}$ & $[{1, 0}]_{S_3}$ \\
\cline{2-4}
 & $[{2, 0}]_{S_3},[{2, 0}]_{S_3},[{0, 0}]_{S_3}$ & $[{1, 0}]_{S_3},[{1, 0}]_{S_3},[{1, 2}]_{S_3}$ & $[{2, 1}]_{S_3}$ \\
\hline
\multirow{2}{*}{$\begin{pmatrix} \times & \times & \times \\ 0 & \times & \times \\ 0 & \times & \times \\ \end{pmatrix}$} & $[{1, 0}]_{S_3},[{0, 0}]_{S_3},[{0, 0}]_{S_3}$ & $[{0, 0}]_{S_3},[{1, 0}]_{S_3},[{1, 0}]_{S_3}$ & $[{1, 0}]_{S_3}$ \\
\cline{2-4}
 & $[{2, 0}]_{S_3},[{0, 0}]_{S_3},[{0, 0}]_{S_3}$ & $[{1, 0}]_{S_3},[{1, 2}]_{S_3},[{1, 2}]_{S_3}$ & $[{2, 1}]_{S_3}$ \\
\hline
\multirow{2}{*}{$\begin{pmatrix} \times & \times & \times \\ \times & \times & \times \\ \times & 0 & \times \\ \end{pmatrix}$} & $[{1, 0}]_{S_3},[{1, 0}]_{S_3},[{0, 0}]_{S_3}$ & $[{1, 0}]_{S_3},[{0, 0}]_{S_3},[{1, 0}]_{S_3}$ & $[{1, 0}]_{S_3}$ \\
\cline{2-4}
 & $[{2, 0}]_{S_3},[{2, 0}]_{S_3},[{0, 0}]_{S_3}$ & $[{1, 2}]_{S_3},[{1, 0}]_{S_3},[{1, 2}]_{S_3}$ & $[{2, 1}]_{S_3}$ \\
\hline
$\begin{pmatrix} \times & \times & \times \\ \times & \times & \times \\ \times & \times & \times \\ \end{pmatrix}$ & $[{2, 0}]_{S_3},[{1, 0}]_{S_3},[{1, 0}]_{S_3}$ & $[{2, 0}]_{S_3},[{1, 0}]_{S_3},[{1, 0}]_{S_3}$ & $[{1, 0}]_{S_3}$ \\
\hline
\hline
\end{tabular}
\caption{The quark Yukawa textures generated by the $S_3$ gauging of the $\mathbb{Z}_5\times\mathbb{Z}'_5$ symmetry and the corresponding field assignments.}
\label{tab:s3n5-yukawa-textures}
\end{table}

\section{Symmetry realization and phenomenology of quark Yukawa textures}\label{sec:pheno-analysis}

As demonstrated in Appendix~\ref{app:10-zeros}, the experimentally measured quark mixing angles and CP-violating phase cannot be simultaneously reproduced when the quark Yukawa matrices $Y_u$ and $Y_d$ contain ten or more zero entries in total. We thus turn to the nine-zero textures of the up- and down-type quark Yukawa matrices that can arise from the $\mathbb{Z}_3$ or $S_3$ gauging of $\mathbb{Z}_N\times\mathbb{Z}'_N$ symmetries, and examine whether they can accommodate the experimental data. After eliminating singular Yukawa matrices, the nine zeros can be distributed between the two quark sectors in only two ways: six zeros in one sector and three in the other, or five zeros in one sector and four in the other.

\subsection{Diagonal $Y_u$ and three zeros in $Y_d$}\label{sec:6+3tex}

We first consider the case in which the up-type quark mass matrix is diagonal and the down-type quark mass matrix contains three zeros. In this basis, the CKM matrix arises entirely from the down-type quark sector. The case in which the roles of the up- and down-type sectors are interchanged can be treated analogously. A diagonal up-type quark mass matrix requires the three generations of left-handed quark doublets, as well as the three generations of right-handed up-type quarks, to be assigned to distinct classes. Distinct class assignments for the left-handed quark doublets can always be supplemented by suitable assignments for the Higgs fields and the right-handed up-type quarks, thereby making the up-type mass matrix diagonal.

As can be seen from section~\ref{subsec:N3-Delta3N2}, and section~\ref{subsec:N3-Delta6N2}, the three-zero textures can not be generated from neither $\mathbb{Z}_3$ gauging of $\mathbb{Z}_N\times \mathbb{Z}_N'$ nor $S_3$ gauging of $\mathbb{Z}_N\times \mathbb{Z}_N'$ when $N=2,3$. For models with only one Higgs field, we have found in $\mathbb{Z}_3$ gauging of $\mathbb{Z}_N\times \mathbb{Z}_N'$ and $S_3$ gauging of $\mathbb{Z}_N\times \mathbb{Z}_N'$ for $N=4,5$ some assignments that can generate a diagonal up-type quark mass matrix and a down-type quark mass matrix with three zeros. They are listed below up to row and column permutations.

For the $\mathbb{Z}_3$ gauging of $\mathbb{Z}_4\times \mathbb{Z}_4'$ symmetry, the representative assignments which can generate a diagonal $Y_u$ as well as a $Y_d$ with three-zero textures are given by
\begin{eqnarray}
&&Y_u=\begin{pmatrix}
    \times & 0 & 0 \\
    0 & \times & 0 \\
    0 & 0 & \times
\end{pmatrix}\,,\quad Y_d=\begin{pmatrix}
    \times & \times & 0 \\
    0 & \times & \times \\
    \times & 0 & \times
\end{pmatrix} \quad\text{when}\quad (\bar{Q}_L;u_R;d_R;H)\sim \nonumber\\
&&([{2, 0}]_{\mathbb{Z}_3},[{2, 1}]_{\mathbb{Z}_3},[{1, 0}]_{\mathbb{Z}_3};
[{3, 0}]_{\mathbb{Z}_3},[{1, 0}]_{\mathbb{Z}_3},[{0, 0}]_{\mathbb{Z}_3};
[{1, 0}]_{\mathbb{Z}_3},[{1, 2}]_{\mathbb{Z}_3},[{2, 0}]_{\mathbb{Z}_3};
[{1, 0}]_{\mathbb{Z}_3})\dots \label{eq:udiag-dthree-z3N4-1} \\
&&Y_u=\begin{pmatrix}
    \times & 0 & 0 \\
    0 & \times & 0 \\
    0 & 0 & \times
\end{pmatrix}\,,\quad Y_d=\begin{pmatrix}
    \times & \times & \times \\
    \times & \times & 0 \\
    0 & 0 & \times \\
\end{pmatrix} \quad\text{when}\quad (\bar{Q}_L;u_R;d_R;H)\sim \nonumber\\
&&([{1, 0}]_{\mathbb{Z}_3},[{2, 0}]_{\mathbb{Z}_3},[{2, 1}]_{\mathbb{Z}_3};
[{0, 0}]_{\mathbb{Z}_3},[{3, 0}]_{\mathbb{Z}_3},[{1, 0}]_{\mathbb{Z}_3};
[{1, 0}]_{\mathbb{Z}_3},[{1, 0}]_{\mathbb{Z}_3},[{2, 0}]_{\mathbb{Z}_3};
[{1, 0}]_{\mathbb{Z}_3})\dots \label{eq:udiag-dthree-z3N4-2}
\end{eqnarray}
For the $\mathbb{Z}_3$ gauging of $\mathbb{Z}_5\times \mathbb{Z}_5'$ symmetry, the textures of $Y_u$ and $Y_d$ and the corresponding representative assignments are given by:
\begin{eqnarray}
&&Y_u=\begin{pmatrix}
    \times & 0 & 0 \\
    0 & \times & 0 \\
    0 & 0 & \times
\end{pmatrix}\,,\quad Y_d=\begin{pmatrix}
    \times & \times & 0 \\
    0 & \times & \times \\
    \times & 0 & \times
\end{pmatrix} \quad\text{when}\quad (\bar{Q}_L;u_R;d_R;H)\sim \nonumber\\
&&([{3, 0}]_{\mathbb{Z}_3},[{3, 1}]_{\mathbb{Z}_3},[{1, 0}]_{\mathbb{Z}_3};
[{3, 0}]_{\mathbb{Z}_3},[{2, 0}]_{\mathbb{Z}_3},[{0, 0}]_{\mathbb{Z}_3};
[{1, 0}]_{\mathbb{Z}_3},[{1, 3}]_{\mathbb{Z}_3},[{3, 0}]_{\mathbb{Z}_3};
[{1, 0}]_{\mathbb{Z}_3})\dots \label{eq:udiag-dthree-z3N5-1}\\
&&Y_u=\begin{pmatrix}
\times & 0 & 0 \\
0 & \times & 0 \\
0 & 0 & \times
\end{pmatrix}\,,\quad Y_d=\begin{pmatrix}
\times & \times & \times \\
\times & \times & 0 \\
0 & 0 & \times
\end{pmatrix} \quad\text{when}\quad (\bar{Q}_L;u_R;d_R;H)\sim \nonumber\\
&&([{3, 1}]_{\mathbb{Z}_3},[{1, 0}]_{\mathbb{Z}_3},[{2, 0}]_{\mathbb{Z}_3};
[{2, 0}]_{\mathbb{Z}_3},[{0, 0}]_{\mathbb{Z}_3},[{4, 0}]_{\mathbb{Z}_3};
[{3, 0}]_{\mathbb{Z}_3},[{3, 0}]_{\mathbb{Z}_3},[{1, 2}]_{\mathbb{Z}_3};
[{1, 0}]_{\mathbb{Z}_3})\dots \label{eq:udiag-dthree-z3N5-2}
\end{eqnarray}
In the following, we proceed to investigate whether the $S_3$ gauging of $\mathbb{Z}_N\times \mathbb{Z}'_N$ symmetry can yield a diagonal $Y_u$ and a $Y_d$ texture containing three zeros. In the $S_3$ gauging of $\mathbb{Z}_4\times \mathbb{Z}_4'$ symmetry, the three-zero texture can be generated as follows,
\begin{eqnarray}
&&Y_u=\begin{pmatrix}
    \times & 0 & 0 \\
    0 & \times & 0 \\
    0 & 0 & \times
\end{pmatrix}\,,\quad Y_d=\begin{pmatrix}
    \times & \times & 0 \\
    0 & \times & \times \\
    \times & 0 & \times
\end{pmatrix} \quad\text{when}\quad (\bar{Q}_L;u_R;d_R;H)\sim \nonumber\\
&&([{2, 0}]_{S_3},[{2, 1}]_{S_3},[{1, 0}]_{S_3};
[{1, 2}]_{S_3},[{0, 0}]_{S_3},[{2, 0}]_{S_3};
[{1, 0}]_{S_3},[{2, 1}]_{S_3},[{2, 0}]_{S_3};
[{2, 1}]_{S_3})\dots \\
&&Y_u=\begin{pmatrix}
    \times & 0 & 0 \\
    0 & \times & 0 \\
    0 & 0 & \times
\end{pmatrix}\,,\quad Y_d=\begin{pmatrix}
    \times & \times & \times \\
    \times & \times & 0 \\
    0 & 0 & \times
\end{pmatrix} \quad\text{when}\quad (\bar{Q}_L;u_R;d_R;H)\sim \nonumber\\
&&([{1, 0}]_{S_3},[{2, 0}]_{S_3},[{2, 1}]_{S_3};
[{2, 0}]_{S_3},[{1, 2}]_{S_3},[{0, 0}]_{S_3};
[{1, 0}]_{S_3},[{1, 0}]_{S_3},[{2, 0}]_{S_3};
[{2, 1}]_{S_3})\dots
\end{eqnarray}
We find that the $\mathbb{Z}_3$ gauging of $\mathbb{Z}_4\times \mathbb{Z}_4'$ symmetry, the $\mathbb{Z}_3$ gauging of $\mathbb{Z}_5\times \mathbb{Z}_5'$ symmetry, and the $S_3$ gauging of $\mathbb{Z}_4\times \mathbb{Z}_4'$ symmetry yields the same set of of $Y_u$ and $Y_d$ textures. While the $S_3$ gauging of the $\mathbb{Z}_5\times\mathbb{Z}_5'$ symmetry can generate all textures in which $Y_d$ has three zeros and $Y_u$ is diagonal. The corresponding assignments for the quark and Higgs fields are given by
\begin{eqnarray}
&&Y_u=\begin{pmatrix}
\times & 0 & 0 \\
0 & \times & 0 \\
 0 & 0 & \times
\end{pmatrix}\,,\quad
Y_d=\begin{pmatrix}
\times & \times & 0 \\
0 & \times & \times \\
\times & 0 & \times
\end{pmatrix} \quad\text{when}\quad (\bar{Q}_L;u_R;d_R;H)\sim \nonumber\\
&&([{2, 0}]_{S_3},[{3, 1}]_{S_3},[{1, 0}]_{S_3};
[{2, 1}]_{S_3},[{1, 2}]_{S_3},[{2, 0}]_{S_3};
[{1, 0}]_{S_3},[{3, 1}]_{S_3},[{2, 0}]_{S_3};
[{2, 1}]_{S_3})\dots \label{eq:udiag-dthree-S3N5-1}\\
&&Y_u=\begin{pmatrix}
\times & 0 & 0 \\
0 & \times & 0 \\
 0 & 0 & \times
\end{pmatrix}\,,\quad
Y_d=\begin{pmatrix}
\times & \times & \times \\
0 & \times & \times \\
0 & 0 & \times
\end{pmatrix} \quad\text{when}\quad (\bar{Q}_L;u_R;d_R;H)\sim \nonumber\\
&&([{1, 0}]_{S_3},[{2, 1}]_{S_3},[{1, 3}]_{S_3};
[{0, 0}]_{S_3},[{1, 3}]_{S_3},[{3, 1}]_{S_3};
[{0, 0}]_{S_3},[{1, 0}]_{S_3},[{2, 0}]_{S_3};
[{1, 0}]_{S_3})\dots \label{eq:udiag-dthree-S3N5-2} \\
&&Y_u=\begin{pmatrix}
\times & 0 & 0 \\
0 & \times & 0 \\
0 & 0 & \times
\end{pmatrix}\,,\quad
Y_d=\begin{pmatrix}
\times & \times & \times \\
\times & \times & 0 \\
0 & 0 & \times
\end{pmatrix} \quad\text{when}\quad (\bar{Q}_L;u_R;d_R;H)\sim \nonumber\\
&&([{1, 0}]_{S_3},[{2, 1}]_{S_3},[{1, 3}]_{S_3};
[{0, 0}]_{S_3},[{1, 3}]_{S_3},[{3, 1}]_{S_3};
[{1, 0}]_{S_3},[{1, 0}]_{S_3},[{2, 1}]_{S_3};
[{1, 0}]_{S_3})\,,\nonumber\\
&&([{1, 0}]_{S_3},[{2, 0}]_{S_3},[{2, 1}]_{S_3};
[{3, 1}]_{S_3},[{2, 1}]_{S_3},[{0, 0}]_{S_3};
[{1, 0}]_{S_3},[{1, 0}]_{S_3},[{1, 3}]_{S_3};
[{2, 1}]_{S_3})\dots \label{eq:udiag-dthree-S3N5-3} \\
&&Y_u=\begin{pmatrix}
    \times & 0 & 0 \\
    0 & \times & 0 \\
    0 & 0 & \times
\end{pmatrix}\,,\quad Y_d=\begin{pmatrix}
    \times & \times & 0 \\
    \times & \times & 0 \\
    \times & 0 & \times
\end{pmatrix} \quad\text{when}\quad (\bar{Q}_L;u_R;d_R;H)\sim \nonumber\\
&&([{2, 1}]_{S_3},[{1, 0}]_{S_3},[{1, 3}]_{S_3};
[{1, 3}]_{S_3},[{0, 0}]_{S_3},[{3, 1}]_{S_3};
[{2, 0}]_{S_3},[{1, 0}]_{S_3},[{3, 1}]_{S_3};
[{1, 0}]_{S_3})\dots \label{eq:udiag-dthree-S3N5-4}
\end{eqnarray}

The three-zero textures presented in this section all contain, up to row and column permutations, a $2\times2$ submatrix of the form $\begin{pmatrix}\times&\times\\ \times&0\end{pmatrix}$. Such a structure cannot be enforced by conventional flavor symmetries, highlighting the additional possibilities offered by non-invertible symmetries.

\subsubsection{Phenomenological implications}

Exploiting the freedom of field redefinitions of the left-handed quark doublets and right-handed up-type quark singlets, one can always cast the diagonal up-type quark Yukawa matrix $Y_u$ into a form with its eigenvalues ordered according to the $u$, $c$, and $t$ quarks. Consequently, after electroweak symmetry breaking, the diagonal entries $\left(Y_u^{\mathrm{diag}}\right)_{11}$, $\left(Y_u^{\mathrm{diag}}\right)_{22}$, and $\left(Y_u^{\mathrm{diag}}\right)_{33}$ correspond to the $u$, $c$, and $t$ quark masses, respectively. Since permutations of the right-handed down-type quarks leave the quark masses, mixing angles, and CP-violating phase invariant, textures of $Y_d$ related by column permutations are physically equivalent. Consequently, up to such permutations, the down-type quark Yukawa matrix with three zeros can be classified into the following classes:
\begin{enumerate}[(i)]

\item{All three zeros lie in the same column. Since this texture of $Y_d$ has rank two, one of the down-type quarks is massless, and the texture is therefore excluded.}

\item{Two zeros lie in one column, while the third lies in another. Up to column permutations, there are nine distinct textures of this type:}
\begin{eqnarray}
M_1&=&\begin{pmatrix}
0 & \times & 0 \\
0 & \times & \times \\
\times & \times & \times
\end{pmatrix}\,,\quad
M_2=\begin{pmatrix}
0 & \times & \times \\
0 & 0 & \times \\
\times & \times & \times
\end{pmatrix}\,,\quad
M_3=\begin{pmatrix}
\times & \times & 0 \\
\times & \times & 0 \\
\times & 0 & \times
\end{pmatrix}\,,\nonumber\\
M_4&=&\begin{pmatrix}
\times & 0 & 0 \\
\times & \times & \times \\
\times & 0 & \times
\end{pmatrix}\,,\quad
M_5=\begin{pmatrix}
\times & 0 & \times \\
\times & \times & 0 \\
\times & 0 & \times
\end{pmatrix}\,,\quad
M_6=\begin{pmatrix}
\times & 0 & \times \\
\times & \times & \times \\
0 & 0 & \times
\end{pmatrix}\,,\nonumber\\
M_7&=&\begin{pmatrix}
\times & \times & 0 \\
0 & \times & \times \\
0 & \times & \times
\end{pmatrix}\,,\quad
M_8=\begin{pmatrix}
\times & \times & \times \\
0 & \times & 0 \\
0 & \times & \times
\end{pmatrix}\,,\quad
M_9=\begin{pmatrix}
\times & \times & \times \\
0 & \times & \times \\
0 & 0 & \times
\end{pmatrix}\,.
\label{eq:threezero1}
\end{eqnarray}

The representative constructions in Eqs.~\eqref{eq:udiag-dthree-S3N5-2} and \eqref{eq:udiag-dthree-S3N5-4} show that all nine textures $M_1\sim M_9$ are realizable through the $S_3$ gauging of $\mathbb{Z}_5\times\mathbb{Z}'_5$ symmetry. Notice that if the representative textures don't not coincide directly with the forms in Eq.~\eqref{eq:threezero1}, they can be brought into the latter forms by permuting the class assignments of the three generations of left- and right-handed quarks.

All textures in this class, together with their column permutations, have been studied in Ref.~\cite{tanimoto:2016rqy}. In particular, after accounting for permutations of the right-handed down-type quarks, $M_1$ corresponds to $M_d^{(3)}$, $M_2$ to $M_d^{(13)}$, $M_3$ to $M_d^{(6)}$, $M_4$ to $M_d^{(1)}$, $M_5$ to $M_d^{(12)}$, $M_6$ to $M_d^{(4)}$, $M_7$ to $M_d^{(2)}$, $M_d^{(16)}$, or $M_d^{(17)}$, $M_8$ to $M_d^{(11)}$ or $M_d^{(15)}$, and $M_9$ to $M_d^{(5)}$ or $M_d^{(14)}$. These textures are all consistent with the current experimental data. Moreover, the combination of a diagonal $Y_u$ and a $Y_d$ texture of type $M_1$, $M_2$, $M_4$, $M_6$, $M_8$, or $M_9$ can potentially provide a solution to the strong CP problem, as discussed in Appendix~\ref{app:strong-cp}.

\item{The three zeros are located in different columns. After accounting for row permutations, seven inequivalent full-rank textures of this type can be identified:}
\begin{eqnarray}
M_{10}&=&\begin{pmatrix}
0 & 0 & \times \\
\times & \times & 0 \\
\times & \times & \times
\end{pmatrix}\,,\quad
M_{11}=\begin{pmatrix}
0 & \times & 0 \\
\times & \times & \times \\
\times & 0 & \times
\end{pmatrix}\,,\quad
M_{12}=\begin{pmatrix}
\times & \times & 0 \\
0 & 0 & \times \\
\times & \times & \times
\end{pmatrix}\,,\quad
M_{13}=\begin{pmatrix}
\times & \times & \times \\
0 & \times & 0 \\
\times & 0 & \times
\end{pmatrix}\,,\nonumber\\
M_{14}&=&\begin{pmatrix}
\times & \times & 0 \\
\times & \times & \times \\
0 & 0 & \times
\end{pmatrix}\,,\quad
M_{15}=\begin{pmatrix}
\times & \times & \times \\
\times & \times & 0 \\
0 & 0 & \times
\end{pmatrix}\,,\quad
M_{16}=\begin{pmatrix}
\times & \times & 0 \\
0 & \times & \times \\
\times & 0 & \times
\end{pmatrix}\,,
\end{eqnarray}
\end{enumerate}
The case in which all three zeros are located in the same row is omitted here. Each of the textures $M_{10}\sim M_{15}$ can be transformed into a texture with one additional zero by a right-handed weak-basis transformation. Furthermore, these textures give rise to one vanishing off-diagonal element in $M_i M_i^\dagger$ ($i=10,\dots,15$), i.e.,
\begin{eqnarray}
M_i M_i^\dagger=\left\{\begin{array}{cc}
\begin{pmatrix}
\times & 0 &\times \\
0 & \times &\times \\
\times & \times &\times \\
\end{pmatrix}\,, & i=10,12 \,, \\
\begin{pmatrix}
\times & \times &0 \\
\times & \times &\times \\
0 & \times &\times \\
\end{pmatrix}\,, & i=11,14 \,, \\
\begin{pmatrix}
\times & \times &\times \\
\times & \times &0 \\
\times & 0 &\times \\
\end{pmatrix}\,, & i=13,15 \,.
\end{array}\right.
\end{eqnarray}

By redefining the phases of the left-handed quark doublets, each of these matrices can be made real and symmetric. Since the up-type mass matrix is diagonal, the CKM matrix is consequently real and orthogonal, implying $\delta_{CP}=0$ or $\pi$. The remaining texture, $M_{16}$, has been studied in Refs.~\cite{tanimoto:2016rqy,Emmanuel-Costa:2016gdp}, where it is denoted by $M_d^{(10)}$ or, after a column permutation, by $M_d^{(20)}$. This texture is incompatible with the experimental data.

\subsection{Diagonal $Y_d$ and three zeros in $Y_u$}\label{sec:3+6tex}

The configurations considered in this subsection, in which $Y_d$ is diagonal and $Y_u$ has three texture zeros, are directly related to the textures listed in section~\ref{sec:6+3tex}. Specifically, for each field assignment in that section, interchanging the class assignments of the right-handed up- and down-type quarks and replacing the Higgs field assignment by its inverse class would map $Y_u$ onto $Y_d$ and vice versa, while leaving the class assignments of the left-handed quark doublets unchanged.

When the down-type quark mass matrix is diagonal, all full-rank three-zero textures of the up-type mass matrix are represented by $M_1\sim M_{16}$, up to column permutations. As discussed in Ref.~\cite{Emmanuel-Costa:2016gdp}, for each texture $M_i$ with $i=1,\ldots,9$, the configuration with diagonal $Y_d$ and $Y_u=M_i$ can be transformed, by weak-basis transformations of the left- and right-handed quark fields, into the corresponding configuration with diagonal $Y_u$ and $Y_d=M_i$. Since weak-basis-equivalent configurations yield identical physical predictions, the compatibility of the configurations with diagonal $Y_u$ and $Y_d=M_1\sim M_9$ with the experimental data implies the compatibility of their counterparts with diagonal $Y_d$ and $Y_u=M_1\sim M_9$.

For the up-type quark Yukawa matrix $Y_u$ taking one of the textures $M_{10}\sim M_{15}$, the Hermitian combination $Y_uY_u^\dagger$ can be rendered real and symmetric by appropriate field redefinitions. Since $Y_d$ is diagonal, the CKM matrix is consequently real and orthogonal, and the CP-violating phase is restricted to $\delta_{CP}=0$ or $\pi$. This is incompatible with the experimentally measured CP violation, and hence the textures $M_{10}\sim M_{15}$ are excluded. The texture $M_{16}$, which has also been analyzed in Ref.~\cite{Emmanuel-Costa:2016gdp}, is similarly incompatible with the experimental data.

\subsection{Five zeros in $Y_u$ ($Y_d$) and four zeros in $Y_d$ ($Y_u$)}

In the following, we shall firstly present the explicit form of the quark Yukawa matrix with five texture zero and four texture zero respectively. Then we investigate the possible combinations of $Y_u (Y_d)$ with five zeros and $Y_d (Y_u)$ with four zeros which can be obtained from the $\mathbb{Z}_3$ and $S_3$ gauging of $\mathbb{Z}_N\times\mathbb{Z}_N$ symmetry. Up to row and column permutations, there is a unique full-rank Yukawa texture with five texture zeros,
\begin{eqnarray}
T_5=\begin{pmatrix}
\times & \times & 0 \\
0 & \times & 0 \\
0 & 0 & \times
\end{pmatrix}\,, \label{eq:fullrankfive0}
\end{eqnarray}
while the full-rank Yukawa textures with four texture zeros can be classified into the following four inequivalent forms:
\begin{eqnarray}
T_1=\begin{pmatrix}
\times & 0 & 0 \\
\times & \times & 0 \\
\times & 0 & \times
\end{pmatrix} \,, \quad
T_2=\begin{pmatrix}
\times & 0 & 0 \\
0 & \times & 0 \\
\times & \times & \times
\end{pmatrix} \,, \quad
T_3=\begin{pmatrix}
\times & 0 & 0 \\
\times & \times & 0 \\
0 & \times & \times
\end{pmatrix} \,,\quad
T_4=\begin{pmatrix}
\times & \times & 0 \\
\times & \times & 0 \\
0 & 0 & \times
\end{pmatrix}\,.~~~~\label{eq:fullrankfour0}
\end{eqnarray}

We next consider the case in which one quark mass matrix has five texture zeros and the other has four. By suitable permutations of the left- and right-handed quark fields, the five-zero matrix can be fixed to $T_5$. The four-zero matrix can then be chosen from $T_1$, $T_2$, and $T_3$, together with their row permutations. The texture $T_4$ is excluded by the experimental data. In fact, both $T_5T_5^\dagger$ and $T_4T_4^\dagger$ are block diagonal, and hence their diagonalization can generate at most two mixing angles in the CKM matrix. As a result, at least one CKM matrix element vanishes, and the CP-violating phase is necessarily zero.

Since the ordering of the three generations of left-handed quarks is fixed by the five-zero matrix, row permutations of the four-zero matrix must be considered separately. Therefore, up to column permutations, the four-zero textures are given by
\begin{eqnarray}
\label{eq:T1}&&T_1^{(1)}=\begin{pmatrix}
    \times & 0 & 0 \\
    \times & \times & 0 \\
    \times & 0 & \times
\end{pmatrix} \,,\quad
T_1^{(2)}=\begin{pmatrix}
    \times & \times & 0 \\
    0 & \times & 0 \\
    0 & \times & \times
\end{pmatrix} \,,\quad
T_1^{(3)}=\begin{pmatrix}
    \times & 0 & \times \\
    0 & \times & \times \\
    0 & 0 & \times
\end{pmatrix} \,,\\
\label{eq:T2}&&T_2^{(1)}=\begin{pmatrix}
    \times & 0 & 0 \\
    0 & \times & 0 \\
    \times & \times & \times
\end{pmatrix} \,,\quad
T_2^{(2)}=\begin{pmatrix}
    \times & 0 & 0 \\
    \times & \times & \times \\
    0 & 0 & \times
\end{pmatrix} \,,\quad
T_2^{(3)}=\begin{pmatrix}
    \times & \times & \times \\
    0 & \times & 0 \\
    0 & 0 & \times
\end{pmatrix} \,,\\
\label{eq:T3123}&&T_3^{(1)}=\begin{pmatrix}
    \times & 0 & 0 \\
    \times & \times & 0 \\
    0 & \times & \times
\end{pmatrix} \,,\quad
T_3^{(2)}=\begin{pmatrix}
    \times & 0 & 0 \\
    0 & \times & \times \\
    \times & 0 & \times
\end{pmatrix} \,,\quad
T_3^{(3)}=\begin{pmatrix}
    \times & \times & 0 \\
    0 & \times & 0 \\
    \times & 0 & \times
\end{pmatrix} \,,\\
\label{eq:T3456}&&T_3^{(4)}=\begin{pmatrix}
    \times & 0 & \times \\
    0 & \times & 0 \\
    0 & \times & \times
\end{pmatrix} \,,\quad
T_3^{(5)}=\begin{pmatrix}
    \times & \times & 0 \\
    0 & \times & \times \\
    0 & 0 & \times
\end{pmatrix} \,,\quad
T_3^{(6)}=\begin{pmatrix}
    \times & 0 & \times \\
    \times & \times & 0 \\
    0 & 0 & \times
\end{pmatrix} \,.
\end{eqnarray}
It is worth noting that $T_3^{(2)}$ can be brought into the nearest-neighbor-interaction form discussed in Ref.~\cite{PhysRevD.39.3443} by interchanging its first and second columns.

\subsubsection{Realization of $5+4$ quark Yukawa textures from $\mathbb{Z}_3$ and $S_3$ gauging}

In the following, we proceed to consider the quark textures arising from the $\mathbb{Z}_3$ and $S_3$ gaugings of $\mathbb{Z}_N\times\mathbb{Z}'_N$ for $N=4,5$. Notice that neither of the two gaugings considered here generates a five-zero texture when $N=2,3$, and these cases are therefore not considered. In the following, we take the case where $Y_u$ has five zeros and $Y_d$ has four zeros as an example, while the case where $Y_u$ has four zeros and $Y_d$ has five zeros can be obtained by exchanging the assignments of the right-handed up- and down-type quarks and taking the assignment of the Higgs field to be its inverse. Since row and column permutations can be used to fix the five-zero texture,  we take $Y_u=T_5$ without loss of generality and consider $Y_d$ to be one of the four-zero textures shown in Eqs.~(\ref{eq:T1}, \ref{eq:T2}, \ref{eq:T3123}, \ref{eq:T3456}).

For the $\mathbb{Z}_3$ and $S_3$ gaugings of the $\mathbb{Z}_4\times\mathbb{Z}'_4$ symmetry, the resulting quark Yukawa textures and representative charge assignments of the quark and Higgs fields are summarized in Table~6. The $\mathbb{Z}_3$ gauging realizes all twelve four-zero textures in combination with $T_5$, whereas the $S_3$ gauging yields only the following six:
\begin{eqnarray}
T_2^{(2)},\quad T_2^{(3)},\quad T_3^{(1)},\quad T_3^{(3)},\quad T_3^{(5)},\quad T_3^{(6)}.
\end{eqnarray}
The corresponding results for the $\mathbb{Z}_5\times\mathbb{Z}'_5$ symmetry are presented in Table~7. In this case, the $\mathbb{Z}_3$ gauging again realizes all twelve four-zero textures combined with $T_5$, while the $S_3$ gauging produces nine of them:
\begin{eqnarray}
T_1^{(1)},\quad T_1^{(2)},\quad T_1^{(3)},\quad
T_2^{(2)},\quad T_2^{(3)},\quad
T_3^{(1)},\quad T_3^{(3)},\quad T_3^{(5)},\quad T_3^{(6)}.
\end{eqnarray}

\begin{table}[hptb!]
\centering
\begin{tabular}{|c|c|c|c|c|c|}
\hline\hline
\multicolumn{6}{|c|}{$\mathbb{Z}_3$ gauging of the $\mathbb{Z}_4\times\mathbb{Z}'_4$ symmetry} \\
\hline
$Y_u$ & $Y_d$ & $\bar{Q}_L$ & $u_R$ & $d_R$ & $H$ \\
\hline
\multirow{12}{*}{$T_5$} & $T_1^{(1)}$ & $([2,1]_{\mathbb{Z}_3},[3,0]_{\mathbb{Z}_3},[2,0]_{\mathbb{Z}_3})$ & $([1,0]_{\mathbb{Z}_3},[2,0]_{\mathbb{Z}_3},[2,1]_{\mathbb{Z}_3})$ & $([1,2]_{\mathbb{Z}_3},[0,0]_{\mathbb{Z}_3},[1,0]_{\mathbb{Z}_3})$ & $[1,0]_{\mathbb{Z}_3}$ \\
\cline{2-6}
& $T_1^{(2)}$ & $([2,1]_{\mathbb{Z}_3},[3,0]_{\mathbb{Z}_3},[2,0]_{\mathbb{Z}_3})$ & $([1,0]_{\mathbb{Z}_3},[2,0]_{\mathbb{Z}_3},[2,1]_{\mathbb{Z}_3})$ & $([2,0]_{\mathbb{Z}_3},[1,2]_{\mathbb{Z}_3},[1,0]_{\mathbb{Z}_3})$ & $[1,0]_{\mathbb{Z}_3}$ \\
\cline{2-6}
& $T_1^{(3)}$ & $([2,1]_{\mathbb{Z}_3},[3,0]_{\mathbb{Z}_3},[2,0]_{\mathbb{Z}_3})$ & $([1,0]_{\mathbb{Z}_3},[2,0]_{\mathbb{Z}_3},[2,1]_{\mathbb{Z}_3})$ & $([2,0]_{\mathbb{Z}_3},[0,0]_{\mathbb{Z}_3},[1,2]_{\mathbb{Z}_3})$ & $[1,0]_{\mathbb{Z}_3}$ \\
\cline{2-6}
& $T_2^{(1)}$ & $([1,0]_{\mathbb{Z}_3},[2,0]_{\mathbb{Z}_3},[2,1]_{\mathbb{Z}_3})$ & $([0,0]_{\mathbb{Z}_3},[2,1]_{\mathbb{Z}_3},[1,0]_{\mathbb{Z}_3})$ & $([2,0]_{\mathbb{Z}_3},[1,2]_{\mathbb{Z}_3},[3,0]_{\mathbb{Z}_3})$ & $[1,0]_{\mathbb{Z}_3}$ \\
\cline{2-6}
& $T_2^{(2)}$ & $([1,0]_{\mathbb{Z}_3},[2,0]_{\mathbb{Z}_3},[2,1]_{\mathbb{Z}_3})$ & $([0,0]_{\mathbb{Z}_3},[2,1]_{\mathbb{Z}_3},[1,0]_{\mathbb{Z}_3})$ & $([1,0]_{\mathbb{Z}_3},[2,1]_{\mathbb{Z}_3},[1,2]_{\mathbb{Z}_3})$ & $[1,0]_{\mathbb{Z}_3}$ \\
\cline{2-6}
& $T_2^{(3)}$ & $([2,1]_{\mathbb{Z}_3},[0,0]_{\mathbb{Z}_3},[1,0]_{\mathbb{Z}_3})$ & $([2,0]_{\mathbb{Z}_3},[1,0]_{\mathbb{Z}_3},[0,0]_{\mathbb{Z}_3})$ & $([1,2]_{\mathbb{Z}_3},[3,0]_{\mathbb{Z}_3},[2,0]_{\mathbb{Z}_3})$ & $[1,0]_{\mathbb{Z}_3}$ \\
\cline{2-6}
& $T_3^{(1)}$ & $([3,0]_{\mathbb{Z}_3},[2,0]_{\mathbb{Z}_3},[1,0]_{\mathbb{Z}_3})$ & $([2,0]_{\mathbb{Z}_3},[3,0]_{\mathbb{Z}_3},[0,0]_{\mathbb{Z}_3})$ & $([2,1]_{\mathbb{Z}_3},[1,0]_{\mathbb{Z}_3},[2,0]_{\mathbb{Z}_3})$ & $[1,0]_{\mathbb{Z}_3}$ \\
\cline{2-6}
& $T_3^{(2)}$ & $([1,0]_{\mathbb{Z}_3},[2,0]_{\mathbb{Z}_3},[2,1]_{\mathbb{Z}_3})$ & $([0,0]_{\mathbb{Z}_3},[2,1]_{\mathbb{Z}_3},[1,0]_{\mathbb{Z}_3})$ & $([2,0]_{\mathbb{Z}_3},[2,1]_{\mathbb{Z}_3},[1,2]_{\mathbb{Z}_3})$ & $[1,0]_{\mathbb{Z}_3}$ \\
\cline{2-6}
& $T_3^{(3)}$ & $([2,1]_{\mathbb{Z}_3},[0,0]_{\mathbb{Z}_3},[1,0]_{\mathbb{Z}_3})$ & $([2,0]_{\mathbb{Z}_3},[1,0]_{\mathbb{Z}_3},[0,0]_{\mathbb{Z}_3})$ & $([2,0]_{\mathbb{Z}_3},[3,0]_{\mathbb{Z}_3},[1,0]_{\mathbb{Z}_3})$ & $[1,0]_{\mathbb{Z}_3}$ \\
\cline{2-6}
& $T_3^{(4)}$ & $([2,0]_{\mathbb{Z}_3},[1,0]_{\mathbb{Z}_3},[2,1]_{\mathbb{Z}_3})$ & $([3,0]_{\mathbb{Z}_3},[2,1]_{\mathbb{Z}_3},[1,0]_{\mathbb{Z}_3})$ & $([2,1]_{\mathbb{Z}_3},[2,0]_{\mathbb{Z}_3},[1,2]_{\mathbb{Z}_3})$ & $[1,0]_{\mathbb{Z}_3}$ \\
\cline{2-6}
& $T_3^{(5)}$ & $([3,0]_{\mathbb{Z}_3},[2,0]_{\mathbb{Z}_3},[1,0]_{\mathbb{Z}_3})$ & $([2,0]_{\mathbb{Z}_3},[3,0]_{\mathbb{Z}_3},[0,0]_{\mathbb{Z}_3})$ & $([0,0]_{\mathbb{Z}_3},[2,1]_{\mathbb{Z}_3},[1,0]_{\mathbb{Z}_3})$ & $[1,0]_{\mathbb{Z}_3}$ \\
\cline{2-6}
& $T_3^{(6)}$ & $([1,0]_{\mathbb{Z}_3},[2,0]_{\mathbb{Z}_3},[2,1]_{\mathbb{Z}_3})$ & $([0,0]_{\mathbb{Z}_3},[2,1]_{\mathbb{Z}_3},[1,0]_{\mathbb{Z}_3})$ & $([1,0]_{\mathbb{Z}_3},[2,1]_{\mathbb{Z}_3},[2,0]_{\mathbb{Z}_3})$ & $[1,0]_{\mathbb{Z}_3}$ \\
\hline
\multicolumn{6}{|c|}{$S_3$ gauging of the $\mathbb{Z}_4\times\mathbb{Z}'_4$ symmetry} \\
\hline
$Y_u$ & $Y_d$ & $\bar{Q}_L$ & $u_R$ & $d_R$ & $H$ \\
\hline
\multirow{6}{*}{$T_5$}& $T_2^{(2)}$ & $([2,1]_{S_3},[1,0]_{S_3},[1,2]_{S_3})$ & $([0,0]_{S_3},[1,0]_{S_3},[1,2]_{S_3})$ & $([2,0]_{S_3},[1,2]_{S_3},[1,0]_{S_3})$ & $[2,1]_{S_3}$ \\
\cline{2-6}
& $T_2^{(3)}$ & $([1,0]_{S_3},[0,0]_{S_3},[2,0]_{S_3})$ & $([2,0]_{S_3},[2,1]_{S_3},[1,2]_{S_3})$ & $([2,0]_{S_3},[1,2]_{S_3},[1,0]_{S_3})$ & $[2,1]_{S_3}$ \\
\cline{2-6}
& $T_3^{(1)}$ & $([2,1]_{S_3},[1,0]_{S_3},[1,2]_{S_3})$ & $([0,0]_{S_3},[1,0]_{S_3},[1,2]_{S_3})$ & $([2,0]_{S_3},[1,0]_{S_3},[0,0]_{S_3})$ & $[2,1]_{S_3}$ \\
\cline{2-6}
& $T_3^{(3)}$ & $([1,0]_{S_3},[0,0]_{S_3},[2,0]_{S_3})$ & $([2,0]_{S_3},[2,1]_{S_3},[1,2]_{S_3})$ & $([1,0]_{S_3},[1,2]_{S_3},[2,1]_{S_3})$ & $[2,1]_{S_3}$ \\
\cline{2-6}
& $T_3^{(5)}$ & $([2,1]_{S_3},[1,0]_{S_3},[1,2]_{S_3})$ & $([0,0]_{S_3},[1,0]_{S_3},[1,2]_{S_3})$ & $([2,1]_{S_3},[2,0]_{S_3},[1,0]_{S_3})$ & $[2,1]_{S_3}$ \\
\cline{2-6}
& $T_3^{(6)}$ & $([1,0]_{S_3},[1,2]_{S_3},[2,1]_{S_3})$ & $([2,1]_{S_3},[2,0]_{S_3},[0,0]_{S_3})$ & $([1,0]_{S_3},[0,0]_{S_3},[2,0]_{S_3})$ & $[2,1]_{S_3}$ \\
\hline\hline
\end{tabular}
\caption{Representative assignments of $\bar{Q}_L$, $u_R$, $d_R$, and $H$ under the $\mathbb{Z}_3$ and $S_3$ gauging of the $\mathbb{Z}_4\times\mathbb{Z}'_4$ symmetry that generate the texture combinations with $Y_u=T_5$ and $Y_d=T_i^{(j)}$ which is one of the four-zero textures in Eqs.~\eqref{eq:T1}, \eqref{eq:T2}, \eqref{eq:T3123} and \eqref{eq:T3456}.}
\label{tab:N=4-5plus4}
\end{table}

\begin{table}[hptb!]
\centering
\begin{tabular}{|c|c|c|c|c|c|}
\hline\hline
\multicolumn{6}{|c|}{$\mathbb{Z}_3$ gauging of the $\mathbb{Z}_5\times\mathbb{Z}'_5$ symmetry} \\
\hline
$Y_u$ & $Y_d$ & $\bar{Q}_L$ & $u_R$ & $d_R$ & $H$ \\
\hline
\multirow{12}{*}{$T_5$}& $T_1^{(1)}$ & $([1,3]_{\mathbb{Z}_3},[3,1]_{\mathbb{Z}_3},[1,0]_{\mathbb{Z}_3})$ & $([3,1]_{\mathbb{Z}_3},[2,0]_{\mathbb{Z}_3},[0,0]_{\mathbb{Z}_3})$ & $([3,0]_{\mathbb{Z}_3},[1,2]_{\mathbb{Z}_3},[1,0]_{\mathbb{Z}_3})$ & $[1,0]_{\mathbb{Z}_3}$ \\
\cline{2-6}
& $T_1^{(2)}$ & $([1,3]_{\mathbb{Z}_3},[3,1]_{\mathbb{Z}_3},[1,0]_{\mathbb{Z}_3})$ & $([3,1]_{\mathbb{Z}_3},[2,0]_{\mathbb{Z}_3},[0,0]_{\mathbb{Z}_3})$ & $([2,1]_{\mathbb{Z}_3},[3,0]_{\mathbb{Z}_3},[1,0]_{\mathbb{Z}_3})$ & $[1,0]_{\mathbb{Z}_3}$ \\
\cline{2-6}
& $T_1^{(3)}$ & $([1,3]_{\mathbb{Z}_3},[3,1]_{\mathbb{Z}_3},[1,0]_{\mathbb{Z}_3})$ & $([3,1]_{\mathbb{Z}_3},[2,0]_{\mathbb{Z}_3},[0,0]_{\mathbb{Z}_3})$ & $([2,1]_{\mathbb{Z}_3},[1,2]_{\mathbb{Z}_3},[3,0]_{\mathbb{Z}_3})$ & $[1,0]_{\mathbb{Z}_3}$ \\
\cline{2-6}
& $T_2^{(1)}$ & $([1,0]_{\mathbb{Z}_3},[3,0]_{\mathbb{Z}_3},[3,1]_{\mathbb{Z}_3})$ & $([0,0]_{\mathbb{Z}_3},[2,1]_{\mathbb{Z}_3},[2,0]_{\mathbb{Z}_3})$ & $([3,0]_{\mathbb{Z}_3},[1,3]_{\mathbb{Z}_3},[1,2]_{\mathbb{Z}_3})$ & $[1,0]_{\mathbb{Z}_3}$ \\
\cline{2-6}
& $T_2^{(2)}$ & $([1,0]_{\mathbb{Z}_3},[3,1]_{\mathbb{Z}_3},[2,0]_{\mathbb{Z}_3})$ & $([0,0]_{\mathbb{Z}_3},[1,2]_{\mathbb{Z}_3},[4,0]_{\mathbb{Z}_3})$ & $([3,0]_{\mathbb{Z}_3},[1,3]_{\mathbb{Z}_3},[1,2]_{\mathbb{Z}_3})$ & $[1,0]_{\mathbb{Z}_3}$ \\
\cline{2-6}
& $T_2^{(3)}$ & $([2,1]_{\mathbb{Z}_3},[0,0]_{\mathbb{Z}_3},[2,0]_{\mathbb{Z}_3})$ & $([3,0]_{\mathbb{Z}_3},[1,0]_{\mathbb{Z}_3},[4,0]_{\mathbb{Z}_3})$ & $([1,3]_{\mathbb{Z}_3},[4,0]_{\mathbb{Z}_3},[2,0]_{\mathbb{Z}_3})$ & $[1,0]_{\mathbb{Z}_3}$ \\
\cline{2-6}
& $T_3^{(1)}$ & $([1,0]_{\mathbb{Z}_3},[3,1]_{\mathbb{Z}_3},[2,0]_{\mathbb{Z}_3})$ & $([0,0]_{\mathbb{Z}_3},[1,2]_{\mathbb{Z}_3},[4,0]_{\mathbb{Z}_3})$ & $([3,0]_{\mathbb{Z}_3},[1,2]_{\mathbb{Z}_3},[2,0]_{\mathbb{Z}_3})$ & $[1,0]_{\mathbb{Z}_3}$ \\
\cline{2-6}
& $T_3^{(2)}$ & $([1,0]_{\mathbb{Z}_3},[3,0]_{\mathbb{Z}_3},[3,1]_{\mathbb{Z}_3})$ & $([0,0]_{\mathbb{Z}_3},[2,1]_{\mathbb{Z}_3},[2,0]_{\mathbb{Z}_3})$ & $([3,0]_{\mathbb{Z}_3},[3,1]_{\mathbb{Z}_3},[1,3]_{\mathbb{Z}_3})$ & $[1,0]_{\mathbb{Z}_3}$ \\
\cline{2-6}
& $T_3^{(3)}$ & $([2,1]_{\mathbb{Z}_3},[0,0]_{\mathbb{Z}_3},[2,0]_{\mathbb{Z}_3})$ & $([3,0]_{\mathbb{Z}_3},[1,0]_{\mathbb{Z}_3},[4,0]_{\mathbb{Z}_3})$ & $([2,0]_{\mathbb{Z}_3},[4,0]_{\mathbb{Z}_3},[2,1]_{\mathbb{Z}_3})$ & $[1,0]_{\mathbb{Z}_3}$ \\
\cline{2-6}
& $T_3^{(4)}$ & $([3,0]_{\mathbb{Z}_3},[1,0]_{\mathbb{Z}_3},[3,1]_{\mathbb{Z}_3})$ & $([3,0]_{\mathbb{Z}_3},[2,1]_{\mathbb{Z}_3},[2,0]_{\mathbb{Z}_3})$ & $([3,1]_{\mathbb{Z}_3},[3,0]_{\mathbb{Z}_3},[1,3]_{\mathbb{Z}_3})$ & $[1,0]_{\mathbb{Z}_3}$ \\
\cline{2-6}
& $T_3^{(5)}$ & $([1,0]_{\mathbb{Z}_3},[3,1]_{\mathbb{Z}_3},[2,0]_{\mathbb{Z}_3})$ & $([0,0]_{\mathbb{Z}_3},[1,2]_{\mathbb{Z}_3},[4,0]_{\mathbb{Z}_3})$ & $([1,0]_{\mathbb{Z}_3},[3,0]_{\mathbb{Z}_3},[1,2]_{\mathbb{Z}_3})$ & $[1,0]_{\mathbb{Z}_3}$ \\
\cline{2-6}
& $T_3^{(6)}$ & $([3,1]_{\mathbb{Z}_3},[2,0]_{\mathbb{Z}_3},[1,0]_{\mathbb{Z}_3})$ & $([2,0]_{\mathbb{Z}_3},[1,3]_{\mathbb{Z}_3},[0,0]_{\mathbb{Z}_3})$ & $([1,2]_{\mathbb{Z}_3},[2,0]_{\mathbb{Z}_3},[3,0]_{\mathbb{Z}_3})$ & $[1,0]_{\mathbb{Z}_3}$ \\
\hline
\multicolumn{6}{|c|}{$S_3$ gauging of the $\mathbb{Z}_5\times\mathbb{Z}'_5$ symmetry} \\
\hline
$Y_u$ & $Y_d$ & $\bar{Q}_L$ & $u_R$ & $d_R$ & $H$ \\
\hline
\multirow{18}{*}{$T_5$} & \multirow{2}{*}{$T_1^{(1)}$} & $([1,0]_{S_3},[2,1]_{S_3},[1,3]_{S_3})$ & $([0,0]_{S_3},[1,0]_{S_3},[3,1]_{S_3})$ & $([2,0]_{S_3},[1,3]_{S_3},[3,1]_{S_3})$ & $[1,0]_{S_3}$ \\
\cline{3-6}
& & $([1,0]_{S_3},[1,3]_{S_3},[2,0]_{S_3})$ & $([3,1]_{S_3},[2,0]_{S_3},[2,1]_{S_3})$ & $([1,0]_{S_3},[2,1]_{S_3},[1,2]_{S_3})$ & $[2,1]_{S_3}$ \\
\cline{2-6}
& \multirow{2}{*}{$T_1^{(2)}$} & $([2,0]_{S_3},[2,1]_{S_3},[1,0]_{S_3})$ & $([3,1]_{S_3},[1,3]_{S_3},[0,0]_{S_3})$ & $([3,1]_{S_3},[1,0]_{S_3},[0,0]_{S_3})$ & $[1,0]_{S_3}$ \\
\cline{3-6}
& & $([1,0]_{S_3},[1,3]_{S_3},[2,0]_{S_3})$ & $([3,1]_{S_3},[2,0]_{S_3},[2,1]_{S_3})$ & $([2,0]_{S_3},[1,0]_{S_3},[1,2]_{S_3})$ & $[2,1]_{S_3}$ \\
\cline{2-6}
& \multirow{2}{*}{$T_1^{(3)}$} & $([1,0]_{S_3},[2,1]_{S_3},[1,3]_{S_3})$ & $([0,0]_{S_3},[1,0]_{S_3},[3,1]_{S_3})$ & $([0,0]_{S_3},[1,3]_{S_3},[2,0]_{S_3})$ & $[1,0]_{S_3}$ \\
\cline{3-6}
& & $([1,0]_{S_3},[1,3]_{S_3},[2,0]_{S_3})$ & $([3,1]_{S_3},[2,0]_{S_3},[2,1]_{S_3})$ & $([2,0]_{S_3},[2,1]_{S_3},[1,0]_{S_3})$ & $[2,1]_{S_3}$ \\
\cline{2-6}
& \multirow{2}{*}{$T_2^{(2)}$} & $([2,1]_{S_3},[1,0]_{S_3},[1,3]_{S_3})$ & $([1,3]_{S_3},[1,0]_{S_3},[3,1]_{S_3})$ & $([1,0]_{S_3},[0,0]_{S_3},[2,1]_{S_3})$ & $[1,0]_{S_3}$ \\
\cline{3-6}
& & $([2,0]_{S_3},[1,0]_{S_3},[2,1]_{S_3})$ & $([2,1]_{S_3},[1,0]_{S_3},[0,0]_{S_3})$ & $([1,0]_{S_3},[2,0]_{S_3},[1,3]_{S_3})$ & $[2,1]_{S_3}$ \\
\cline{2-6}
& \multirow{2}{*}{$T_2^{(3)}$} & $([1,0]_{S_3},[0,0]_{S_3},[3,1]_{S_3})$ & $([0,0]_{S_3},[1,0]_{S_3},[1,3]_{S_3})$ & $([0,0]_{S_3},[1,0]_{S_3},[2,0]_{S_3})$ & $[1,0]_{S_3}$ \\
\cline{3-6}
& & $([2,0]_{S_3},[0,0]_{S_3},[1,0]_{S_3})$ & $([1,3]_{S_3},[2,1]_{S_3},[2,0]_{S_3})$ & $([3,1]_{S_3},[1,2]_{S_3},[1,0]_{S_3})$ & $[2,1]_{S_3}$ \\
\cline{2-6}
& \multirow{2}{*}{$T_3^{(1)}$} & $([2,1]_{S_3},[1,0]_{S_3},[1,3]_{S_3})$ & $([1,3]_{S_3},[1,0]_{S_3},[3,1]_{S_3})$ & $([1,0]_{S_3},[2,1]_{S_3},[3,1]_{S_3})$ & $[1,0]_{S_3}$ \\
\cline{3-6}
& & $([2,0]_{S_3},[1,0]_{S_3},[2,1]_{S_3})$ & $([2,1]_{S_3},[1,0]_{S_3},[0,0]_{S_3})$ & $([1,0]_{S_3},[1,3]_{S_3},[2,1]_{S_3})$ & $[2,1]_{S_3}$ \\
\cline{2-6}
& \multirow{2}{*}{$T_3^{(3)}$} & $([1,0]_{S_3},[0,0]_{S_3},[3,1]_{S_3})$ & $([0,0]_{S_3},[1,0]_{S_3},[1,3]_{S_3})$ & $([2,0]_{S_3},[1,0]_{S_3},[1,3]_{S_3})$ & $[1,0]_{S_3}$ \\
\cline{3-6}
& & $([2,0]_{S_3},[0,0]_{S_3},[1,0]_{S_3})$ & $([1,3]_{S_3},[2,1]_{S_3},[2,0]_{S_3})$ & $([1,0]_{S_3},[1,2]_{S_3},[2,0]_{S_3})$ & $[2,1]_{S_3}$ \\
\cline{2-6}
& \multirow{2}{*}{$T_3^{(5)}$} & $([2,1]_{S_3},[1,0]_{S_3},[1,3]_{S_3})$ & $([1,3]_{S_3},[1,0]_{S_3},[3,1]_{S_3})$ & $([1,3]_{S_3},[1,0]_{S_3},[2,1]_{S_3})$ & $[1,0]_{S_3}$ \\
\cline{3-6}
& & $([2,0]_{S_3},[1,0]_{S_3},[2,1]_{S_3})$ & $([2,1]_{S_3},[1,0]_{S_3},[0,0]_{S_3})$ & $([1,2]_{S_3},[1,0]_{S_3},[1,3]_{S_3})$ & $[2,1]_{S_3}$ \\
\cline{2-6}
& \multirow{2}{*}{$T_3^{(6)}$} & $([1,0]_{S_3},[2,1]_{S_3},[1,3]_{S_3})$ & $([0,0]_{S_3},[1,0]_{S_3},[3,1]_{S_3})$ & $([1,0]_{S_3},[1,3]_{S_3},[2,1]_{S_3})$ & $[1,0]_{S_3}$ \\
\cline{3-6}
& & $([1,0]_{S_3},[2,0]_{S_3},[2,1]_{S_3})$ & $([3,1]_{S_3},[1,0]_{S_3},[0,0]_{S_3})$ & $([1,0]_{S_3},[1,2]_{S_3},[1,3]_{S_3})$ & $[2,1]_{S_3}$ \\
\hline\hline
\end{tabular}
\caption{Representative assignments under the $\mathbb{Z}_3$ and $S_3$ gauging of the $\mathbb{Z}_5\times\mathbb{Z}'_5$ symmetry that generate the texture combinations with $Y_u=T_5$ and $Y_d=T_i^{(j)}$ which is a four-zero texture in Eqs.~\eqref{eq:T1}, \eqref{eq:T2}, \eqref{eq:T3123} and \eqref{eq:T3456}.}
\label{tab:N=5-5plus4}
\end{table}

As with the three-zero textures, the full-rank textures with five or four zeros discussed above, except for the four-zero texture $T_4$, cannot be enforced by conventional flavor symmetries, but can be realized through the non-invertible symmetries considered here.

\subsubsection{Phenomenological implications}

Although the twelve four-zero textures in Eqs.~\eqref{eq:T1}, \eqref{eq:T2}, \eqref{eq:T3123} and \eqref{eq:T3456} are distinct under column permutations, they are not all inequivalent under general weak-basis transformations of the right-handed quark fields. Such transformations leave the CKM matrix invariant and can therefore relate apparently different texture structures. For instance, a unitary rotation in the 2--3 subspace of the right-handed quark fields can move the zero in the $(2,3)$ entry of $T_3^{(1)}$ to any of the positions $(2,2)$, $(2,3)$, $(3,2)$, or $(3,3)$. When the zero is transformed to $(3,3)$, the resulting texture is equivalent to $T_2^{(2)}$ up to a column permutation. More generally, the right-handed weak-basis transformations establish the following equivalence relations:
\begin{eqnarray}
\label{eq:equal1} T_3^{(1)}\sim T_2^{(2)}\sim T_3^{(5)} \,,\quad T_3^{(2)}\sim T_2^{(1)}\sim T_3^{(4)}\,,\quad T_3^{(3)}\sim T_2^{(3)}\sim T_3^{(6)}\,,
\end{eqnarray}
where the equivalence relation $\sim$ indicates that the corresponding textures are related by right multiplication with a unitary matrix, corresponding to a weak-basis transformation of the right-handed quark fields.

For combinations of five- and four-zero textures, an arbitrary rotation in the 1--2 row space can be compensated by a corresponding rotation in the 1--2 column space, such that $T_5$ is restored to its original form. Hence, arbitrary rotations in the 1--2 row space are admissible in defining the equivalence classes of the four-zero textures. In particular, interchanging the first two rows gives
\begin{eqnarray}
\label{eq:equal4}T_1^{(1)}\sim T_1^{(2)} \,,\quad T_2^{(2)}\sim T_2^{(3)} \,.
\end{eqnarray}
Similarly, the following equivalence relations can be established among the four-zero textures:
\begin{eqnarray}
\label{eq:equal5}
T_1^{(1)}\sim T_3^{(1)}\sim T_2^{(1)} \,,\quad
T_1^{(3)}\sim T_3^{(5)} \,.
\end{eqnarray}
Combining these relations with Eqs.~\eqref{eq:equal1} and \eqref{eq:equal4}, we find that all four-zero textures belong to a single equivalence class for the present purpose. Hence, it suffices to consider one representative, which we choose to be $T_1^{(1)}$. Moreover, the textures $(Y_u,Y_d)=(T_5,T_1^{(1)})$ and $(T_1^{(1)},T_5)$ are related by unitary transformations of the left- and right-handed quark fields and are therefore physically equivalent. Hence, for combinations of five- and four-zero textures, it is sufficient to consider the representative choice
$Y_u=T_5$ and $Y_d=T_1^{(1)}$. Exploiting the freedom of field redefinitions, all unphysical phases can be removed, and the up- and down-type quark Yukawa matrices can be parametrized as
\begin{eqnarray}
Y_d=\begin{pmatrix}
a ~&~ 0 ~&~ 0 \\
b ~&~ c ~&~ 0 \\
d ~&~ 0 ~&~ f
\end{pmatrix},~\quad
Y_u=\begin{pmatrix}
a' ~&~ b'e^{-i\phi} ~&~ 0 \\
0 ~&~ c' ~&~ 0 \\
0 ~&~ 0 ~&~ d'
\end{pmatrix} \,.
\label{eq:para-T5T11}
\end{eqnarray}
where $a$, $b$, $c$, $d$, $f$, $a'$, $b'$, $c'$, and $d'$ are real and positive, and $\phi$ is the sole remaining physical phase responsible for CP violation. We test the viability of this texture by performing a conventional $\chi^2$ fit to the experimental data, accounting for their uncertainties. The fit uses the quark masses and CKM parameters at the $M_Z$ scale in the $\overline{\text{MS}}$ renormalization scheme, as provided in Ref.~\cite{Antusch:2025fpm}:
\begin{eqnarray}
\begin{array}{cc}
m_u=(7.09^{+1.56}_{-0.88})\times10^{-6}\times\frac{v}{\sqrt{2}} \,, ~~&~~ m_c=(3.55^{+0.10}_{-0.09})\times10^{-3}\times\frac{v}{\sqrt{2}} \,, \\
m_t=(0.968\pm0.004)\times\frac{v}{\sqrt{2}} \,, ~~&~~ m_d=(1.55^{+0.14}_{-0.07})\times10^{-5}\times\frac{v}{\sqrt{2}} \,, \\
m_s=(3.10^{+0.26}_{-0.14})\times10^{-4}\times\frac{v}{\sqrt{2}} \,, ~~&~~ m_b=(1.63^{+0.02}_{-0.01})\times10^{-2}\times\frac{v}{\sqrt{2}}\,, \\
\end{array}
\label{eq:quarkconstraintmass}
\end{eqnarray}
and
\begin{eqnarray}
&&\sin\theta_{12}=0.2251\pm0.0008 \,,\qquad
\sin\theta_{13}=0.00370\pm0.00008 \,,\nonumber\\
&&\sin\theta_{23}=0.04193\pm0.00041 \,,\qquad
\delta_{CP}=1.139\pm0.023 \,.
\label{eq:quarkconstraintckm}
\end{eqnarray}
Here the quoted errors represent the corresponding $1\sigma$ uncertainties, and $v=248.408~\mathrm{GeV}$ denotes the vacuum expectation value of the Higgs field. At the best-fit point, the free parameters in the Yukawa couplings of Eq.~\eqref{eq:para-T5T11} take the values
\begin{eqnarray}
&&a=1.486\times10^{-4} \,,~\quad b=6.736\times10^{-4} \,,~\quad
c=3.017\times10^{-4} \,,\nonumber\\
&&d=1.619\times10^{-2} \,,~\quad f=1.747\times10^{-3} \,,~\quad
a'=7.296\times10^{-6} \,,\nonumber\\
\label{eq:T5T11_input}&&b'=8.367\times10^{-4} \,,~\quad c'=3.450\times10^{-3} \,,~\quad
d'=9.680\times10^{-1} \,,~\quad \phi=0.392 \,.\label{eq:parameter-set}
\end{eqnarray}
The resulting quark masses and mixing parameters are in excellent agreement with the experimental data,
\begin{eqnarray}
&&m_u=1.245 ~~\text{MeV}\,,~~ m_c=0.6236 ~~\text{GeV}\,,~~ m_t=170.0 ~~\text{GeV} \,,\nonumber\\
&&m_d=2.723 ~~\text{MeV}\,,~~ m_s=54.45 ~~\text{MeV}\,,~~ m_b=2.863 ~~\text{GeV} \,,\nonumber\\
\sin\theta_{12}&&=0.22507 \,,~~ \sin\theta_{13}=0.003704 \,,~~ \sin\theta_{23}=0.041930 \,,~~ \delta_{CP}=1.1392 \,,\label{eq:result-observable}
\end{eqnarray}
with all the fitted observables lying very close to their respective central values.

\begin{figure}[htbp!]
\centering
\includegraphics[width=0.49\textwidth]{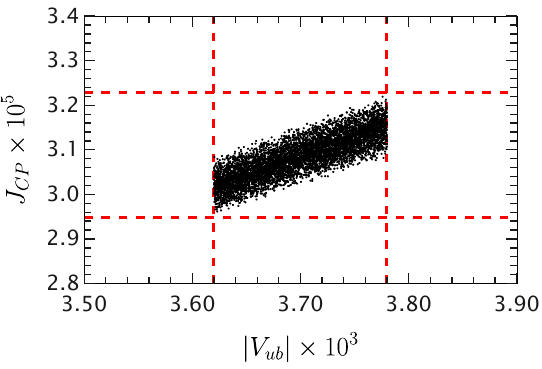}
\includegraphics[width=0.49\textwidth]{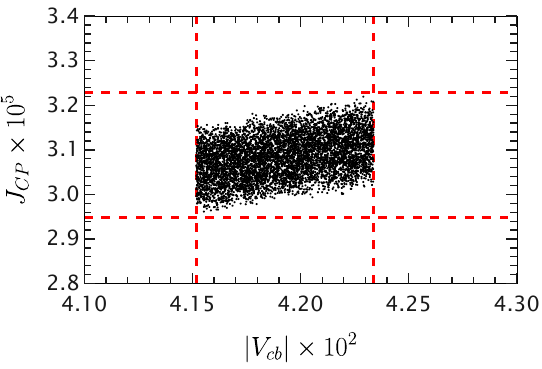}
\includegraphics[width=0.49\textwidth]{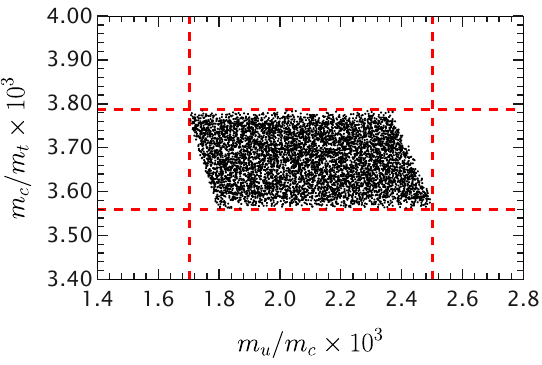}
\includegraphics[width=0.49\textwidth]{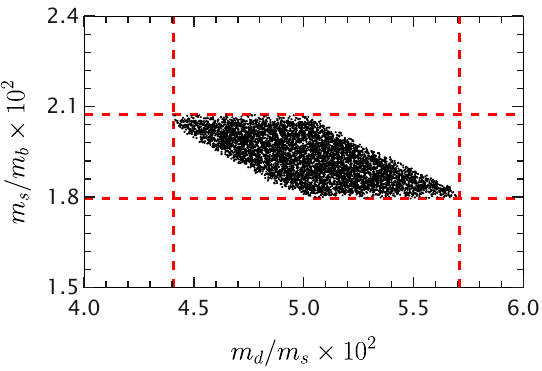}
\includegraphics[width=0.49\textwidth]{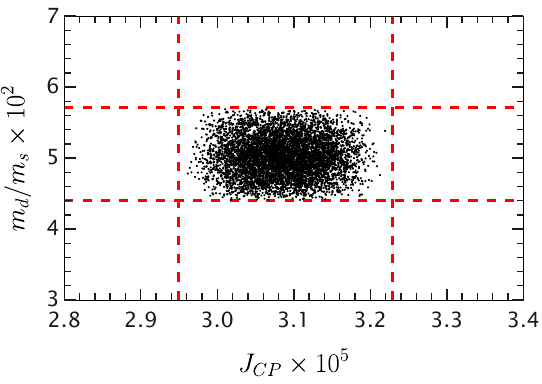}
\caption{Predicted correlations among quark-flavor observables for the texture $Y_u=T_5$, $Y_d=T_1^{(1)}$. The red dashed lines denote the ranges of the relevant CKM moduli, $J_{CP}$, and quark-mass ratios allowed by the current experimental constraints, as given in Eqs.~\eqref{eq:quarkconstraintmass} and~\eqref{eq:quarkconstraintckm}.}
\label{fig:T5T11}
\end{figure}

Furthermore, we scan the parameter space subject to the experimental constraints in Eqs.~\eqref{eq:quarkconstraintmass} and~\eqref{eq:quarkconstraintckm}. The allowed parameter points are shown in figure~\ref{fig:T5T11}, where several pronounced correlations among the observables can be clearly seen. These correlations originate from the texture zeros, which impose nontrivial algebraic relations among the Yukawa matrix elements and, consequently, among the physical quark masses and mixing parameters. The first two panels display clear positive correlations between $J_{CP}$ and $|V_{ub}|$, and between $J_{CP}$ and $|V_{cb}|$, respectively. Here,
$J_{CP}=\cos\theta_{12}\cos^2\theta_{13}\cos\theta_{23}\sin\theta_{12}\sin\theta_{13}\sin\theta_{23}\sin\delta_{CP}$
denotes the Jarlskog invariant, while $V_{ub}$ and $V_{cb}$ are the $(1,3)$ and $(2,3)$ elements of the CKM matrix, respectively. The middle two panels exhibit negative correlations between $m_u/m_c$ and $m_c/m_t$, and between $m_d/m_s$ and $m_s/m_b$. The last panel shows the correlation between $J_{CP}$ and $m_d/m_s$. Therefore, this texture not only provides a viable description of the current data but also predicts characteristic correlations among quark-mass ratios, CKM matrix elements, and CP violation. Moreover, it follows from Eq.~\eqref{eq:para-T5T11} that $\text{det}(Y_u Y_d)=acfa'c'd'$ is real and positive. This implies that the texture naturally yields a vanishing contribution to the strong CP phase at this level, and hence may provide a solution to the strong CP problem, as discussed in Appendix~\ref{app:strong-cp}.

\section{Conclusion}

In this work, we have investigated quark Yukawa texture zeros arising from the $\mathbb{Z}_3$ and $S_3$ gauging of $\mathbb{Z}_N\times\mathbb{Z}'_N$. We have found out the equivalence classes and derived their fusion rules, which determine whether a product of the quark and Higgs field classes contains the identity, or equivalently, whether a Yukawa coupling is allowed. The resulting zeros are therefore consequences of the symmetry-based selection rules, rather than assumptions imposed directly on the mass matrices. In addition, the equivalence class structure depends crucially on whether $N$ is divisible by three. For the $\mathbb{Z}_3$ gauging, when $3\nmid N$, only the identity element forms a one-element class and all nontrivial classes contain three elements, as can be seen from Eq.~\eqref{eq:class-Nneq3}. When $3\mid N$, there are three one-element classes and the remaining classes contain three elements, as shown in Eq.~\eqref{eq:class3N}. For the $S_3$ gauging, when $3\nmid N$, Eq.~\eqref{eq:s3class-Nneq3int} demonstrates that nontrivial classes contain six elements or three elements. When $3\mid N$, two additional nontrivial one-element classes also appear alongside the class containing the identity element as shown in Eq.~\eqref{eq:s3class-Neq3int}. Moreover, an $S_3$ class combines the two $\mathbb{Z}_3$ classes $[k,\ell]_{\mathbb{Z}_3}$ and $[-\ell,-k]_{\mathbb{Z}_3}$. Consequently, the $S_3$ fusion rules are generally less restrictive and can allow more Yukawa interactions than the corresponding $\mathbb{Z}_3$ fusion rules.

For $N=2,3,4,$ and $5$, we take account of equivalent field assignments up to row and column permutations and systematically enumerate the inequivalent assignments of the three generations of left- and right-handed quarks and of the Higgs field. This analysis yields a broad variety of full-rank Yukawa textures with zero entries which depend on both $N$ and the gauged automorphism subgroup. For $N=2$, the two discrete gaugings give the same set of full-rank textures. For $N=3$, the $S_3$ gauging admits additional one- and two-zero textures that cannot be realized by the $\mathbb{Z}_3$ gauging. For $N=4$, the two gaugings generate the same set of full-rank textures, as show in tables~\ref{tab:z3n4-yukawa-textures} and \ref{tab:s3n4-yukawa-textures}. When $N=5$, the $S_3$ gauging produces an additional two-zero texture than the $\mathbb{Z}_3$ gauging of $\mathbb{Z}_5\times\mathbb{Z}_5'$ symmetry which can be seen in tables~\ref{tab:z3n5-yukawa-textures} and \ref{tab:s3n5-yukawa-textures}.

We then considered the most restrictive texture pairs that are compatible with quark data. Our analysis in Appendix~\ref{app:10-zeros} shows that the full-rank  quark Yukawa matrices with totally ten texture-zeros fail to reproduce either a physical CP-violating phase or the observed quark mixing angles. Thus, for the nonsingular three-generation quark Yukawa matrices studied here, nine is the largest phenomenologically viable total number of zeros. The nonsingular nine-zero configurations fall into the $6+3$ and $5+4$ distributions between the up- and down-type sectors, together with the cases obtained by interchanging the two sectors. In section~\ref{sec:6+3tex}, we investigated the symmetry realizations of texture pairs with $6+3$ distribution. This pattern can not be generated from neither $\mathbb{Z}_3$ gauging nor $S_3$ of $\mathbb{Z}_N\times \mathbb{Z}_N'$ when $N=2,3$. Eqs.~\eqref{eq:udiag-dthree-z3N4-1}$\sim$\eqref{eq:udiag-dthree-S3N5-4} display the textures of this type generated by the $\mathbb{Z}_3$ gauging of the $\mathbb{Z}_4\times\mathbb{Z}'_4$ symmetry, the $\mathbb{Z}_3$ gauging of the $\mathbb{Z}_5\times\mathbb{Z}'_5$ symmetry and the corresponding $S_3$ gauging. In particular, we note that the $S_3$ gauging of the $\mathbb{Z}_5\times\mathbb{Z}'_5$ symmetry can generate all the four possible patterns of $6+3$ texture-zero distribution, as shown in Eqs.~\eqref{eq:udiag-dthree-S3N5-1}$\sim$\eqref{eq:udiag-dthree-S3N5-4}. Regarding the $5+4$ distribution, it cannot be realized through either of the two gaugings for $N=2,3$, but is realizable for $N=4,5$. The realizable textures, together with the corresponding symmetries and field assignments are summarized in tables~\ref{tab:N=4-5plus4} and~\ref{tab:N=5-5plus4}. Note that the $\mathbb{Z}_3$ gauging of $\mathbb{Z}_4\times\mathbb{Z}'_4$ or $\mathbb{Z}_5\times\mathbb{Z}'_5$ symmetry realizes the complete set of texture pairs with $5+4$ distribution, whereas the corresponding $S_3$ gauging realizes only a subset of them.

The phenomenological implications of $6+3$ distributions have already been numerically investigated in other studies~\cite{tanimoto:2016rqy}, with some of them shown to be viable. In the $5+4$ case, it is sufficient to only consider the five-zero matrix of $T_5$ in Eq.~\eqref{eq:fullrankfive0} by using the freedom of field redefinition. The four-zero matrix in the other sector may still be subjected to arbitrary column permutations, since these correspond to relabelings of the right-handed quark fields and do not change the physical observables. Furthermore, the weak-basis transformations relate the viable $5+4$ texture pairs, allowing their physical predictions to be studied through representative configurations which are parameterized in Eq.~\eqref{eq:para-T5T11}. We identify a set of parameter values in Eq.~\eqref{eq:parameter-set} that reproduces the experimental observables. We further scan the parameter space subject to the $1\sigma$ experimental constraints, revealing correlations among the quark-mass ratios, CKM matrix elements, and CP-violating observables. Improved determinations of quark masses and CKM parameters can therefore test these texture-zero structures.

The freedom of field redefinitions allows some of the nine-zero textures to be cast into the forms given in Eqs.~\eqref{eq:realdet-6zero}$\sim$\eqref{eq:realdet-4zero}, in which the quark mass-matrix determinants are real without eliminating the physical CP-violating phase in the CKM matrix. This feature suggests that non-invertible flavor symmetries may provide a new route toward an axionless solution to the strong CP problem. Establishing a complete solution, however, requires a dynamical ultraviolet realization of the texture, together with an analysis of radiative corrections and the stability of the determinant condition. It would also be worthwhile to extend the classification to larger values of $N$, additional Higgs fields, and the lepton sector. Such extensions may reveal a broader class of phenomenologically interesting textures.

Finally, although non-invertible symmetries provide a mechanism for generating texture zeros, they do not, by themselves, explain the magnitudes of the nonzero entries in the Yukawa matrices. As illustrated in Eq.~\eqref{eq:T5T11_input}, a pronounced hierarchy among these parameters is necessary to reproduce the observed quark mass spectrum. Explaining the origin of this hierarchy therefore requires an additional mechanism beyond the texture-zero structure considered here~\cite{Otsuka:2026fpy,Ding:2026sfm}. A detailed investigation of such a mechanism is left for future work.

\section*{Acknowledgements}

This work is supported by the National Natural Science Foundation of China under Grant Nos. 12375104, 12547106 and Guizhou Provincial Major Scientific and Technological Program XKBF (2025)010.

\clearpage

\begin{appendix}

\section{\label{app:10-zeros} Quark Yukawa coupling textures with a total of ten texture zeros }

There are two types of textures with ten zeros: five zeros in both the up- and down-type quark Yukawa couplings, or four zeros in one sector and six zeros in the other. Here and in what follows, we restrict our analysis to Yukawa matrices of full rank.

\subsection{5+5 texture zeros}

In this case, the rank-three quark Yukawa matrices, up to column permutations, can be obtained from Eq.~\eqref{eq:fullrankfive0} as follows:
\begin{eqnarray}
Y_{u,d}=\begin{pmatrix}
\times & \times & 0 \\
0 & \times & 0 \\
0 & 0 & \times
\end{pmatrix} \,,\,
\begin{pmatrix}
\times & 0 & 0 \\
\times & \times & 0 \\
0 & 0 & \times
\end{pmatrix} \,,\,
\begin{pmatrix}
\times & 0 & \times \\
0 & \times & 0 \\
0 & 0 & \times
\end{pmatrix} \,,\,
\begin{pmatrix}
\times & 0 & 0 \\
0 & \times & 0 \\
\times & 0 & \times
\end{pmatrix} \,,\,
\begin{pmatrix}
\times & 0 & 0 \\
0 & \times & \times \\
0 & 0 & \times
\end{pmatrix} \,,\,
\begin{pmatrix}
\times & 0 & 0 \\
0 & \times & 0 \\
0 & \times & \times
\end{pmatrix} \,,~~~~
\label{eq:A}
\end{eqnarray}
which gives
\begin{eqnarray}
Y_{u,d}Y_{u,d}^\dagger=\begin{pmatrix}
\times & \times & 0 \\
\times & \times & 0 \\
0 & 0 & \times
\end{pmatrix} \,,\,
\begin{pmatrix}
\times & 0 & \times \\
0 & \times & 0 \\
\times & 0 & \times
\end{pmatrix} \,,\,
\begin{pmatrix}
\times & 0 & 0 \\
0 & \times & \times \\
0 & \times & \times
\end{pmatrix} \,,
\label{eq:AAdagger}
\end{eqnarray}
where the diagonal elements are real, while the off-diagonal elements are generically complex. There are two possibilities for $Y_uY_u^\dagger$ and $Y_dY_d^\dagger$. If they share the same texture in Eq.~\eqref{eq:AAdagger}, two quark mixing angles vanish. If they correspond to two different textures, all phases can be absorbed into the left-handed quark doublets, rendering both matrices real and symmetric. Consequently, CP violation vanishes and one CKM matrix element is exactly zero. In either case, the scenario with five zeros in each quark sector is incompatible with experimental observations.

\subsection{6+4 texture zeros}

In the case with four zeros in the up-type (down-type) quark Yukawa couplings and six zeros in the down-type (up-type) quark Yukawa coupling the six-zero quark mass matrix can be brought into a real diagonal form by appropriate permutations and rephasings of the right-handed quark fields.

As shown in Eq.~\eqref{eq:fullrankfour0}, a full-rank quark mass matrix with four texture zeros admits four patterns, $T_1$, $T_2$, $T_3$ and $T_4$, up to row and column permutations. The textures $T_1$, $T_2$ and $T_3$ are incompatible with the observed CP violation because all phases of their non-zero entries can be eliminated by field redefinitions. The remaining texture $T_4$ is block diagonal and hence predicts two vanishing mixing angles in the CKM matrix. Hence, none of the four-zero textures is phenomenologically viable.

In summary, quark Yukawa matrices $Y_u$ and $Y_d$ with ten or more texture zeros are ruled out by experimental data. Consequently, the maximum number of texture zeros that can be accommodated is nine.

\section{Parameterization of $Y_u$ and $Y_d$ with nine texture-zeros and real determinant}\label{app:strong-cp}

The strong CP problem concerns the striking smallness of the effective QCD vacuum angle
$\bar{\theta}=\theta+\arg\det(Y_uY_d)$, where $\theta$ is the coefficient of $\alpha_s/(8\pi) G^{\mu\nu}\tilde{G}_{\mu\nu}$ , $G_{\mu\nu}$ is the field strength tensor of QCD, $\tilde{G}_{\mu\nu}$ its
dual,. Although $\bar{\theta}$ is naturally expected to be of order $\mathcal{O}(1)$, experimental bounds require $|\bar{\theta}|<10^{-10}$~\cite{Abel:2020pzs}. Explaining this enormous suppression is the essence of the strong CP problem. A simple starting point is to impose CP conservation at the fundamental level, which sets $\theta=0$. The observed CP violation in the quark sector, however, requires complex Yukawa matrices. Once CP is spontaneously broken, the resulting phases in the Yukawa couplings generally induce a nonzero $\arg\det(Y_uY_d)$ and hence shift $\bar{\theta}$ away from zero. A viable mechanism must therefore accommodate weak CP violation while preserving a real and positive determinant for $Y_uY_d$.

Motivated by the Nelson--Barr mechanism~\cite{Nelson:1983zb,Physrevlett.53.329}, one may use texture zeros in the Yukawa matrices to achieve this goal. A suitable pattern of zeros can render $\det(Y_uY_d)$ real independently of the complex phases in the nonzero Yukawa entries. It is then possible to satisfy $\arg\det(Y_uY_d)=0$ and hence $\bar{\theta}=0$, while retaining the phases required to reproduce the observed weak CP violation in the CKM matrix. This texture-zero strategy has been explored in several recent studies of the strong CP problem. Refs.~\cite{Barr:1996wx,Feruglio:2023uof,Bonnefoy:2025rvo,Kobayashi:2025rpx} considered three-zero textures in both the up- and down-type quark mass matrices. A three-zero up-type matrix combined with a four-zero down-type matrix, with both determinants real, was analyzed in Ref.~\cite{Antusch:2013rla}. Four-zero textures in both quark sectors were studied in Ref.~\cite{Hall:2024qqe}, while Ref.~\cite{Tanimoto:2025fnj} investigated a diagonal up-type matrix together with a down-type matrix containing two texture zeros.

In section~\ref{sec:pheno-analysis}, we have discussed the textures with six zeros and three zeros, as well as those with five zeros and four zeros, for $Y_u$ and $Y_d$, which were obtained from non-invertible symmetries.  Among these, a subset of these textures admits parameterizations with real determinants. In particular, the diagonal six-zero texture and the three-zero textures in Eq.~\eqref{eq:threezero1} can be parameterized as follows so that both determinants are real:
\begin{eqnarray}
\label{eq:realdet-6zero}M=\begin{pmatrix}
a' ~&~ 0 ~&~ 0 \\
0 ~&~ b' ~&~ 0 \\
0 ~&~ 0 ~&~ c'
\end{pmatrix}
\end{eqnarray}
with
\begin{eqnarray}
M_1&=&\begin{pmatrix}
0 ~&~ a ~&~ 0 \\
0 ~&~ b ~&~ c \\
d ~&~ fe^{-i\phi} ~&~ g
\end{pmatrix}\,,\quad
M_2=\begin{pmatrix}
0 ~&~ a ~&~ be^{-i\phi} \\
0 ~&~ 0 ~&~ c \\
d ~&~ f ~&~ g
\end{pmatrix}\,,\quad
M_4=\begin{pmatrix}
a ~&~ 0 ~&~ 0 \\
be^{-i\phi} ~&~ c ~&~ d \\
f ~&~ 0 ~&~ g
\end{pmatrix}\,,\nonumber\\
M_6&=&\begin{pmatrix}
a ~&~ 0 ~&~ b \\
ce^{-i\phi} ~&~ d ~&~ f \\
0 ~&~ 0 ~&~ g
\end{pmatrix}\,,\quad
M_8=\begin{pmatrix}
a ~&~ b ~&~ ce^{-i\phi} \\
0 ~&~ d ~&~ 0 \\
0 ~&~ f ~&~ g
\end{pmatrix}\,,\quad
M_9=\begin{pmatrix}
a ~&~ be^{-i\phi} ~&~ c \\
0 ~&~ d ~&~ f \\
0 ~&~ 0 ~&~ g
\end{pmatrix}\,,\label{eq:realdet-3zero}
\end{eqnarray}
where field redefinitions have been used to remove all unphysical phases, leaving only the physical phase. By contrast, the textures $M_3$, $M_5$, and $M_7$ do not admit parameterizations of this form with real determinants. Several of these textures have recently been considered in connection with the strong CP problem. For example, Ref.~\cite{Petcov:2024vph} studied a diagonal up-type quark mass matrix together with a down-type matrix of the $M_1$ form, up to an interchange of the second and third columns. This construction yields both a nonzero weak CP-violating phase and real determinant of quark mass matrices. Refs.~\cite{Liang:2024wbb,Liang:2025dkm,Kobayashi:2025thd} employed an $M_4$ texture for the down-type matrix and a diagonal up-type matrix, and investigated the dynamical origin of these textures.

The five-zero texture $T_5$ and the four-zero textures in Eqs.~\eqref{eq:T1}, \eqref{eq:T2}, \eqref{eq:T3123} and \eqref{eq:T3456} likewise admit parameterizations with real determinants as follows,
\begin{eqnarray}
    \label{eq:realdet-5zero}T_5=\begin{pmatrix}
        a' ~&~ b'e^{-i\phi} ~&~ 0 \\
        0 ~&~ c' ~&~ 0 \\
        0 ~&~ 0 ~&~ d'
    \end{pmatrix}
\end{eqnarray}
and
\begin{eqnarray}
    &&T_1^{(1)}=\begin{pmatrix}
        a ~&~ 0 ~&~ 0 \\
        b ~&~ c ~&~ 0 \\
        d ~&~ 0 ~&~ f
    \end{pmatrix}\,,\quad
    T_1^{(2)}=\begin{pmatrix}
        a ~&~ b ~&~ 0 \\
        0 ~&~ c ~&~ 0 \\
        0 ~&~ d ~&~ f
    \end{pmatrix} \,,\quad
    T_1^{(3)}=\begin{pmatrix}
        a ~&~ 0 ~&~ b \\
        0 ~&~ c ~&~ d \\
        0 ~&~ 0 ~&~ f
    \end{pmatrix} \,,\nonumber\\
    &&T_2^{(1)}=\begin{pmatrix}
        a ~&~ 0 ~&~ 0 \\
        0 ~&~ b ~&~ 0 \\
        c ~&~ d ~&~ f
    \end{pmatrix} \,,\quad
    T_2^{(2)}=\begin{pmatrix}
        a ~&~ 0 ~&~ 0 \\
        b ~&~ c ~&~ d \\
        0 ~&~ 0 ~&~ f
    \end{pmatrix} \,,\quad
    T_2^{(3)}=\begin{pmatrix}
        a ~&~ b ~&~ c \\
        0 ~&~ d ~&~ 0 \\
        0 ~&~ 0 ~&~ f
    \end{pmatrix} \,,\nonumber\\
    &&T_3^{(1)}=\begin{pmatrix}
        a ~&~ 0 ~&~ 0 \\
        b ~&~ c ~&~ 0 \\
        0 ~&~ d ~&~ f
    \end{pmatrix} \,,\quad
    T_3^{(2)}=\begin{pmatrix}
        a ~&~ 0 ~&~ 0 \\
        0 ~&~ b ~&~ c \\
        d ~&~ 0 ~&~ f
    \end{pmatrix} \,,\quad
    T_3^{(3)}=\begin{pmatrix}
        a ~&~ b ~&~ 0 \\
        0 ~&~ c ~&~ 0 \\
        d ~&~ 0 ~&~ f
    \end{pmatrix} \,,\nonumber\\
    &&T_3^{(4)}=\begin{pmatrix}
        a ~&~ 0 ~&~ b \\
        0 ~&~ c ~&~ 0 \\
        0 ~&~ d ~&~ f
    \end{pmatrix} \,,\quad
    T_3^{(5)}=\begin{pmatrix}
        a ~&~ b ~&~ 0 \\
        0 ~&~ c ~&~ d \\
        0 ~&~ 0 ~&~ f
    \end{pmatrix} \,,\quad
    T_3^{(6)}=\begin{pmatrix}
        a ~&~ 0 ~&~ b \\
        c ~&~ d ~&~ 0 \\
        0 ~&~ 0 ~&~ f
    \end{pmatrix} \,.\label{eq:realdet-4zero}
\end{eqnarray}

\end{appendix}

\clearpage


\providecommand{\href}[2]{#2}\begingroup\raggedright\endgroup

\end{document}